\documentclass{revtex4-2}
\usepackage{lineno}
\usepackage[utf8]{inputenc}
\usepackage{amsbsy}
\usepackage{amsopn}
\usepackage{amstext}
\usepackage{amsmath,amsthm,amsfonts,amssymb}
\usepackage[mathcal]{eucal}
\usepackage{mathrsfs}
\usepackage{braket}
\usepackage[english]{babel}
\usepackage{color}
\usepackage{esint}
\usepackage{booktabs}
\usepackage{subfigure}
\usepackage{hologo}
\usepackage{graphicx}
\usepackage{float}
\usepackage{units}
\usepackage{textcomp}
\usepackage{orcidlink}

\DeclareGraphicsExtensions{.png,PNG,.pdf,.PDF}
\usepackage{hyperref}
\usepackage{slashed}
\newcommand{\ie}{\begin{equation}}
\newcommand{\fe}{\end{equation}}
\newcommand{\se}{\begin{eqnarray}}
\newcommand{\ff}{\end{eqnarray}}
\begin{document}
\title{Bound-state solutions for the charged Dirac oscillator in a rotating frame in the Bonnor-Melvin-Lambda spacetime}
\author{R. R. S. Oliveira\,\orcidlink{0000-0002-6346-0720}}
\email{rubensrso@fisica.ufc.br}
\affiliation{Departamento de F\'isica, Universidade Federal da Para\'iba, Caixa Postal 5008, 58051-900, João Pessoa, PB, Brazil}

%%%%%%%%%%%%%%%%%%%%%%%%%%%%%%%%%%%%%%%%%%%%%%%%%%%%%%%%%%%%%%%%%%%%%%

\date{\today}

\begin{abstract}

In this paper, we determine the relativistic bound-state solutions for the charged (DO) Dirac oscillator in a rotating frame in the Bonnor-Melvin-Lambda spacetime in $(2+1)$-dimensions, where such solutions are given by the two-component normalizable Dirac spinor and by the relativistic energy spectrum. To analytically solve our problem, we consider two approximations, where the first is that the cosmological constant is very small (conical approximation), and the second is that the linear velocity of the rotating frame is much less than the speed of light (slow rotation regime). After solving a second-order differential equation, we obtain a generalized Laguerre equation, whose solutions are the generalized Laguerre polynomials. Consequently, we obtain the energy spectrum, which is quantized in terms of the radial and total magnetic quantum numbers $n$ and $m_j$, and depends on the angular frequency $\omega$ (describes the DO), cyclotron frequency $\omega_c$ (describes the external magnetic field), angular velocity $\Omega$ (describes the rotating frame), spin parameter $s$ (describes the ``spin''), spinorial parameter $u$ (describes the components of the spinor), effective rest mass $m_{eff}$ (describes the rest mass modified by the spin-rotation coupling), and on a real parameter $\sigma$ and cosmological constant $\Lambda$ (describes the Bonnor-Melvin-Lambda spacetime). In particular, we note that this spectrum is asymmetrical (due to $\Omega$) and has its degeneracy broken (due to $\sigma$ and $\Lambda$). Besides, we also graphically analyze the behavior of the spectrum and of the probability density as a function of the parameters of the system for different values of $n$ and $m_j$.

\end{abstract}

\maketitle

%%%%%%%%%%%%%%%%%%%%%%%%%%%%%%%%%%%%%%%%%%%%%%%%%%%%%%%%%%%%%%%%%%%%%%%%%%%%%%%%%%
\section{Introduction}

In 1989, Moshinsky and Szczepaniak developed the first consistent relativistic version of the well-known quantum harmonic oscillator (QHO) for spin-1/2 particles (or Dirac fermions), which became known as the Dirac oscillator (DO) \cite{Moshinsky}. Basically, the DO is a model of QHO for spin-1/2 fermions in a high-energy regime. To build the DO, it needs to insert in the free Dirac equation (DE) a nonminimal coupling/substitution given by: $\Vec{p}\to\Vec{p}-im_0\omega\beta\Vec{r}$, where $\Vec{p}$ is the linear momentum operator/vector, $i=\sqrt{-1}$ is the imaginary unit, $m_0>0$ is the rest mass of the oscillator with angular frequency of oscillation $\omega>0$, $\Vec{r}$ is the position operator/vector, and $\beta$ is one of the usual Dirac matrices (i.e. the Hamiltonian of the DO is linear in both $\Vec{p}$ and $\Vec{r}$) \cite{Moshinsky,Bentez,Romero}. Consequently, in the nonrelativistic limit (or low-energy regime), the DO reduces to the QHO (whose Hamiltonian is quadratic in both $\Vec{p}$ and $\Vec{r}$) with a strong spin-orbit coupling \cite{Moshinsky}. Thus, in addition to being an exactly solvable model introduced in the context of many-particle models in relativistic quantum mechanics (RQM), the DO can be interpreted as an interaction of the anomalous magnetic moment of a Dirac neutral fermion with an electric field generated by a uniformly charged dielectric sphere \cite{Bentez,Romero}. In 2013, the DO was verified experimentally by Franco-Villafañe et al. \cite{Villafane}, where the experimental setup was based on a microwave system consisting of a chain of coupled dielectric disks. Since it was introduced into the literature, the DO has been studied in different areas of physics, such as in thermodynamics \cite{Frassino,Oli5}, physics-mathematics \cite{Bentez,Villalba1,Villalba2,Andrade}, nuclear physics \cite{Grineviciute}, quantum optics \cite{Bermudez2}, and graphene physics \cite{Quimbay,Belouad}. Besides, the DO has already been studied in connection with the Aharonov-Bohm-Coulomb system \cite{Oli6}, quantum phase transitions \cite{Bermudez3}, nuclear structure calculations \cite{Yang}, spin and pseudo-spin symmetries \cite{deOliveira}, noncommutative geometry \cite{Oli7}, Aharonov-Casher effect \cite{Oli8}, 2D quantum ring \cite{Bakke1}, and under the influence of noninertial effects (with and without cosmic strings) \cite{Strange,Bakke2,Oli9,Oli10}. 

In nonrelativistic quantum mechanics (NRQM) \cite{Bhuiyan,Li1999,Wakamatsu,Ribeiro,Muniz} as well as in RQM \cite{Haldane,Schakel,Miransky,Bermudez}, the effects generated by external magnetic fields (in special, by uniform magnetic fields) are without a doubt something very important for the physical description of the system under study, where one of the effects generated are the well-known Landau levels (LLs) \cite{Landau}. In particular, these energy levels are a consequence of the so-called Landau quantization, which states that the energies of cyclotron orbits of charged particles in a uniform magnetic field are quantized \cite{Landau}. Recently, the LLs have been investigated in the noncommutative quantum Hall effect in different relativistic scenarios \cite{Oli2,Oli3}, thermodynamics \cite{Oli4}, rainbow gravity \cite{rainbowgravity}, quantum chromodynamics (QCD) \cite{Bruckmann}, quantum electrodynamic (QED) \cite{Jentschura}, twisted photons \cite{Karlovets,Pavlov}, photonic crystals \cite{Barczyk}, free-electron vortex \cite{Schattschneider,Bliokh,Lloyd}, etc. In addition, magnetic fields also play an important role on large scales (galactic/cosmological scales) as well as in the early universe and in many astrophysical phenomena \cite{Zeldovich,Giovannini,Neronov,Vachaspati,Grasso,Harrison,Turner,Brandenburg,Kronberg,Pshirkov,Widrow,Beck}. However, as they usually occur in the vicinity of compact massive objects or in strong gravitational fields (or curved spacetimes), it is interesting to study them in the context of general relativity (GR), i.e. through a metric carrying a magnetic field (or something like that). In this way, a model of magnetic/magnetized universe emerged, the so-called Bonnor-Melvin magnetic universe (or Bonnor-Melvin spacetime), which is one exact solution of the Einstein-Maxwell equations that describes a static and cylindrically symmetric inhomogeneous magnetic field immersed in its own gravitational field \cite{Bonnor,Melvin}. In particular, this model is one possible analogy with the classical homogeneous magnetic field. On the other hand, as the magnetic field contributes explicitly to the energy-momentum tensor, it implies that it needs to decrease away from the axis so as not to collapse on itself and, consequently, the scalar invariant $F_{\mu\nu}F^{\mu\nu}$ is thus not constant (unlike in the classical case) and the field is not homogeneous.

In 2012, a generalization was proposed for the Bonnor-Melvin spacetime in the presence of the cosmological
constant ($\Lambda\neq 0$), which was considered an uncharged metric without any curvature singularity (but completely regular) via a solution of Einstein equations with a (negative) cosmological constant, where such solution is supported by a potential in the form $A_\mu=\left(0,0,0,\pm\frac{2B}{1+\rho^2/4}\right)$ \cite{Astorino}. However, some recent works/papers in RQM \cite{Barbosa,Castro,Guvendi,Ahmed,Ahmed2,ARXIV} have been done based on another generalization, which was made by {\v{Z}}ofka in 2019, and also has considered a nonvanishing cosmological constant (but positive, i.e. $\Lambda>0$) \cite{Zofka}. In this case, the generalized Bonnor-Melvin universe became known as the Bonnor-Melvin-Lambda spacetime or Bonnor-Melvin-$\Lambda$ universe \cite{Barbosa,Castro,Guvendi,Ahmed,Ahmed2,ARXIV}. So, such a background maintains its symmetrically cylindrical and static nature but, unlike the original solution, it truly represents a homogeneous magnetic field \cite{Zofka}. In addition, this background also admits a deficit angle due to the presence of a real parameter $\sigma$ in your line element (or metric) and, therefore, forming the spacetime’s only singularity \cite{Zofka} (i.e. a conical or cone-type curvature/singularity such as in \cite{Oli2,Oli3,Oli8,Bakke2,Oli9,Oli10}). It is also interesting to mention here that, although Refs. \cite{Barbosa,Castro,Guvendi,Ahmed,Ahmed2,ARXIV} worked with relativistic particles in the Bonnor-Melvin-Lambda spacetime, only Ref. \cite{Castro} worked (for a charged scalar boson) in the presence of external magnetic fields (or with the magnetic minimal coupling), where one, inclusive, is the magnetic field itself responsible by the LLs (i.e. $\Vec{B}=B_0 \vec{e}_z$). However, all Refs. \cite{Barbosa,Castro,Guvendi,Ahmed,Ahmed2,ARXIV} have one thing in common (a ``limitation''), which is the fact that they worked only in an inertial frame of reference, i.e. free of the noninertial effects generated by rotation or acceleration (or free from Coriolis, centrifugal, and Euler forces, respectively).

The present paper has as its objective to determine the relativistic bound-state solutions for the charged DO in a rotating frame in the Bonnor-Melvin-Lambda spacetime in $(2+1)$-dimensions (i.e. we adopt a purely planar dynamic). In particular, such solutions (which are based on a totally theoretical approach) are given by the two-component Dirac spinor (which are our normalized eigenfunctions/eigenspinors or spinorial wave functions) and by the relativistic energy spectrum (which are our relativistic energy eigenvalues, bound state energies, or relativistic LLs). Besides, with the normalized spinor in hand, we were also able to determine another important result (often neglected in the literature), which is the radial probability density (or ``probability amplitude''). For this case, the graphs were analyzed for the ground state with different values of an angular quantum number and of the parameters of the system, whereas in the case of energies, another quantum number was considered, called the radial quantum number. So, to achieve our objective, we work with the curved DO with minimal coupling (whose coupling constant is the well-known electric charge) in relativistic polar coordinates $(t, r,\theta)$, where the formalism used to write the DO in a curved spacetime was the tetrad(s) formalism of GR. In particular, this formalism is considered an excellent tool for studying Dirac fermions in curved spacetimes because it allows introducing the curved spacetime point-by-point with a flat spacetime, and also appears in the Einstein–Cartan formulation of GR.

To include a rotating frame $S'$ in our problem, we apply a (passive) rotation on the polar coordinate system in the form: $\theta\to\theta+\Omega t$ (this means that the plane rotates by an angle $\theta'=\Omega t$ around its origin) \cite{Oli2,Strange,Bakke2,Oli9,Oli10}, where $0\leq\theta\leq 2\pi$ is the angular coordinate or polar angle, $-\infty\leq t\leq +\infty$ is the temporal coordinate, and $\Omega\geq 0$ (positive/counterclockwise rotation) is the constant angular velocity or rotational velocity (in rad/s) of the rotating frame. So, to analytically solve our problem (i.e. easily solve the second-order differential equation of our problem), we consider two approximations/simplifications: the first is that the cosmological constant is (very) small, i.e. we adopt a conical approximation \cite{Castro}, and the second is that the tangential/linear velocity of the rotating frame is much less than the speed of light, i.e. we adopted a (very) slow/weak rotation regime (or sufficiently slow rotation) \cite{Strange,Oli2,Oli8,Bakke2,Oli9,Oli10}. Last but not least, we also consider the spin effects of the planar fermion (or planar DO), which was introduced in the literature by Hagen in the early 90s \cite{Hagen1,Hagen2,Blum,Hagen3,Hagen4,Hagen5,Hagen6} and is described by a real parameter $s$ (twice the spin value), called the spin parameter (should not be confused with the spin magnetic quantum number $m_s=\pm 1/2$), whose values are $s=+1$, for spin ``up  ($\uparrow$)'', and $s=-1$, for spin ``down ($\downarrow$)'', respectively (other papers that worked with this parameter were Refs. \cite{Oli2,Oli3,Oli9,Oli10,Villalba1,Andrade,Andrade2,CQG,Park,Gavrilov}). Therefore, here we can say that we investigate the relativistic bound state solutions (approximate analytical solutions) for the charged DO under the influence of noninertial effects (due to a rotating frame), magnetic effects (due to an external magnetic field), and spin effects (due to the parameter $s$) in the Bonnor-Melvin-Lambda spacetime modeled by a real parameter $\sigma$ and by the cosmological constant $\Lambda$ \cite{Zofka}.

The structure of this paper is organized as follows. In Sect. \ref{sec2}, we introduce the charged DO with minimal coupling in a rotating frame in the $(2+1)$-dimensional Bonnor-Melvin-Lambda spacetime, where we use the tetrad formalism. To analytically solve our problem (i.e. simplify the calculations), we consider two approximations, where the first is that the cosmological constant is very small (i.e. we adopt a conical approximation), and the second is that the linear velocity of the rotating frame is much less than the speed of light (i.e. we adopted a slow rotation regime), respectively. Once this is done, we then define a stationary ansatz for the spinor, where we obtain two coupled first-order differential equations. So, by substituting one equation into
the other and vice-versa (decoupling the equations), we obtain a second-order differential equation for each component of the spinor. In particular, these differential equations depend on a quantity given by $Ms$ ($Ms>0$ and $Ms<0$), being $M$ an effective total magnetic quantum number (whose origin comes from the angular part of the spinor) and $s$ is the spin parameter. In Sect. \ref{sec3}, we solve analytically the differential equations via a change of variable and the asymptotic behavior. Consequently, we obtain from this a generalized Laguerre equation (whose solutions are the generalized Laguerre polynomials) as well as the relativistic energy spectrum for the DO. Besides, we discuss in detail various aspects/features of this spectrum and also analyze it for $Ms>0$ and $Ms<0$. Subsequently, we graphically analyze the behavior of the spectrum as a function of the magnetic field $B_0$, angular frequency $\omega$, angular velocity $\Omega$, and cosmological constant $\Lambda$ for three different values of the quantum numbers $n$ and $m_j$. In Sect. \ref{sec4}, we graphically analyze the behavior of the probability density (as a function of the radial distance) for six different values of $m_j$, $B_0$, $\omega$, $\Omega$, and $\Lambda$, with $n=0$ (ground state). In Sect. \ref{sec5}, we present our conclusions. Here, we use the system of natural units $(\hslash=c=G=1)$, the spacetime with signature $(+,-,-)$, and the Wolfram Mathematica software to perform most calculations.

%-------------------------------------------------------------------
\section{The charged Dirac oscillator in a rotating frame in the $(2+1)$-dimensional Bonnor-Melvin-Lambda spacetime \label{sec2}}

The $(2+1)$-dimensional DO with minimal coupling (or interacting with an external electromagnetic field) in an arbitrary curved spacetime is given by the following expression (in polar coordinates) \cite{Oli2,Oli3,Bakke2,Oli9,Oli10,CQG,Lawrie,Greiner,Strangebook}
\begin{equation}\label{dirac1}
i\gamma^\mu(x)\left(\bar{\nabla}_\mu(x)+m_0\omega r\gamma^0\delta^r_\mu\right)\psi(t,r,\theta)=m_0\psi(t,r,\theta), \ \ (\mu=t,r,\theta),
\end{equation}
where $\gamma^{\mu}(x)=e^\mu_{\ a}(x)\gamma^a$ are the curved gamma matrices, $e^\mu_{\ a}(x)$ are the tetrads (also called tetrad field or vielbeins), being $\gamma^a=(\gamma^0,\gamma^i)=(\gamma^0,\vec{\gamma})$ ($a,b=0,1,2$) the usual or flat gamma matrices (in Cartesian coordinates), with $\beta=\gamma^0$ and $\vec{\alpha}= \beta\vec{\gamma}$ being the usual Dirac matrices (appeared before gamma matrices), $\bar{\nabla}_\mu(x)=\nabla_\mu(x)+iqA_\mu(x)$ is the generalized covariant derivative or curved minimal coupling (i.e. a generalization of the flat covariant derivative or flat minimal coupling given by $D_a=\partial_a+iqA_a$), being $\nabla_\mu (x)=\partial_\mu+\Gamma_\mu(x)$ the tetrad covariant derivative (or simply covariant derivative), $\partial_\mu=\frac{\partial}{\partial x^\mu}=(\partial_t,\partial_r,\partial_\theta)=\left(\frac{\partial}{\partial t},\frac{\partial}{\partial r},\frac{\partial}{\partial \theta}\right)$ are the usual partial derivatives (with $\partial_\mu\neq e^a_{\ \mu}(x)\partial_a$), $\Gamma_\mu(x)=-\frac{i}{4}\omega_{ab\mu}(x)\sigma^{ab}=(\Gamma_t(x),\Gamma_r(x),\Gamma_\theta(x))$ is the spinorial connection (or spinor affine connection), being $\omega_{ab\mu}(x)=-\omega_{ba\mu}(x)$ the spin connection (an antisymmetric tensor at indices $a$ and $b$), $\sigma^{ab}=\frac{i}{2}[\gamma^a,\gamma^b]=i\gamma^a\gamma^b=-i\gamma^b\gamma^a$ ($a\neq b$) is a flat antisymmetric tensor (``Dirac tensor of rank 2''), $A_\mu(x)=e^b_{\ \mu}(x)A_b=A_\theta (x)\delta^\theta_\mu$ is the curved electromagnetic potential/field (or curved gauge field), $e^b_{\ \mu}(x)$ are inverse tetrads ($e^b_{\ \mu}(x)=(e^\mu_{\ b}(x))^{-1}$), $A_b=(0,0,A_2)=(0,0,-A_\theta)$ is the flat electromagnetic potential/flat (or flat gauge field), that is, $A_b$ the vector potential defined in the inertial frame of the observer (where $\Omega=0$), $\psi=e^{\frac{i\theta\Sigma_3}{2}}\Psi$ is the two-component curvilinear or polar spinor, being $\Psi=\Psi_D$ the original Dirac spinor (i.e. $\psi$ and $\Psi$ are two-element column matrices, and must satisfy: $\psi(\theta\pm 2\pi)=-\psi(\theta)$ and $\Psi(\theta\pm 2\pi)=\Psi(\theta)$ \cite{Villalba1,Villalba2,Schluter}), and the constants $m_0>0$ and $q=\pm e$ ($e>0$) are the rest mass (or mass term) and the electric charge of the DO, respectively. Besides, here we use the Latin indices $(a, b, c, \ldots)$ to label the local coordinates system (local reference frame, locally inertial frame, or the inertial Minkowski spacetime) and the Greek indices $(\mu, \nu, \lambda, \ldots)$ to label the general coordinates system (general reference frame or the noninertial curved spacetime).

Explicitly, we can rewrite Eq. \eqref{dirac1} as follows
\begin{equation}\label{dirac2}
\left\{i\gamma^t(x)\partial_t+i\gamma^r(x)(\partial_r+m_0\omega r\gamma^0)+i\gamma^\theta(x)(\partial_\theta-ieA_\theta (x))+i[\gamma^t(x)\Gamma_t(x)+\gamma^{r}(x)\Gamma_{r}(x)+\gamma^{\theta}(x)\Gamma_{\theta}(x)]\right\}\psi=m_0\psi,
\end{equation}
where we consider a negative electric charge given by $q=-e$, being $e>0$ the elementary charge.

Now, we need to build the line element for the DO in a rotating frame in the $(2+1)$-dimensional Bonnor-Melvin-Lambda spacetime ($dz^2=0$). Once this is done, we can then focus our attention on the form of the metric, tetrads, curved gamma matrices, vector potential, and spinorial and spin connections. Therefore, according to Refs. \cite{Zofka,Barbosa,Castro,Guvendi,Ahmed,Ahmed2}, the line element (or spacetime interval) for the $(2+1)$-dimensional Bonnor-Melvin-Lambda spacetime with signature $(+,-,-)$ is given by
\begin{equation}\label{lineelement}
ds^2=dt^2-dr^2-\sigma^2\sin^2\left(\sqrt{2\Lambda}r\right)d\theta^2,
\end{equation}
where $0\leq r<\infty$ ($r=\sqrt{x^2+y^2}$) is the polar radial coordinate (or polar radial distance), $\sigma>0$ is a constant of integration or a real parameter (with a dimension of length$^2$), and $\Lambda>0$ is the cosmological constant (with a dimension of length$^{-2}$), in which both are related with the following (dimensionless periodic) magnetic field (``Bonnor-Melvin-Lambda magnetic field'') aligned with the axis of symmetry
\begin{equation}\label{magneticfield}
H(r)=\sqrt{\Lambda}\sigma\sin\left(\sqrt{2\Lambda}r\right),
\end{equation}
being the Ricci scalar curvature corresponding given by $R=4\Lambda$ (i.e. a positive or ``sphere-like'' curvature) \cite{Castro,Guvendi}. In particular, this result demonstrates that the spacetime described by the line element \eqref{lineelement} is not asymptotically flat, but exhibits a uniform curvature throughout the universe (with a closed and finite geometry since it has a positive curvature). Here, it is interesting to comment a little on the cosmological constant. So, the cosmological constant is a term introduced by Einstein in his equations of GR to explain a static universe. Although Einstein originally added it to maintain a static universe model, which was later shown to be unnecessary with the discovery of an expanding universe, the concept has become crucial in modern cosmology. For example, a positive cosmological constant directly implies a repulsive force acting on cosmic scales, where this repulsion leads to what we observe as dark energy, which drives the accelerated expansion of the universe (i.e. dark energy is associated with a positive cosmological constant in the most widely accepted cosmological model today, which is the $\Lambda$CDM model). For more information on a positive cosmological constant, we recommend Refs. \cite{Wald,Perlmutter,Zehavi,Sahni,Smolin,Padmanabhan,Steinhardt,Ashtekar,Boonserm}.

On the other hand, to include the noninertial effects of a uniformly rotating frame, we must modify the angular coordinate in the form $\theta\to\theta+\Omega t$ \cite{Oli2,Strange,Bakke2,Oli9,Oli10}. Thus, the line element \eqref{lineelement} becomes (``noninertial curved line element'')
\begin{equation}\label{lineelement2}
ds^2=\left(1-V^2\right)dt^2-2V\sigma\sin\left(\sqrt{2\Lambda}r\right)dtd\theta-dr^2-\sigma^2\sin^2\left(\sqrt{2\Lambda}r\right)d\theta^2,
\end{equation}
or yet (both forms are useful here)
\begin{equation}\label{lineelement3}
ds^2=\left(1-V^2\right)\left(dt-\frac{V\sigma\sin\left(\sqrt{2\Lambda}r\right)}{\left(1-V^2\right)}d\theta\right)^2-dr^2-\frac{\sigma^2\sin^2\left(\sqrt{2\Lambda}r\right)}{\left(1-V^2\right)}d\theta^2,
\end{equation}
where $V$ is a dimensionless parameter defined as $V\equiv\sigma\Omega\sin\left(\sqrt{2\Lambda}r\right)$. Before we move forward, this parameter deserves a little discussion (similar to what was done in \cite{Oli2}). So, in SI units (restoring the speed of light $c$), this parameter is written as: $V=\frac{\sigma\Omega\sin\left(\sqrt{2\Lambda}r\right)}{c}=\frac{v}{c}$, that is the ratio between the (linear) velocity of the rotating frame (given by $v$) and the of light (for this reason it is dimensionless), and must obligatorily satisfy the causality condition, which is $V<1$. In particular, this condition means that the DO is inside the light cone, i.e. that the velocity of the rotating frame is less than the speed of light (as it should be). Consequently, this condition implies in a restriction (or natural limitation) for the range of the sine function, given by: $0\leq\sin\left(\sqrt{2\Lambda}r\right)<\Lambda_0$, where $\Lambda_0\equiv\frac{1}{\sigma\Omega}$, or $\Lambda_0\equiv \frac{c}{\sigma\Omega}$ (in SI units). As we will see shortly, this will lead to a new range for the radial coordinate $r$.

However, starting from the fact that any line element (of a curved spacetime) can be written as $ds^2=g_{\mu\nu}(x)dx^\mu dx^\nu$, where $g_{\mu\nu}(x)$ is the curved metric tensor (or curved metric), being $g^{\mu\nu}(x)$ its inverse (inverse curved metric), implies that such metrics takes the following form (for this, we use \eqref{lineelement2})
\begin{equation}\label{metric1}
g_{\mu\nu}(x)=\left(\begin{array}{ccc}
1-V^2 & \ 0 & -V\sigma\sin\left(\sqrt{2\Lambda}r\right) \\ 0 & -1 &  0 \\ -V\sigma\sin\left(\sqrt{2\Lambda}r\right) & \ 0 & -\sigma^2\sin^2\left(\sqrt{2\Lambda}r\right)
\end{array}\right), 
\ \
g^{\mu\nu}(x)=\left(\begin{array}{ccc}
1 & \ 0 & -\Omega \\ 0 & -1 &  0 \\ -\Omega & \ 0 & -\frac{1-V^2}{\sigma^2\sin^2\left(\sqrt{2\Lambda}r\right)}
\end{array}\right).
\end{equation}

Thus, with the line element well-defined (or metrics well-defined), we now need to build a local reference frame where the observer will be placed (i.e. the inertial laboratory frame). In particular, it is in this frame that we can define/write the gamma matrices (or the spinor) in a curved spacetime \cite{Oli2,Bakke2,Oli9,Oli10,CQG}. In that way, we can use the tetrad formalism to achieve this objective. According to this formalism, a local reference frame can be built through a noncoordinate basis given by $\hat{\theta}^a=(\hat{\theta}^0,\hat{\theta}^1,\hat{\theta}^2)$ (also called dual basis vectors \cite{Panahi} or basis one-forms \cite{Strange}), which is written in terms of inverse tetrads as $\hat{\theta}^a=e^a_{\ \mu}(x)dx^\mu$, where it must also satisfy the relation: $dx^\mu=e^\mu_{\ a}(x)\hat{\theta}^a$. Therefore, to (easily) determine the form of $e^\mu_{\ a}(x)$ and $e^a_{\ \mu}(x)$, one only needs to find $\hat{\theta}^a$, since $dx^\mu=(dt,dr,d\theta)$. However, before finding $\hat{\theta}^a$, it is important to briefly discuss some properties of the tetrads and their inverses. So, the tetrads and their inverses must satisfy the following relations \cite{Oli2,Oli9,Oli10,CQG,Lawrie,Panahi}
\begin{eqnarray}\label{metric2}
&& g_{\mu\nu}(x)=e^a_{\ \mu}(x)e^b_{\ \nu}(x)\eta_{ab},
\nonumber\\
&& g^{\mu\nu}(x)=e^\mu_{\ a}(x)e^\nu_{\ b}(x)\eta^{ab},
\nonumber\\
&& g^{\mu\sigma}(x)g_{\nu\sigma}(x)=\delta^\mu_{\ \nu}=e^a_{\ \nu}(x)e^\mu_{\ a}(x),
\end{eqnarray}
where $\eta_{ab}=\eta^{ab}=$diag$(+1,-1,-1)$ is the Cartesian Minkowski metric tensor (or flat metric), which must satisfy
\begin{eqnarray}\label{metric3}
&& \eta_{ab}=e^\mu_{\ a}(x)e^\nu_{\ b}(x)g_{\mu\nu}(x),
\nonumber\\
&& \eta^{ab}=e^a_{\ \mu}(x)e^b_{\ \nu}(x)g^{\mu\nu}(x),
\nonumber\\
&& \eta^{ac}\eta_{cb}=\delta^a_{\ b}=e^a_{\ \mu}(x)e^\mu_{\ b}(x),
\end{eqnarray}
with $\delta^a_b=\delta^\mu_\nu=$diag$(+1,+1,+1)$ being the $(2+1)$-dimensional Kronecker delta (i.e. the $3\times 3$ identity matrix). Furthermore, we have from this the anticommutation relations of the Clifford algebra for the curved and flat gamma matrices, given as: $\{\gamma^\mu(x),\gamma^\nu(x)\}=2g^{\mu\nu}(x)$, or $\{\gamma_\mu(x),\gamma_\nu(x)\}=2g_{\mu\nu}(x)$, where $\gamma^\mu (x)=g^{\mu\nu}(x)\gamma_\nu (x)$, and $\{\gamma^a,\gamma^b\}=2\eta^{ab}$, or $\{\gamma_a,\gamma_b\}=2\eta_{ab}$, where $\gamma^a=\eta^{ab}\gamma_b=(\gamma^0,\gamma^1,\gamma^2)=(\gamma_0,-\gamma_1,-\gamma_2)$.

With this, let us obtain the form of the tetrads (and their inverses), which can be easily calculated by writing the line element $ds^2=g_{\mu\nu}(x)dx^\mu dx^\nu$ in terms of the noncoordinate basis. Therefore, we have \cite{Strange,Panahi}
\begin{equation}\label{lineelement4}
ds^2=g_{\mu\nu}(x)e^\mu_{\ a}(x)\hat{\theta}^a e^\nu_{\ a}(x)\hat{\theta}^b=ds^2=[g_{\mu\nu}(x)e^\mu_{\ a}(x)e^\nu_{\ a}(x)]\hat{\theta}^a\hat{\theta}^b=\eta_{ab}\hat{\theta}^a\hat{\theta}^b=(\hat{\theta}^0)^2-(\hat{\theta}^1)^2-(\hat{\theta}^2)^2.
\end{equation}

However, comparing \eqref{lineelement4} with \eqref{lineelement3}, implies that the components of the noncoordidate basis are given by
\begin{equation}\label{noncoordidatebasis}
\hat{\theta}^0=\sqrt{1-V^2}dt-\frac{V\sigma\sin\left(\sqrt{2\Lambda}r\right)}{\sqrt{1-V^2}}d\theta, \ \ \hat{\theta}^1=dr, \ \ \hat{\theta}^2=\frac{\sigma\sin\left(\sqrt{2\Lambda}r\right)}{\sqrt{1-V^2}}d\theta.
\end{equation}

Consequently, the tetrads (with $\mu$ rows and $a$ columns) and their inverses (with $a$ rows and $\mu$ columns) take the form
\begin{equation}\label{tetrads}
e^{\mu}_{\ a}(x)=\left(
\begin{array}{ccc}
 \frac{1}{\sqrt{1-V^2}} & 0 & \frac{V}{\sqrt{1-V^2}} \\
 0 & 1 & 0 \\
 0 & 0 &  \frac{\sqrt{1-V^2}}{\sigma\sin\left(\sqrt{2\Lambda}r\right)} \\
\end{array}
\right), 
\ \
e^{a}_{\ \mu}(x)=\left(
\begin{array}{ccc}
 \sqrt{1-V^2} & 0 & -\frac{V\sigma\sin\left(\sqrt{2\Lambda}r\right)}{\sqrt{1-V^2}} \\
 0 & 1 & 0 \\
 0 & 0 &  \frac{\sigma\sin\left(\sqrt{2\Lambda}r\right)}{\sqrt{1-V^2}} \\
\end{array}
\right),
\end{equation}
where we see that the only coordinate that is not changed by the noninertial Bonnor-Melvin-Lambda spacetime (or noninertial curved line element) is the radial coordinate (since $e^r_{\ 1}(x)=e^1_{\ r}(x)=1$).

With the tetrads well-defined, we can then obtain the curved gamma matrices. Explicitly, these matrices are given by
\begin{equation}\label{gammamatrices}
\gamma^\mu(x)= \begin{cases}
\gamma^t(x)=\frac{1}{\sqrt{1-V^2}}(\gamma^0+V\gamma^2), \\
\gamma^r(x)=\gamma^1, \\
\gamma^\theta(x)=\frac{\sqrt{1-V^2}}{\sigma\sin\left(\sqrt{2\Lambda}r\right)}\gamma^2,
\end{cases}
\end{equation}
while the vector potential $A_\theta (x)$ takes the form $A_\theta (x)=e^2_{\ \theta}(x)A_2=\frac{\sigma\sin\left(\sqrt{2\Lambda}r\right)}{\sqrt{1-V^2}}A_2=-\frac{\sigma\sin\left(\sqrt{2\Lambda}r\right)}{\sqrt{1-V^2}}A_\theta(r)$ (since we have $A_b=\eta_{bc}A^c=(A_0,A_1,A_2)=(0,0,-A_\theta(r))$). Here, we make explicit the radial dependence of the angular component of $A_b$, which already preserves/conserves the axial symmetry of the Bonnor-Melvin-Lambda spacetime. With this, $A_\theta (x)$ should not be confused with $A_\theta (r)$, since they are different things. In other words, $A_\theta (x)$ depends on the coordinates/parameters that model the spacetime under study (or its curved geometry), while $A_\theta (r)$ depends only on the functional form of the external magnetic field that generates the LLs. Besides, it is important to note that the inverse tetrad implies that only $A_\theta (x)$ is the only non-null component of $A_\mu (x)$, that is \cite{Oli10}
\begin{equation}
A_\mu (x)=e^b_{\ \mu}(x)A_b=(e^2_{\ t}(x)A_2,e^2_{\ r}(x)A_2,e^2_{\ \theta}(x)A_2)=\left(0,0,-\frac{\sigma\sin\left(\sqrt{2\Lambda}r\right)}{\sqrt{1-V^2}}A_\theta(r)\right)=(0,0,A_\theta (x))=A_\theta (x)\delta^\theta_{\ \mu}.
\end{equation}
 
Now, we are ready to obtain the form of the spinorial and spin connections (i.e. we have all the ``ingredients'' on hand). However, we must first obtain the spin connection since the spinorial connection depends on it. So, according to Refs. \cite{Lawrie,Panahi}, the spin connection is defined as follows (torsion-free)
\begin{equation}\label{spinconnection}
\omega_{ab\mu}(x)=-\omega_{ba\mu}(x)=\eta_{ac}e^c_{\ \nu}(x)\left[e^\sigma_{\ b}(x)\Gamma^\nu_{\ \mu\sigma}(x)+\partial_\mu e^\nu_{\ b}(x)\right], 
\end{equation}
where $\Gamma^\nu_{\ \mu\sigma}(x)$ are the well-known Christoffel symbols of the second type (a symmetric tensor that depends only on the metric and derivatives of the inverse metric), whose expression is given by
\begin{equation}\label{Christoffel}
\Gamma^\nu_{\ \mu\sigma}(x)=\Gamma^\nu_{\ \sigma\mu}(x)=\frac{1}{2}g^{\nu\lambda}(x)\left[\partial_\mu g_{\lambda\sigma}(x)+\partial_\sigma g_{\lambda\mu}(x)-\partial_\lambda g_{\mu\sigma}(x)\right].
\end{equation}

Therefore, using the metrics in \eqref{metric1}, implies that the (non-null) components of the Christoffel symbols are given by
\begin{equation}\label{Christoffelsymbols}
\Gamma^\nu_{\ \mu\sigma}(x)= \begin{cases}
\Gamma^r_{\ tt}(x)=-\frac{\sqrt{\Lambda}\sigma^2\Omega^2\sin\left(2\sqrt{2\Lambda}r\right)}{\sqrt{2}}, 
\\
\Gamma^r_{\ t\theta}(x)=\Gamma^r_{\ \theta t}(x)=-\frac{\sqrt{\Lambda}\sigma^2\Omega\sin\left(2\sqrt{2\Lambda}r\right)}{\sqrt{2}}, 
\\
\Gamma^r_{\ \theta\theta}(x)=-\frac{\sqrt{\Lambda}\sigma^2\sin\left(2\sqrt{2\Lambda}r\right)}{\sqrt{2}},
\\
\Gamma^\theta_{\ tr}(x)=\Gamma^\theta_{\ rt}(x)=\sqrt{2\Lambda}\Omega\cot\left(\sqrt{2\Lambda}r\right),
\\
\Gamma^\theta_{\ \theta r}(x)=\Gamma^\theta_{\ r\theta}(x)=\sqrt{2\Lambda}\cot\left(\sqrt{2\Lambda}r\right),
\end{cases}
\end{equation}
where $\cot\left(\sqrt{2\Lambda}r\right)=\frac{\cos\left(\sqrt{2\Lambda}r\right)}{\sin\left(\sqrt{2\Lambda}r\right)}$.

Consequently, the (non-null) components of the spin connection are written as
\begin{equation}\label{spinconnection2}
\omega_{ab \mu}(x)=\begin{cases}
\omega_{01t}(x)=-\omega_{10t}(x)=\sqrt{\frac{\Lambda}{2}}\frac{\sigma^2\Omega^2\sin\left(2\sqrt{2\Lambda}r\right)}{\sqrt{1-V^2}}, 
\\
\omega_{12t}(x)=-\omega_{21t}(x)=-\frac{\sqrt{2\Lambda}\sigma\Omega\cos\left(\sqrt{2\Lambda}r\right)}{\sqrt{1-V^2}}, 
\\
\omega_{02r}(x)=-\omega_{20r}(x)=-\frac{\sqrt{2\Lambda}\sigma\Omega\cos\left(\sqrt{2\Lambda}r\right)}{1-V^2},
\\
\omega_{01\theta}(x)=-\omega_{10\theta}(x)=\sqrt{\frac{\Lambda}{2}}\frac{\sigma^2\Omega\sin\left(2\sqrt{2\Lambda}r\right)}{\sqrt{1-V^2}},
\\
\omega_{12\theta}(x)=-\omega_{21\theta}(x)=-\frac{\sqrt{2\Lambda}\sigma\cos\left(\sqrt{2\Lambda}r\right)}{\sqrt{1-V^2}}.
\end{cases}
\end{equation}
where implies in the following (non-null) components for the spinorial connection
\begin{equation}\label{spinorialconnection}
\Gamma_\mu(x)= \begin{cases}
\Gamma_t(x)=\sqrt{\frac{\Lambda}{2}}\frac{\sigma\Omega\cos \left(\sqrt{2\Lambda}r\right)\left(\gamma^1\gamma^2-V\gamma^0\gamma^1\right)}{\sqrt{1-V^2}}, 
\\
\Gamma_r(x)=\sqrt{\frac{\Lambda}{2}}\frac{\sigma\Omega\cos\left(\sqrt{2\Lambda}r\right)}{\left(1-V^2\right)}\gamma^0\gamma^2, 
\\
\Gamma_\theta(x)=\sqrt{\frac{\Lambda}{2}}\frac{\sigma\cos\left(\sqrt{2\Lambda }r\right)\left(\gamma^1\gamma^2-V\gamma^0\gamma^1\right)}{\sqrt{1-V^2}}=\frac{\Gamma_t(x)}{\Omega}.
\end{cases}
\end{equation}

Therefore, using the matrices \eqref{gammamatrices} and \eqref{spinorialconnection}, we obtain the following contribution of the spinorial/spin connection for the curved DO
\begin{equation}\label{contributionofthespinorialconnection}
i\gamma^\mu (x)\Gamma_\mu (x)=i[\gamma^t(x)\Gamma_t(x)+\gamma^{r}(x)\Gamma_{r}(x)+\gamma^{\theta}(x)\Gamma_{\theta}(x)]=i\left[\sqrt{\frac{\Lambda}{2}}\cot\left(\sqrt{2\Lambda}r\right)\gamma^1-\sqrt{\frac{\Lambda}{2}}\frac{\sigma\Omega\cos\left(\sqrt{2\Lambda}r\right)}{\left(1-V^2\right)}\gamma^0\gamma^1\gamma^2\right].
\end{equation}

In that way, using \eqref{gammamatrices} and \eqref{contributionofthespinorialconnection}, we have from Eq. \eqref{dirac2} the explicit form for the charged DO in a rotating frame in the Bonnor-Melvin-Lambda spacetime, which is given by
\begin{equation}\label{dirac3}
\left[\frac{i}{\sqrt{1-V^2}}\left(\gamma^0+V\gamma^2\right)\partial_t+i\gamma^1\left(\partial_r+m_0\omega r\gamma^0+\sqrt{\frac{\Lambda}{2}}\cot\left(\sqrt{2\Lambda}r\right)\right)+i\frac{\sqrt{1-V^2}}{\sigma\sin\left(\sqrt{2\Lambda}r\right)}\gamma^2\left(\partial_\theta+iqA_\theta (x)\right)-M_0\right]\psi=0,
\end{equation}
where we define
\begin{equation}\label{define0}
M_0\equiv\sqrt{\frac{\Lambda}{2}}\frac{\sigma\Omega\cos\left(\sqrt{2\Lambda}r\right)}{\left(1-V^2\right)}\gamma^0\Sigma^3+m_0=\sqrt{2\Lambda}\frac{\sigma\cos\left(\sqrt{2\Lambda}r\right)}{\left(1-V^2\right)}\gamma^0\vec{S}\cdot\vec{\Omega}+m_0,
\end{equation}
being the term $-\vec{S}\cdot\vec{\Omega}=-\frac{1}{2}\Sigma_3\Omega$ (with $\Sigma_3=\Sigma^3=i\gamma^1\gamma^2=\Sigma_z$ \cite{Greiner}) the well-known spin-rotation coupling \cite{Oli2,Oli9,Oli10,Bakke2,Strange} (in the nonrelativistic limit of the DO, this coupling describes the Mashhoon effect \cite{Hehl,Mashhoon}. See Appendix \ref{sec6}), which arises when Dirac fermions are in a rotating frame (or better, when the intrinsic spin of the fermion interacts with a rotating frame), being $\vec{S}=\frac{1}{2}\vec{\Sigma}$ is the spin operator/vector, $\vec{\Omega}=\Omega\vec{e}_z$ is the angular velocity (pseudo)vector, with $\Omega=\Omega_z=constant\geq 0$. Consequently, the direction of rotation of the rotating frame is either parallel ($\vec{S}\cdot\vec{\Omega}\geq 0$) or antiparallel ($\vec{S}\cdot\vec{\Omega}\leq 0$) to the fermion spin \cite{Strange}. In particular, the spin-rotation coupling leads to an additional phase shift or energy term and, as we will see shortly, here (relativistic regime), such coupling will modify the rest mass of the fermion (this happens because we are working in $(2+1)$-dimensions where $\gamma^0\Sigma^3=1$).

On the other hand, we see that it is (very) difficult to proceed without a simplification of Eq. \eqref{dirac3}. Therefore, to solve analytically this equation, we consider here two approximations, where the first is that the cosmological constant has very small values, such as $\Lambda\ll 1$. According to Ref. \cite{Castro}, where such simplification is called conical approximation, this is plausible since in a quantum scenario (rather than cosmological) the spatial coordinate $r$ is at a microscopic scale. Consequently, the cosmological constant is no longer able to control the order of magnitude of the argument of the sine function (given by $\sqrt{2\Lambda}r$), thus making it very small, or, equivalently, $\Lambda\ll 1$ \cite{Castro}. Besides, the plausibility of a small cosmological constant has been considered in various scenarios at the quantum scale, such as in non-supersymmetric theories with a naturally light dilaton \cite{Bellazzini}, 5D brane world models \cite{Arvanitaki}, supersymmetric string theory \cite{Bousso,Demirtas}, non-perturbative standard model \cite{Holland}, etc. So, with the conical approximation $\Lambda\ll 1$, where we have $\sin\left(\sqrt{2\Lambda}r\right)\approx \sqrt{2\Lambda}r$ (a first-order approximation), the line element \eqref{lineelement} becomes: $ds^2=dt^2-dr^2-2\Lambda\sigma^2 r^2 d\theta^2$. In this case, the Ricci scalar curvature and the magnetic field are now given by $R=\delta(r)$ and $H(r)=\sqrt{2}\Lambda\sigma r$, i.e. now the curvature is null everywhere (asymptotically flat) except at the point where the source is located ($r=0$), resembling very much the spacetime generated by a cosmic string (a conical gravitational topological defect), where its $(2+1)$-dimensional line element is given by $ds^2=dt^2-dr^2-\alpha^2 r^2 d\theta^2$, with $0<\alpha\leq 1$ being the topological or conical curvature parameter (``angular deficit'') \cite{Oli2,Bakke2,Oli9,Oli10}. Therefore, since for $\Lambda\ll 1$ the line element of the Bonnor-Melvin-Lambda spacetime is very very similar to the line element of a cosmic string, it implies that the name ``conical approximation'' is very suggestive. From this, we can say that the quantity $2\Lambda\sigma^2$ (with $\alpha_{\Lambda}\equiv \sqrt{2\Lambda}\sigma$) describes a spacetime with a conical singularity at $r=0$ \cite{Castro}. So, without further ado regarding the first approximation, the parameter $V$ becomes $V\approx\sqrt{2\Lambda}\sigma\Omega r$. Consequently, Eq. \eqref{dirac3} takes the following form (with $\cos\left(\sqrt{2\Lambda}r\right)\approx 1$)
\begin{equation}\label{dirac4}
\left[\frac{i}{\sqrt{1-2\Lambda\sigma^2\Omega^2 r^2}}\left(\gamma^0+\sqrt{2\Lambda}\sigma\Omega r\gamma^2\right)\partial_t+i\gamma^1\left(\partial_r+\frac{1}{2r}+m_0\omega r\gamma^0\right)+i\frac{\sqrt{1-2\Lambda\sigma^2\Omega^2 r^2}}{\sqrt{2\Lambda}\sigma r}\gamma^2\left(\partial_\theta-ieA_\theta (x)\right)-M_0\right]\psi=0,
\end{equation}
where
\begin{equation}\label{M00}
M_0=\frac{\sqrt{2\Lambda}\sigma}{\left(1-2\Lambda\sigma^2\Omega^2 r^2\right)}\gamma^0\vec{S}\cdot\vec{\Omega}+m_0, \ \ A_\theta (x)=-\frac{\sqrt{2\Lambda}\sigma r}{\sqrt{1-2\Lambda\sigma^2\Omega^2 r^2}}A_\theta (r).
\end{equation}

In addition, the relation $0\leq\sin\left(\sqrt{2\Lambda}r\right)<\Lambda_0$ for $\Lambda\ll 1$, where $\Lambda_0\equiv\frac{1}{\sigma\Omega}$, must be reformulated/redefined such as: $0\leq r<r_0$, where $r_0\equiv\frac{1}{\sqrt{2\Lambda}\sigma\Omega}$, that is, now we have a new definition for the radial coordinate. In other words, now we have a new range for the coordinate $r$, where the values for this range are controlled by both $\sigma$, $\Omega$, and $\Lambda$, respectively. In this case, for the spinor to be normalized (finite, square-integrable, or physically acceptable solution), must vanish at $r\to 0$ and $r\to r_0$ (basically, these limits are the two boundary conditions/asymptotic behavior that the spinor must satisfy for describes the bound-state solutions or physically realizable states) \cite{Oli2,Oli3,Bakke2,Oli9,Oli10,CQG}.

Now, with respect to the second approximation, we consider here a (very) slow rotation compared to the speed of light, such as $\Omega\ll 1$, or better, $\Omega r\ll 1$ \cite{Strange,Oli2,Oli8,Bakke2,Oli9,Oli10}. In that way, we have $V=\sqrt{2\Lambda}\sigma\Omega r\ll 1$ (or yet $v=\sqrt{2\Lambda}\sigma\Omega r\approx constant$, with $0\leq v\ll 1$), where implies that $V^2=2\Lambda\sigma^2\Omega^2 r^2\approx 0$; consequently, linear terms of $V$ are conserved while quadratic terms are eliminated/ignored in Eq. \eqref{dirac4} (indeed, since $\Lambda\ll 1$, this further reinforces the condition $V^2\approx 0$). With this, we have $r_0\gg 1$, or better (without loss of generality) $r_0\to\infty$ for $\Lambda\ll 1$ and $\Omega\ll 1$, where implies $0\leq r<\infty$ (i.e. now the spinor must vanish at $r\to 0$ and $r\to\infty$). Here, it is important to comment that the condition $\Lambda\ll 1$ alone does not guarantee that $V^2\approx 0$, that is, $\Omega$ could be large enough to avoid this. Therefore, to ensure that $V^2\approx 0$ (and $0\leq r<\infty$), we must (really) have both $\Lambda$ and $\Omega$ very small. Besides, the plausibility of a slow-rotation approximation/limit ($V\ll 1$) has been considered in various systems at the quantum scale, such as in gluon plasma \cite{Braguta}, neutrino interaction/oscillations with/in rotating matter \cite{Dvornikov}, scalar bosons \cite{C,Santos}, pairing phase transitions \cite{Jiang}, Aharonov–Bohm quantum ring \cite{ring}, chiral symmetry
breaking \cite{Chernodub}, spinning quantum systems \cite{Brito}, etc. So, from this second approximation, Eq. \eqref{dirac4} takes the following form
\begin{equation}\label{dirac5}
\left[i\left(\gamma^0+\sqrt{2\Lambda}\sigma\Omega r\gamma^2\right)\partial_t+i\gamma^1\left(\partial_r+\frac{1}{2r}+m_0\omega r\gamma^0\right)+\frac{i}{\sqrt{2\Lambda}\sigma r}\gamma^2\left(\partial_\theta-ieA_\theta(x)\right)-\left(\sqrt{\frac{\Lambda}{2}}\sigma\Omega\gamma^0\Sigma^3+m_0\right)\right]\psi=0,
\end{equation}
being $A_\theta (x)=-\sqrt{2\Lambda}\sigma r A_\theta(r)$, and $\sqrt{2\Lambda}\sigma\Omega r (i\partial_t)$ is a term that includes the angular frequency of rotation, radial distance, and the time derivative of the spinor and represents clearly the coupling of the orbital motion of the fermion to the rotation, that is, it is the term that describes the Sagnac effect \cite{Strange,Hehl}. In fact, rewriting this term in SI units as $\sqrt{2\Lambda}\sigma\Omega r (i\hslash\partial_t)/c$, where $(i\hslash\partial_t)$ has energy dimension and $(i\hslash\partial_t)/c$ has linear momentum dimension, implies that such a term has orbital angular momentum times rotation dimension, i.e. $\sqrt{2\Lambda}\sigma\Omega r(i\hslash\partial_t)/c=\sqrt{2\Lambda}\sigma(-\vec{L}\cdot\vec{\Omega})$, where we define $\vec{L}=\vec{L}_{fermion}\equiv -(r(i\hslash\partial_t)/c)\vec{e}_z$. In other words, this term is the orbital angular momentum-rotation coupling of the Sagnac effect. \cite{Strange,Hehl} (furthermore, taking $\Lambda\to 1/2$ and $\sigma\to 1$, we obtain exactly the term of Ref. \cite{Strange}). In particular, the vector potential $A_\theta (x)$ is similar to Ref. \cite{Castro}, where $A_\varphi=\sqrt{2\Lambda}\sigma r f(r)$, with $f(r)$ being the angular component (or functional form) of the flat vector potential (in our case exist a negative sign because we are working with a metric of signature $(+,-,-)$). Furthermore, it is also worth mentioning that we could have (easily) arrived in Eq. \eqref{dirac5} by making the two approximations from the beginning, such as was done in Ref. \cite{Strange} for the case of the rotating DO. With this, the line element \eqref{lineelement2} would be written as: $ds^2=dt^2-2\sqrt{2\Lambda} V\sigma r dt d\theta-dr^2-2\Lambda\sigma^2 r^2 d\theta^2$, while the line element \eqref{lineelement3} would be: $ds^2=\left(dt-\sqrt{2\Lambda} V\sigma r d\theta\right)^2-dr^2-2\Lambda\sigma^2 r^2 d\theta^2$, respectively.

Besides, here we consider a uniform external magnetic field oriented along the $z$-axis, given by $\vec{B}=B_0\vec{e}_z$, where $B_0=B_z\geq 0$ is the field strength/modulus, implies that the associated vector potential takes the form $\vec{A}=\frac{1}{2}\vec{B}\times\vec{r}=A_\theta (r)\vec{e}_\theta$, being $A_\theta (r)=\frac{1}{2}B_0 r$ is the radial or polar symmetric gauge (``symmetric Landau gauge'') \cite{Castro,Oli2,Oli3,Oli10,Villalba2,Li1999,Ribeiro,Pavlov}. With this, we have $A_\theta (x)=-\sqrt{\frac{\Lambda}{2}}\sigma B_0 r^2$, that is, the curved vector potential is a quadratic function (or quadratic interaction) in the radial coordinate, while the flat vector potential is a linear function (or linear interaction), respectively. In that way, Eq. \eqref{dirac5} becomes
\begin{equation}\label{dirac6}
\left[i\left(\gamma^0+\sqrt{2\Lambda}\sigma\Omega r\gamma^2\right)\partial_t+i\gamma^1\left(\partial_r+\frac{1}{2r}+m_0\omega r\gamma^0\right)-\frac{m_0 \omega_c}{2}r\gamma^2+\frac{i}{\sqrt{2\Lambda}\sigma r}\gamma^2\partial_\theta-\left(\sqrt{\frac{\Lambda}{2}}\sigma\Omega\gamma^0\Sigma^3+m_0\right)\right]\psi=0,
\end{equation}
where $\omega_c=\frac{eB_0}{m_0}\geq 0$ is the well-known (classical) cyclotron frequency (gyrofrequency or cyclotron resonance), which is also a type of angular velocity (or circular motion), but of the DO (or Dirac fermion) on the rotating plane/frame. In particular, the Landau quantization implies that the orbits generated by $\omega_c$ have discrete/quantized energies (known as LLs), where the associated cyclotron radius (radius of the circular motion) is given by: $r_c=v_f/\omega_c$ (this comes from the linear velocity $v_f=\omega_c r_c$, being $v_f=\displaystyle v_{\perp}$ the component of the fermion velocity perpendicular to the direction of the magnetic field, or tangential to the circular path, and $\omega_c$ is the strength/modulus of $\vec{\omega}_c=\omega_c\vec{e}_z$). Besides, we could also write Eq. \eqref{dirac6} in terms of the Larmor frequency (describes the Larmor precession), given by $\omega_L=\omega_c/2$ \cite{Mashhoon}, i.e. the Larmor frequency is half the cyclotron frequency \cite{Schattschneider,Bliokh,Lloyd,Wakamatsu,Villalba2}. So, unlike Larmor frequency, which is the frequency/angular velocity at which a charged particle (or magnetic moment) precesses around the magnetic field (i.e. when the axis of rotation changes to describe a cone), cyclotron frequency is the frequency/angular velocity at which the particle completes a full circular orbit in the magnetic field. Here, we prefer to use $\omega_c$ instead of $\omega_L$ because it is ``much more common'' in the literature when studying LLs (however, although they have different origins, both $\omega_c$ and $\omega_L$ are also useful in nonuniform magnetic fields).

On the other hand, since here we are interested in stationary bound states (i.e. states with defined total energy, or whose potential or probability density does not depend explicitly on time \cite{Griffiths}), we can use/define the following ansatz for the spinor $\psi$ \cite{Panahi,Oli2,Oli3,Oli10,Villalba1,Villalba2,Greiner,Strangebook,Schluter}
\begin{equation}\label{spinor}
\psi(t,r,\theta)=\frac{e^{i(m_j\theta-Et)}}{\sqrt{r}}\psi_0(r), \ \ \psi_0(r)=\left(
           \begin{array}{c}
            \psi^+(r) \\
            i\psi^-(r) \\
           \end{array}
         \right), \ \ (\psi^+(r)\neq \psi^-(r)),
\end{equation}
where $\psi^\pm (r)$ are real radial functions (for simplicity/convenience, we will label them as the components of the spinor), $E=E^\pm=\pm\vert E^\pm\vert$ is the relativistic total energy (particle/antiparticle), and $m_j=m_{\theta}=\pm 1/2,\pm 3/2,\pm 5/2,\ldots$ is the total magnetic quantum number (and are eigenvalues of $J_z=-i\partial_\theta$), whose values come from the fact that $\psi(\theta\pm 2\pi)=-\psi(\theta)$. In particular, this angular quantum number can also be written in terms of two others, i.e. $m_j=m_l+m_s$, being $m_l=0,\pm 1,\pm 2,\ldots$ and $m_s=\pm 1/2$ (=``$s/2$'') the orbital and spin magnetic quantum numbers (and are eigenvalues of $L_z$ and $S_z$ \cite{Griffiths}). With this, for a given $m_j$ ($m_j>0$ or $m_j<0$), we have two spin states, i.e. $m_j=m_l+m_s>0$ as well as $m_j=m_l+m_s<0$ describes two spin states each one (if $m_j>0$, so $m_l\geq 0$ and $m_s=+1/2$ or $m_l>0$ and $m_s=-1/2$, and if $m_j<0$, so $m_l<0$ and $m_s=+1/2$ or $m_l\leq 0$ and $m_s=-1/2$, respectively \cite{Oli2,Griffiths}). Thus, each component accommodates the two spin states, that is, $\psi=(\psi_{E>0}^{\uparrow\downarrow},\psi_{E<0}^{\uparrow\downarrow})^T$ (in the case of a four-component spinor, each component accommodates only one spin state, that is, $\Psi=(\Psi_{E>0}^{\uparrow},\Psi_{E>0}^{\downarrow},\Psi_{E<0}^{\uparrow},\Psi_{E<0}^{\downarrow})^T$ \cite{Greiner,Strangebook}).

Therefore, using the spinor \eqref{spinor} in Eq. \eqref{dirac6}, we obtain the following time-independent DO (or stationary DO)
\begin{equation}\label{dirac7}
\left\{\gamma^0 E+i\gamma^1\left(\frac{d}{dr}+m_0\omega r\gamma^0\right)-\gamma^2\left[\frac{m_j}{\sqrt{2\Lambda}\sigma r}+m_0\left(\frac{\omega_c}{2}-\frac{\sqrt{2\Lambda}\sigma\Omega E}{m_0}\right)r\right]-\left(\sqrt{\frac{\Lambda}{2}}\sigma\Omega\gamma^0\Sigma^3+m_0\right)\right\}\psi_0(r)=0,
\end{equation}
or written as an eigenvalue equation (or eigenequation), such as
\begin{equation}\label{genvaluesequation}
H_{DO}\psi_0(r)= E\psi_0(r),
\end{equation}
where $H_{DO}$ is the Dirac Hamiltonian for the DO (or simply the DO Hamiltonian), given by
\begin{equation}\label{H}
H_{DO}=\left\{-i\gamma^0\gamma^1\left(\frac{d}{dr}+m_0\omega r\gamma^0\right)+\gamma^0\gamma^2\left[\frac{m_j}{\sqrt{2\Lambda}\sigma r}+m_0\left(\frac{\omega_c}{2}-\frac{\sqrt{2\Lambda}\sigma\Omega E}{m_0}\right)r\right]+\left(\sqrt{\frac{\Lambda}{2}}\sigma\Omega\Sigma^3+m_0 \gamma^0 \right)\right\},
\end{equation}
being $\psi_0 (r)$ the eigenfuctions (eigenvectors or eigenstates), and $E$ are the correspondent eigenvalues (eigenenergies).

It is important to mention that, similar to the case of a cosmic string \cite{Oli2,Bakke2,V} and of the Gödel-type spacetime \cite{CQG}, here, the total angular momentum is also modified/shifted by the parameters of the Bonnor-Melvin-Lambda spacetime, that is, both quantities $\sigma$ and $\Lambda$ have the function of modifying the total angular momentum (along the $z$-direction) and, consequently, we can define an effective total magnetic quantum number (controlled by $\sigma$ and $\Lambda$), given by $M=M_{eff}=M(\sigma,\Lambda)\equiv\frac{m_j}{\sqrt{2\Lambda}\sigma}$. With this, the operator $J_z$ takes the form $\mathbb{J}=\mathbb{J}(\sigma,\Lambda)=-\frac{i}{\sqrt{2\Lambda}\sigma}\partial_\theta$.

Now, knowing that in $(2+1)$-dimensions the flat gamma matrices $\gamma^a=(\gamma^0,\gamma^1,\gamma^2)=(\gamma_0,-\gamma_1,-\gamma_2)$, as well as $\Sigma_3$, takes the form (Dirac/standard representation): $\gamma_1=\sigma_3\sigma_1=i\sigma_2$, $\gamma_2=s\sigma_3\sigma_2=-is\sigma_1$, and $\gamma_0=\Sigma_3=\sigma_3$, where $\sigma_1=$off-diag$(+1,+1)$, $\sigma_2=$off-diag$(-i,+i)$ and $\sigma_3=$diag$(+1,-1)$ are the Pauli spin matrices and $s=\pm 1$ is the spin parameter \cite{Hagen1,Hagen2,Blum,Hagen3,Hagen4,Hagen5,Hagen6,CQG,Oli2,Sousa}, implies that we can obtain from Eq. \eqref{dirac7} two coupled first-order differential equations (i.e. equations that contain both two components of the spinor). Explicitly, these two equations are given by
\begin{equation}\label{EDO1}
\left(\sqrt{\frac{\Lambda}{2}}\sigma\Omega+m_0-E\right)\psi^+(r)=\left(\frac{d}{dr}-m_0\bar{\Omega}r+\frac{Ms}{r}\right)\psi^-(r),
\end{equation}
and
\begin{equation}\label{EDO2}
\left(\sqrt{\frac{\Lambda}{2}}\sigma\Omega+m_0+E\right)\psi^-(r)=\left(\frac{d}{dr}+m_0\bar{\Omega}r-\frac{Ms}{r}\right)\psi^+(r),
\end{equation}
where implies
\begin{equation}\label{EDO3}
\psi^+(r)=\frac{1}{\left(\sqrt{\frac{\Lambda}{2}}\sigma\Omega+m_0-E\right)}\left(\frac{d}{dr}-m_0\bar{\Omega}r+\frac{Ms}{r}\right)\psi^-(r),
\end{equation}
and
\begin{equation}\label{EDO4}
\psi^-(r)=\frac{1}{\left(\sqrt{\frac{\Lambda}{2}}\sigma\Omega+m_0+E\right)}\left(\frac{d}{dr}+m_0\bar{\Omega}r-\frac{Ms}{r}\right)\psi^+(r),
\end{equation}
being $\bar{\Omega}=\Omega_{eff}\equiv\left(\omega-s\frac{\omega_c}{2}+s\frac{\sqrt{2\Lambda}\sigma\Omega E}{m_0}\right)>0$ a type of effective angular frequency (for simplicity, here we consider positive), the last term would be a ``noninertial Bonnor-Melvin-Lambda frequency'', and we use $M$ instead of $\frac{m_j}{\sqrt{2\Lambda}\sigma}$ for convenience and simplicity. Besides, to avoid having non-physical/non-normalizable solutions, it is necessary to ensure that the term in the denominator is non-zero, i.e. $\left(\sqrt{\frac{\Lambda}{2}}\sigma\Omega+m_0\pm E\right)\neq 0$; otherwise, implies in $\psi^\pm (r)\to\infty$.

Therefore, substituting \eqref{EDO4} in \eqref{EDO3}, and vice-versa (i.e. decoupling the equations), we obtain the following second-order differential equation for the charged DO in a rotating frame in the Bonnor-Melvin-Lambda spacetime (i.e. we have the ``quadratic DO'' or ``Schrödinger-type oscillator'')
\begin{equation}\label{dirac8}
\left[\frac{d^2}{dr^2}-\frac{Ms(Ms\mp 1)}{r^2}-(m_0\bar{\Omega}r)^2+E^2-\left(m_0+\sqrt{\frac{\Lambda}{2}}\sigma\Omega\right)^2+m_0\bar{\Omega}(2Ms\pm 1)\right]\psi^\pm(r)=0,
\end{equation}
or better (compacting the two equations into just one)
\begin{equation}\label{dirac9}
\left[\frac{d^2}{dr^2}-\frac{Ms(Ms-u)}{r^2}+(m_0\bar{\Omega}r)^2+E^u_{Ms}\right]\psi^u_{Ms}(r)=0,
\end{equation}
where we define
\begin{equation}\label{define1}
E^u_{Ms}\equiv E^2-\left(m_0+\sqrt{\frac{\Lambda}{2}}\sigma\Omega\right)^2+m_0\bar{\Omega}(2Ms+u),
\end{equation}
being $u=\pm 1$ a parameter (``spinorial parameter'') which describes/labels the two components of the spinor: $u=+1$ is for the upper component ($\psi^+_{Ms}=\psi^{upper}_{Ms}$=``$\psi^{particle}_{Ms}$'') and $u=-1$ is for the lower component ($\psi^-_{Ms}=\psi^{lower}_{Ms}$=``$\psi^{antiparticle}_{Ms}$''), respectively (in the nonrelativistic limit of the DO, this parameter will describe the two components of the Pauli spinor, where $u=+1$ is for a particle with spin up and $u=-1$ is for a particle with spin down. See Appendix \ref{sec6}). In particular, the Eq. \eqref{dirac9} is similar to that of Refs. \cite{Villalba1,Villalba2}. Indeed, for $\sigma=1$, $\Lambda=1/2$, and $\Omega=0$, we obtain the equation for the DO with and without electric charge in the $(2+1)$-dimensional Minkowski spacetime \cite{Villalba1,Villalba2}. In this way, our solutions will also be obtained/analyzed for both $Ms>0$ and $Ms<0$, in which each case depends on the values of $M$ and $s$, that is, depending on the values of $M$ and $s$ (or how they are combined), we have two possible configurations for $Ms>0$ as well as two for $Ms<0$. In summary, we have
\begin{equation}\label{Ms}
Ms= \begin{cases}
Ms>0 \begin{cases}
M>0, \ s=+1,
\\
M<0, \ s=-1,
\end{cases}
\\
Ms<0 \begin{cases}
M>0, \ s=-1, 
\\
M<0, \ s=+1,
\end{cases}
\end{cases}
\end{equation}
where $Ms>0$ can be ``interpreted'' as a fermion/DO with both positive angular momentum and spin, or with both negative angular momentum and spin, while $Ms<0$ can be ``interpreted'' as a fermion/DO with positive angular momentum and negative spin, or with negative angular momentum and positive spin, respectively.

%-------------------------------------------------------------------------
\section{Bound-state solutions: Two-component Dirac spinor and the relativistic energy spectrum \label{sec3}}

To analytically solve Eq. \eqref{dirac9}, let us first introduce a new variable in our
problem, given by: $w=m_0\bar{\Omega}r^2>0$ (i.e. a dimensionless variable). In that way, making a simple change of variable, Eq. \eqref{dirac9} becomes
\begin{equation}\label{dirac10}
\left[w\frac{d^2}{dw^2}+\frac{1}{2}\frac{d}{dw}-\frac{Ms(Ms-u)}{4w}-\frac{w}{4}+\frac{E^u_{Ms}}{4m_0\bar{\Omega}}\right]\psi^u_{Ms}(w)=0.
\end{equation}

Now, we need to analyze the asymptotic behavior/limit of Eq. \eqref{dirac10} for both small and large $w$. So, we must have
\begin{itemize}
\item For $w\to 0$, we have $\left[\frac{d^2}{dw^2}+\frac{1}{2w}\frac{d}{dw}-\frac{Ms(Ms-u)}{4w^2}\right]\psi^u_{Ms}(w)\approx 0$, where implies that $\psi^u_{Ms}(w)\approx c_1 w^{\frac{(1+(2Ms-u)}{4}}+c_2 w^{\frac{(1-(2Ms-u))}{4}}$ (i.e. a solution written in terms of two polynomials), or better, $\psi^u_{Ms}(w)\approx c_3 w^{\frac{(1\pm(2Ms-u))}{4}}$ (compacting the two polynomials into just one),
\item For $w\to\infty$, we have $\left[\frac{d^2}{dw^2}-\frac{1}{4}\right]\psi^u_{Ms}(w)\approx 0$, where implies that $\psi^u_{Ms}(w)\approx c_4 e^{-\frac{w}{2}}+c_5 e^{\frac{w}{2}}$ (i.e. a solution written in terms of two exponentials),
\end{itemize}
with $c_i>0$ ($i=1,2,3,4,5$) being arbitrary constants.

However, as the spinor must vanish at $w\to 0$ and $w\to\infty$ (i.e. $\psi^u_{Ms}(w\to 0)=\psi^u_{Ms}(w\to\infty)=0$), implies that for this to be satisfied we need to consider the following ansatz as a (regular/asymptotic) general solution for Eq. \eqref{dirac10}
\begin{equation}\label{solution1}
\psi^u_{Ms}(w)=c^u_{Ms} e^{-\frac{w}{2}}w^{\frac{(1\pm(2Ms-u))}{4}}F^u_{Ms}(w), \ \ (1\pm(2Ms-u)>0),
\end{equation}
where $c^u_{Ms}\equiv c_3 c_4>0$ is a new constant, and $F^u_{Ms}(w)$ are unknown functions to be determined (we can say that it is a proportionality factor, whose goal is to accommodate the results of the two asymptotic limits in only a solution). So, we see that, unlike the solution for $w\to\infty$ (where only one exponential was considered), the solution for $w\to 0$ is perfectly plausible for the two polynomials, however, the value of $Ms$ must be chosen in such a way that it guarantees the condition $\psi^u_{Ms}(w\to 0)=0$ (i.e. a regular solution in the origin). In other words, we must have $(1+(2Ms-u))>0$ for $Ms>0$ and $(1-(2Ms-u))>0$ for $Ms<0$, respectively.

So, through the solution \eqref{solution1}, we can have two possible configurations for each component of the spinor depending on the values of $Ms$. In summary, these configurations are given by
\begin{equation}\label{solution2}
\psi^u_{Ms}(w)= \begin{cases}
\psi^+_{Ms}(w)=\begin{cases}
c^+_{Ms} e^{-\frac{w}{2}}w^{\frac{Ms}{2}}F^+_{Ms}(w), \ \ \ \ \ \ \ \ Ms>0, 
\\
c^+_{Ms} e^{-w/2}w^{\frac{(1-Ms)}{2}}F^+_{Ms}(w), \ \ Ms<0,
\end{cases}
\\
\psi^-_{Ms}(w)=\begin{cases}
c^-_{Ms} e^{-\frac{w}{2}}w^{\frac{(1+Ms)}{2}}F^u_{Ms}(w), \ \ \ \ Ms>0,
\\
c^-_{Ms} e^{-\frac{w}{2}}w^{-\frac{Ms}{2}}F^u_{Ms}(w), \ \ \ \ \ \ Ms<0,
\end{cases}
\end{cases}
\end{equation}
where each specific configuration satisfies the condition $\psi^u_{Ms}(w\to 0)=\psi^u_{Ms}(w\to\infty)=0$ (as it should). In particular, these configurations are in perfect agreement with those obtained from Ref. \cite{Villalba1} (and even with \cite{Villalba2}, although the parameter $s$ was not used). Here (unlike \cite{Villalba1} or \cite{Villalba2}), we just prefer to start from Eq. \eqref{dirac9} instead of Eq. \eqref{dirac8} to be less tiring and laborious (i.e. if we had started from Eq. \eqref{dirac8}, it would be necessary to solve the equation separately for each component, what we consider as something redundant).

Therefore, replacing the the solution \eqref{solution1} in \eqref{dirac10} (or each configuration in \eqref{solution2}), we obtain a equation for $F^u_{Ms}(w)$
\begin{equation}\label{dirac11}
\left[w\frac{d^2}{dw^2}+\left(\Sigma^u_{Ms}-w\right)\frac{d}{dw}-\bar{E}^u_{Ms}\right]F^u_{Ms}(w)=0,
\end{equation}
where we define
\begin{equation}\label{define2}
\Sigma^u_{Ms}\equiv \frac{2\pm(2Ms-u)}{2}, \ \ \bar{E}^u_{Ms}\equiv\frac{\Sigma^u_{Ms}}{2}-\frac{E^u_{Ms}}{4m_0\bar{\Omega}}.
\end{equation}

According to the literature \cite{Villalba1,Villalba2,CQG,Oli2,Bhuiyan,Arfken,Brandt,Li1999,Griffiths}, Eq. \eqref{dirac11} have the form of the well-known associated/generalized Laguerre equation, whose solutions are the associated/generalized Laguerre polynomials, given by $F^u_{Ms}(w)=c_6 L^{\Sigma^u_{Ms}-1}_n(w)=c_6 L^{\pm(Ms-\frac{u}{2})}_n(w)$, being $c_6>0$ a constant, $n$ is a positive integer ($n=0,1,2,\ldots$), i.e. our second quantum number, and the Rodrigues formula for such polynomials is given by $L^\alpha_n (x)=\frac{x^{-\alpha}e^x}{n!}\frac{d^n}{dx^n}(e^{-x}x^{\alpha+n})$. In many applications, particularly in quantum mechanics, we need to work with such polynomials (e.g. the hydrogen atom and the spherically symmetric QHO) \cite{Arfken,Brandt,Griffiths}. Thus, the components \eqref{solution2} takes the form
\begin{equation}\label{solution3}
\psi^u_{Ms}(w)= \begin{cases}
\psi^+_{Ms}(w)=\begin{cases}
C^+_{Ms} e^{-\frac{w}{2}}w^{\frac{Ms}{2}}L^{Ms-\frac{1}{2}}_n(w), \ \ \ \ \ \ \ \ \ \ \ Ms>0, 
\\
C^+_{Ms} e^{-\frac{w}{2}}w^{\frac{(1-Ms)}{2}}L^{-Ms+\frac{1}{2}}_n(w), \ \ \ \ \ Ms<0,
\end{cases}
\\
\psi^-_{Ms}(w)=\begin{cases}
C^-_{Ms} e^{-\frac{w}{2}}w^{\frac{(Ms+1)}{2}}L^{Ms+\frac{1}{2}}_n(w), \ \ \ \ \ \ \ Ms>0,
\\
C^-_{Ms} e^{-\frac{w}{2}}w^{-\frac{Ms}{2}}L^{-Ms-\frac{1}{2}}_n(w), \ \ \ \ \ \ \ Ms<0,
\end{cases}
\end{cases}
\end{equation}
where $C^u_{Ms}$ is a new constant, defined as $C^u_{Ms}\equiv c^u_{Ms}c_6>0$, and it will be the normalization constant(s) of the system.

Now, with respect to the energy spectrum, it can be obtained from a relation (or quantization condition) given by $\bar{E}^u_{Ms}=-n$ (i.e. the last term of the Laguerre equation must be equal to a nonpositive integer), in which it is a mandatory requirement that the Laguerre polynomials must obey to be a finite series (with this, the spinor can then be normalized). Thus, through this quantization condition, we obtain the following relativistic energy spectrum (or relativistic LLs) for the charged DO in a rotating frame in the Bonnor-Melvin-Lambda spacetime
\begin{equation}\label{spectrum}
E^{\lambda}_{n,M,s,u}=s\sqrt{8\Lambda}\sigma\Omega N+\lambda\sqrt{\left(s\sqrt{8\Lambda}\sigma\Omega N\right)^2+m_{eff}^2+4m_0\left(\omega-s\frac{\omega_c}{2}\right)N},
\end{equation}
where we define
\begin{equation}\label{N}
N\equiv\left(n+\frac{1}{4}+\frac{1\pm(2Ms-u)}{4}-\frac{2Ms+u}{4}\right)\geq 0, \ \ m_{eff}=\left(m_0+\sqrt{\frac{\Lambda}{2}}\sigma\Omega\right)>0,
\end{equation}
being $\lambda=\pm 1$ a parameter (``energy parameter'') which represents the positive-energy states/solutions ($\lambda=+1$), or the particle states (electron/DO), whose associated spectrum is given by $E_{particle}=E^+=\vert E^+\vert>0$, as well as the negative-energy states/solutions ($\lambda=-1$), or the antiparticle states (positron/anti-DO), whose associated spectrum is given by $E_{antiparticle}=-E^-=\vert E^-\vert>0$ (i.e. a particle with negative energy is interpreted as an antiparticle with positive energy \cite{Greiner,Strangebook}), and $N=N_{eff}=N(\sigma,\Lambda)$ is an effective quantum number (already that depends on all other quantum numbers as well as of $\sigma$ and $\Lambda$ and of the parameters $s$ and $u$). In particular, the positive integer $n\geq 0$ is often called the radial quantum number (not to be confused with the principal quantum number $n\geq 1$), which can also be symbolized by $n_r$ \cite{Bhuiyan,Schattschneider,CQG}, or even called the Landau (level) index \cite{Miransky,Bruckmann}, and represents, basically (or counts), the number of nodes (``Dirac eigenmodes'') that the radial/spatial part of the spinor has, that is, that $\psi^u_{Ms}(r)$ has. So, we see that the spectrum is quantized in terms of the quantum numbers $n$ and $m_j$, and explicitly depends on the angular and cyclotron frequencies $\omega$ and $\omega_c$, angular velocity $\Omega$, spin, spinorial and curvature parameters $s$, $u$ and $\sigma$, effective rest mass $m_{eff}$ (or effective rest energy $E_{eff}$), and on the cosmological constant $\Lambda$, respectively.

With respect to the first term of the spectrum (or first quadratic term inside of the square root), we can say that it is one noninertial contribution of the rotating frame for the spectrum (i.e. the contribution of the Sagnac effect \cite{Strange,C,Santos}), where such contribution can be represented by a (quantized) rotational spectrum, given by: $E_\Omega=E_{rot}(\Omega)\equiv s\sqrt{8\Lambda}\sigma\Omega N$ (we can call it the ``Sagnac spectrum'', i.e. $E_{Sagnac}=E_{\Omega}$). However, unlike Ref. \cite{Oli2}, where such spectrum is negative ($E_{Sagnac}=-E_{\Omega}$), here, is positive because of the angular frequency of the DO. Consequently, this implies that the spectrum for the particle and antiparticle (or DO and anti-DO) are different ($\vert E^+\vert\neq\vert E^-\vert$), that is, the spectrum is asymmetric (or shifted) about $E=0$ (a ``point of reference'' that separates $E>0$ of $E<0$). In other words, the presence of noninertial effects of a rotating frame breaks the symmetry of energy levels (i.e. an ``observer'' can distinguish the particle by its respective antiparticle) \cite{Oli2,Santos,Castro,Bakke2,Oli9,Oli10,Strange}. Therefore, this asymmetry does not emphasize the equilibrium/equality between particle and antiparticle in the system (indeed, this is something common when a spectrum has an extra term (or more) outside the square root \cite{CQG,Oli3,Oli2,Oli4}). For example, if the pair creation process were applied here, one of them would need more energy than the other to be created. However, for $\Omega=0$ (absence of the rotating frame), this asymmetry disappears, that is, we now have a symmetrical spectrum ($\vert E^+\vert=\vert E^-\vert$) about $E=0$ and, therefore, now exists an equilibrium/equality between particle and antiparticle in the system (i.e. now an ``observer'' cannot distinguish the particle by its respective antiparticle) \cite{Castro}. Still, with respect to the noninertial contribution, we see that the quadratic term in \eqref{spectrum}, given by $E_\Omega^2$, only makes sense to be used for high quantum numbers ($N\gg 1$), since we have the condition $\Lambda\ll1 $ and $\Omega\ll 1$. In this way, we can safely highlight/ignore this term for low quantum numbers. Also, another noninertial contribution is in the rest mass term, which is due to the spin-rotation coupling term. That is, the presence of $\Omega$ also modifies/shifts the rest mass such as $m_0\to m_0+\sqrt{\frac{\Lambda}{2}}\sigma\Omega$ (i.e. rest mass modified by the spin-rotation coupling), where we can consider this noninertial term as a ``zero-point energy'' (in analogy with the ground state of the QHO, i.e. $E_0=\sqrt{\frac{\Lambda}{2}}\sigma\Omega$) \cite{Strange}. Furthermore, it is also important to comment that the parameter $s$ allows defining two ranges for $\omega_c$ (or $B_0$) as well as for $\omega$, i.e. depending on the value of $s$, we have a specific range for $\omega_c$ and $\omega$. For example, ensuring that the condition $(\omega-s\omega_c/2)>0$ is always satisfied (this even avoids possible complex/imaginary energies), implies that for $s=+1$ we have $0\leq\omega_c<2\omega$ and $\omega_c/2<\omega<\infty$, and for $s=-1$, we have $0\leq\omega_c<\infty$ and $0\leq\omega<\infty$, respectively.

On the other hand, through the spectrum \eqref{spectrum} we can have two possible configurations for each component of the spinor depending on the values of $Ms$ (this can be easily analyzed knowing that $(1+(2Ms-u))>0$ for $Ms>0$, and $(1-(2Ms-u))>0$ for $Ms<0$, such as was done to obtain \eqref{solution2}). In summary, these configurations are given by
\begin{equation}\label{spectrum2}
E^{\lambda}_{n,M,s,u}= \begin{cases}
E^{\lambda}_{n,M,s,+}=\begin{cases}
s\sqrt{8\Lambda}\sigma\Omega n+\lambda\sqrt{\left(s\sqrt{8\Lambda}\sigma\Omega n\right)^2+\left(m_0+\sqrt{\frac{\Lambda}{2}}\sigma\Omega\right)^2+4m_0\bar{\omega}n}, \ \ \ \ \ \ \ \ \ \ \ \ \ \ \ \ \ \ \ \ \ \ \ \ \ \ \ Ms>0, 
\\
s\sqrt{8\Lambda}\sigma\Omega N^++\lambda\sqrt{\left(s\sqrt{8\Lambda}\sigma\Omega N^+\right)^2+\left(m_0+\sqrt{\frac{\Lambda}{2}}\sigma\Omega\right)^2+4m_0\bar{\omega}N^+}, \ \ \ \ \ \ \ \ \ \ \ \ \ \ \ \ \ \ Ms<0,
\end{cases}
\\
E^{\lambda}_{n,M,s,-}=\begin{cases}
s\sqrt{8\Lambda}\sigma\Omega (n+1)+\lambda\sqrt{\left(s\sqrt{8\Lambda}\sigma\Omega (n+1)\right)^2+\left(m_0+\sqrt{\frac{\Lambda}{2}}\sigma\Omega\right)^2+4m_0\bar{\omega}(n+1)}, \ \ \ \ Ms>0,
\\
s\sqrt{8\Lambda}\sigma\Omega N^-+\lambda\sqrt{\left(s\sqrt{8\Lambda}\sigma\Omega N^-\right)^2+\left(m_0+\sqrt{\frac{\Lambda}{2}}\sigma\Omega\right)^2+4m_0\bar{\omega}N^-}, \ \ \ \ \ \ \ \ \ \ \ \ \ \ \ \ \ \ Ms<0,
\end{cases}
\end{cases}
\end{equation}
where $Ms>0$ can be ``interpreted'' as the spectrum for a DO with both positive angular momentum and spin ($M>0$ and $s=+1$), or with both negative angular momentum and spin ($M<0$ and $s=-1$), described by $\psi^u_{Ms>0}(w)$, while $Ms<0$ can be ``interpreted'' as the spectrum for a DO with positive angular momentum and negative spin ($M>0$ and $s=-1$), or with negative angular momentum and positive spin ($M<0$ and $s=+1$), described by $\psi^u_{Ms<0}(w)$, and we define for simplicity that $N^\pm=N^u=\left(n+\frac{1}{2}-Ms\right)$, and $\bar{\omega}\equiv(\omega-s\frac{\omega_c}{2})>0$. So, we see that for $Ms<0$, the two spectra generated are the same ($E^\lambda_{n,Ms<0,+}=E^\lambda_{n,Ms<0,-}$), that is, the spectrum is the same for both two components (or is independent of the chosen component). In this case, as the parameter $\sigma$ (or better, $\sqrt{2\Lambda}\sigma$) is ``tied/linked'' to the quantum number $m_j$, we say that such spectra have their degeneracies broken (such as happens in the case of cosmic strings \cite{Oli2,Santos,Bakke2} or Gödel-type spacetime \cite{Oli3,CQG}). In other words, the presence of $\sigma$ (or $\sqrt{2\Lambda}\sigma$) breaks/destroys the degeneracy of all the energy levels (i.e. there is no longer a ``well-defined degeneracy''). Already for $Ms>0$, the two spectra generated now are not the same ($E^\lambda_{n,Ms>0,+}\neq E^\lambda_{n,Ms>0,-}$), that is, the spectrum is different for each component (or dependent on the chosen component). In this case, the spectrum for $Ms>0$ and $u=+1$ is the only one whose ground state ($n=0$) does not depend on $m_j$, $\omega$, and $\omega_c$, which is given by $E^\lambda_{0,Ms>0,+}=\lambda\left(m_0+\sqrt{\frac{\Lambda}{2}}\sigma\Omega\right)$, i.e. it does not matter whether such quantities increase or decrease, the state ground remains unchanged (therefore, we have $E^\lambda_{n,Ms>0,+}<E^\lambda_{n,Ms>0,-}$ for all $n$ and $m_j$). However, this ground state should not be accepted here (should be prohibited) because it implies in a non-physical/non-normalizable solution, i.e. $\psi^\pm(r)\to\infty$ (see \eqref{EDO3} and \eqref{EDO4}).

Besides, comparing the spectrum \eqref{spectrum}, or \eqref{spectrum2}, with other works/papers, we verified that our spectrum generalizes several (particular) cases in the literature. For example, for $\omega_c=0$ (absence of the external magnetic field) and $\sigma=1$ and $\Lambda=1/2$ (absence of the Bonnor-Melvin-Lambda spacetime), with $s=u=+1$ and $m_j>0$ (i.e. first configuration in \eqref{spectrum2}), we have (with $c$ and $\hbar$ restored)
\begin{equation}\label{spectrum3}
E^{\lambda}_{n}=2n\hbar\Omega+\lambda\sqrt{(2n\hbar\Omega)^2+\left(m_0 c^2+\frac{\hbar\Omega}{2}\right)^2+4n m_0c^2\hbar\omega},
\end{equation}
which is exactly the spectrum of Ref. \cite{Strange} (i.e. the spectrum of the rotating DO in the Minkowski spacetime).

For $\Omega=0$ (absence of the rotating frame), $\sigma=1$ and $\Lambda=0$ (absence of the Bonnor-Melvin-Lambda spacetime), with $u=+1$ and $\omega_c\to -\omega_c$, being $\omega_c\geq 0$ (i.e. first and second configuration in \eqref{spectrum2}), we have
\begin{equation}\label{spectrum4}
E^{\lambda}_{n,m_j,s}=\lambda\sqrt{m_0^2+4 m_0\left(\omega+s\frac{\omega_c}{2}\right)\left(n+\frac{1\mp 1}{4}+\frac{\pm m_j s-m_j s}{2}\right)},
\end{equation}
or in terms of the Heaviside step function $\Theta (-m_j s)$ (with $\Theta (-m_j s)=0$ for $m_j s>0$ and $\Theta (-m_j s)=1$ for $m_j s<0$), such as
\begin{equation}\label{spectrum5}
E^{\lambda}_{n,m_j,s}=\lambda\sqrt{m_0^2+4 m_0\left(\omega+s\frac{\omega_c}{2}\right)\left(n-\Theta (-m_j s)\left(m_j s-\frac{1}{2}\right)\right)}.
\end{equation}

In particular, the spectrum above for $\omega_c=0$ reduces exactly to the spectrum of Ref. \cite{Villalba1} (i.e. the spectrum of the chargeless DO), while for $s=+1$, reduces exactly to the spectrum of Ref. \cite{Villalba2} (i.e. the spectrum of the charged DO). On the other hand, using the fact that $m_j=m_l+s/2$ ($m_l=0,\pm 1,\pm 2,\ldots$), we can then rewrite the spectrum \eqref{spectrum4}, or \eqref{spectrum5}, with $\omega_c=0$, in the following form
\begin{equation}\label{spectrum6}
E^{\lambda}_{n,m_l,s}=\lambda\sqrt{m_0^2+4 m_0\omega(n+\vert m_l\vert-m_l s)},
\end{equation}
which is exactly the spectrum of Ref. \cite{Andrade} (i.e. the spectrum of the chargeless DO written in terms of $m_l$). 

Already, for $\Omega=\omega=0$, $\sigma=1$ and $\Lambda=1/2$, with $\omega_c\to -\omega_c$ ($\omega_c\geq 0$), $s=+1$ and $m_j>0$ (i.e. first and third configuration in \eqref{spectrum2}), we have
\begin{equation}\label{spectrum7}
E^{\lambda}_{n,u}=\lambda\sqrt{m_0^2+2 m_0\omega_c\left(n+\frac{1-u}{2}\right)}.
\end{equation}

In particular, the spectrum above for $u=+1$ reduces exactly to the spectrum of Ref. \cite{Haldane,Schakel,Miransky}, while for $u=-1$ reduces exactly to the spectrum of Ref. \cite{Bermudez}, i.e. we have the LLs for the relativistic quantum Hall effect. However, for $m_0=0$ and $u/2\to s_z$, we have the LLs for the massless relativistic quantum Hall effect (or massless LLs) \cite{Bruckmann}.

Now, let us graphically analyze the behavior of the spectrum as a function of the magnetic field $B_0$ ($0\leq B_0<\infty$ and $0\leq B_0<2\omega$), angular frequency $\omega$ ($0\leq\omega<\infty$ and $\omega_c/2<\omega<\infty$), angular velocity $\Omega$ ($0\leq\Omega\leq 0.05$), and cosmological constant $\Lambda$ ($0<\Lambda\leq 0.05$) for three different values of $n$ and $m_j$. Regarding the values of these quantum numbers, they allow us to analyze the behavior of the spectrum in two different scenarios, that are: while $n$ varies ($n=0,1,2$), $m_j$ remains fixed ($m_j=\pm 1/2$), and while $m_j$ varies ($m_j=\pm 1/2, \pm 3/2, \pm 5/2$), $n$ remains fixed ($n=0$), respectively. Besides, we also consider the maximum spectrum for the system (or the spectrum with the maximum energies), whose spectrum is given by the second or fourth configuration in \eqref{spectrum2}. However, as we mentioned previously, for low quantum numbers (our case), we can safely discard the first quadratic term inside the square root, and even the spin-orbit coupling term, that is, $\left(m_0+\sqrt{\frac{\Lambda}{2}}\sigma\Omega\right)^2\approx m_0^2$. So, adopting for simplicity that $m_0=1$ and $e=1$, i.e. an ``unit mass and charge'' (a ``general system of natural units'') and $\sigma=1$ \cite{Castro}, we have the following spectrum
\begin{equation}\label{spectrum8}
E^{\lambda}_{n,m_j,s}=s\sqrt{8\Lambda}\Omega\left(n+\frac{1}{2}-\frac{m_j s}{\sqrt{2\Lambda}}\right)+\lambda\sqrt{1+4\left(\omega-s\frac{B_0}{2}\right)\left(n+\frac{1}{2}-\frac{m_j s}{\sqrt{2\Lambda}}\right)}, \ \ \ \ (\omega-sB_0/2>0),
\end{equation}
where we consider $m_j>0$ and $s=-1$ as well as $m_j<0$ and $s=+1$. With this, we have two configurations for the spectrum \eqref{spectrum8}, given as follows (and both will be used for each specific scenario)
\begin{equation}\label{spectrum9}
E^{\lambda}_{n,m_j,s}= \begin{cases}
-\sqrt{8\Lambda}\Omega\left(n+\frac{1}{2}+\frac{\vert m_j\vert}{\sqrt{2\Lambda}}\right)+\lambda\sqrt{1+4\left(\omega+\frac{B_0}{2}\right)\left(n+\frac{1}{2}+\frac{\vert m_j\vert}{\sqrt{2\Lambda}}\right)}, \ \ \ \ m_j>0, \ s=-1,
\\
\sqrt{8\Lambda}\Omega\left(n+\frac{1}{2}+\frac{\vert m_j\vert}{\sqrt{2\Lambda}}\right)+\lambda\sqrt{1+4\left(\omega-\frac{B_0}{2}\right)\left(n+\frac{1}{2}+\frac{\vert m_j\vert}{\sqrt{2\Lambda}}\right)},  \ \ \ \ \ \ \ m_j<0, \ s=+1.
\end{cases}
\end{equation}

\begin{figure}[ht]
    \centering
    \subfigure{%
        \includegraphics[width=0.49\textwidth]{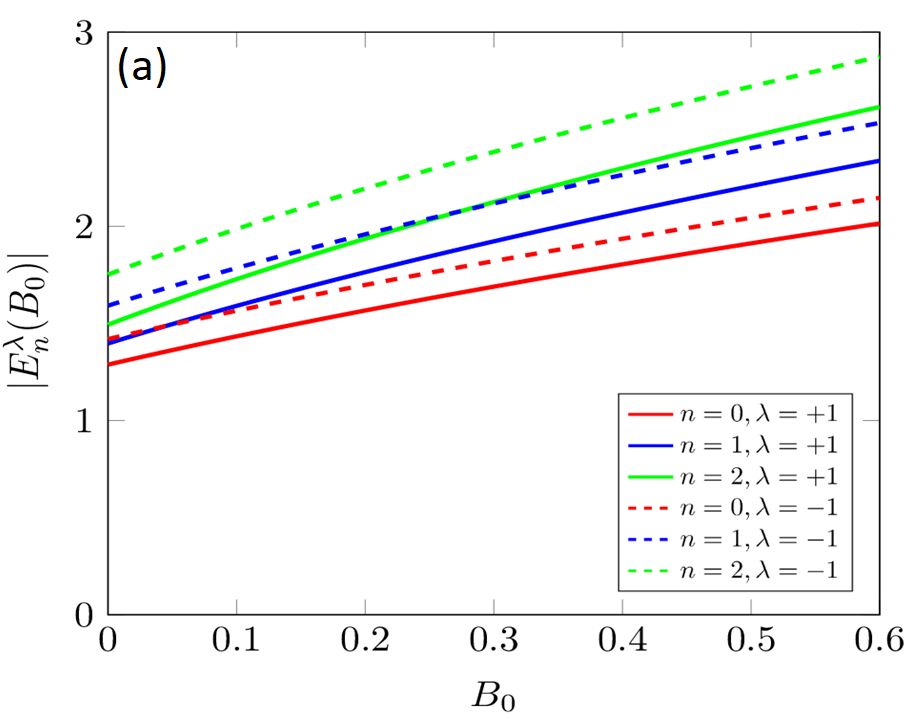}
    }\hfill
    \subfigure{%
        \includegraphics[width=0.49\textwidth]{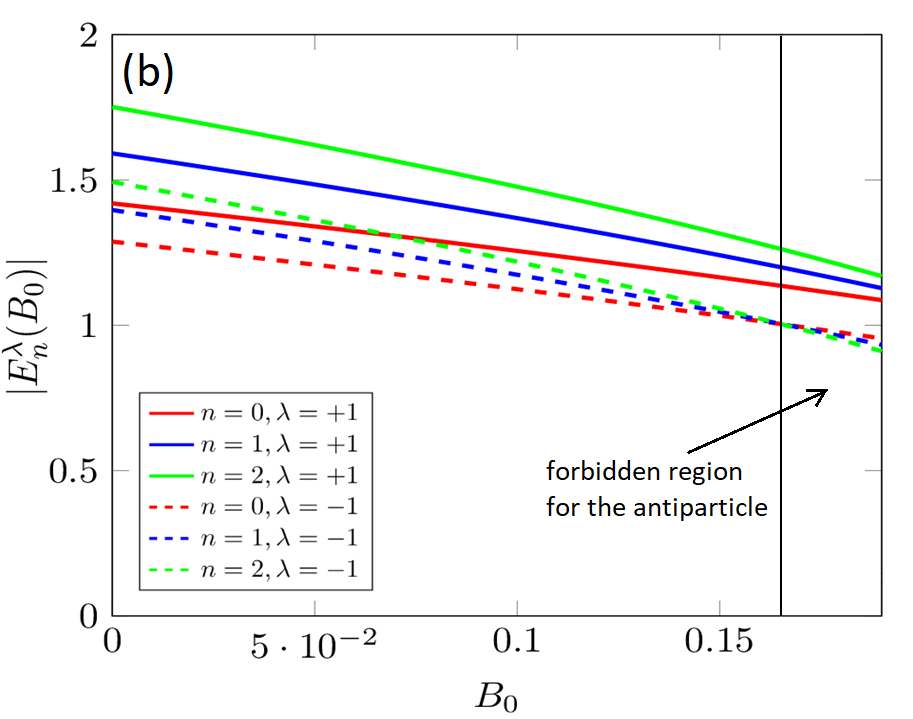}
    }
    \caption{Behavior of $\vert E_n^\lambda (B_0)\vert$ vs. $B_0$ for three different values of $n$, where (a) is for $m_j=+1/2$ and $s=-1$ (with $0\leq B_0\leq 0.6$), and (b) is for $m_j=-1/2$ and $s=+1$ (with $0\leq B_0\leq 0.19$).}
    \label{fig1}
\end{figure}

So, in Fig. \ref{fig1} we have the behavior of $\vert E_n^\lambda (B_0)\vert$ vs. $B_0$ for three different values of $n$, with $n=0$ for the ground state, $n=1$ for the first excited state, and $n=2$ for the second excited state, where (a) is for $m_j=+1/2$ and $s=-1$, with $0\leq B_0\leq 0.6$, and (b) is for $m_j=-1/2$ and $s=+1$, with $0\leq B_0\leq 0.19$ (we use $\Lambda=\Omega=0.05$, and $\omega=0.1$). According to Fig. \ref{fig1}-(a), we see that the energies of the particle ($\lambda=+1$) and antiparticle ($\lambda=-1$) increase with the increase of $n$ (as it should be); consequently, the energy difference (or spacing) between two consecutive levels is positive 
(i.e. $\Delta E_n^\pm=\vert E_{n+1}^\pm\vert-\vert E_n^\pm \vert>0$). However, the energies of the antiparticle are greater than those of the particle ($\vert E^-_n\vert>\vert E^+_n\vert$), where the energy difference between them (for the same energy level) is constant in all the range of the magnetic field (i.e. $\Delta E=\vert E^-_n (B_0)\vert-\vert E^+_n (B_0)\vert=constant>0$). As a matter of curiosity, in the range $0\leq B_0\leq 0.1$, the energy of the particle for $n=1$ is practically equal to the energy of the antiparticle for $n=0$ ($\vert E^+_1\vert\approx\vert E^-_{0}\vert$, i.e. the solid blue and dashed red lines coincide/overlap), while in the range $0\leq B_0\leq 0.2$, the energy of the particle for $n=2$ is less than the energy of the antiparticle for $n=1$ ($\vert E^+_2\vert<\vert E^-_{1}\vert$, i.e. the dashed blue line is above the solid green line). Furthermore, we also see that the energies increase as a function of $B_0$ and, therefore, both the energies of the particle and antiparticle increase with the increase of $B_0$ (here, the function of $B_0$ is to increase the energies); consequently, the energy variation (not to be confused with energy difference) for a given range of $B_0$ is always positive (i.e. $\Delta E(B_0)=E(B_0+\delta B_0)-E(B_0)=E_{final}(B_0)-E_{initial}(B_0)>0$). On the other hand, the energy difference (for the particle or antiparticle) also increases as a function of $B_0$ (i.e. $\Delta E_n^\pm(B_0+\delta B_0)>\Delta E_n^\pm(B_0)$). In other words, for a given value of $B_0$, we have a specific/unique value for $\Delta E_n^\pm(B_0)$.

In Fig. \ref{fig1}-(b), we see that the behavior of the energies presents similarities and differences with those of Fig. \ref{fig1}-(a). For example, similar to Fig. \ref{fig1}-(a), the energies of the particle increase with the increase of $n$ (as it should be). However, unlike Fig. \ref{fig1}-(a), the energies of the antiparticle can increase or decrease with the increase of $n$, that is, in the range $0\leq B_0\approx 0.165$, the energies of the antiparticle increase with the increase of $n$ (as it should be), while in the range $0.165\approx B_0\leq 0.19$ (region posterior to the black vertical line), the energies decrease with the increase of $n$ (i.e. $\vert E^-_2\vert <\vert E^-_1\vert <\vert E^-_0\vert $). Therefore, in the range $0.165\approx B_0\leq 0.19$, we have an anomaly (quantumly incoherent or unrealizable), or better, a forbidden region for the antiparticle. Still regarding the value $B_0\approx 0.165$, we can say that it is a ``convergence point'', that is, all the energy levels of the antiparticle converge to this point (and after this convergence, the anomaly appears). Besides, the energies of the particle are greater than those of the antiparticle ($\vert E^+_n\vert>\vert E^-_n\vert$), where the energy difference between them is also constant (i.e. $\Delta E=\vert E^+_n (B_0)\vert-\vert E^-_n (B_0)\vert=constant>0$). As a matter of curiosity, in all the range of $B_0$, the energy of the particle for $n=0$ is greater than the energy of the antiparticle for $n=0,1$ ($\vert E^+_0\vert>\vert E^-_{0,1}\vert$, i.e. the solid red line is above the dashed blue and red lines), while for $B_0>0.075$, the energy of the particle for $n=0$ is greater than the energy of the antiparticle for $n=2$ ($\vert E^+_0\vert>\vert E^-_{2}\vert$, i.e. the solid red line is above the dashed green line), respectively. Furthermore, unlike Fig. \ref{fig1}-(a), we see that the energies decrease as a function of $B_0$ and, therefore, both the energies of the particle and antiparticle decrease with the increase of $B_0$ (here, the function of $B_0$ is to decrease the energies); consequently, the energy variation for a given range of $B_0$ is always negative (i.e. $\Delta E(B_0)=E(B_0+\delta B_0)-E(B_0)=E_{final}(B_0)-E_{initial}(B_0)<0$). On the other hand, the energy difference also decreases as a function of $B_0$ (i.e. $\Delta E_n^\pm(B_0+\delta B_0)<\Delta E_n^\pm(B_0)$). In addition, it is also important to comment that in both Fig \eqref{fig1}-(a) and \eqref{fig1}-(b), the condition $\left(\omega-s\frac{\omega_c}{2}+s\frac{\sqrt{2\Lambda}\sigma\Omega E}{m_0}\right)>0$ is satisfied (as it should be) according to the values used/adopted for $\omega$, $\Omega$, and $\Lambda$ (in fact, these quantities must be adjusted/chosen so that the condition is always satisfied). Now, comparing (a) with (b), we see that the energies are greater for (a), that is, when the particle/antiparticle has positive angular momentum and negative spin.

\begin{figure}[ht]
    \centering
    \subfigure{%
        \includegraphics[width=0.49\textwidth]{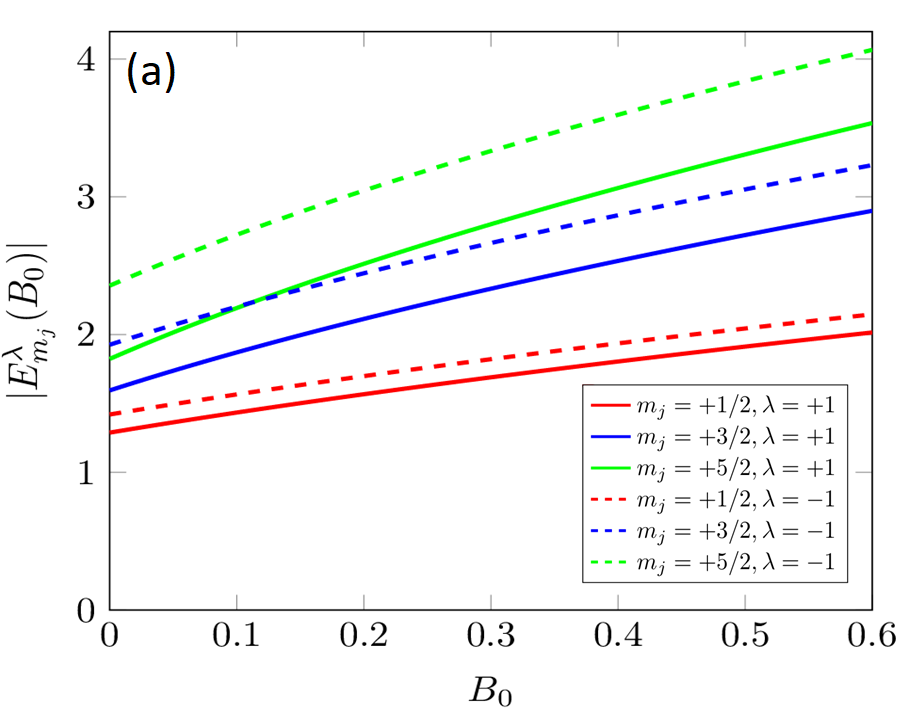}
    }\hfill
    \subfigure{%
        \includegraphics[width=0.49\textwidth]{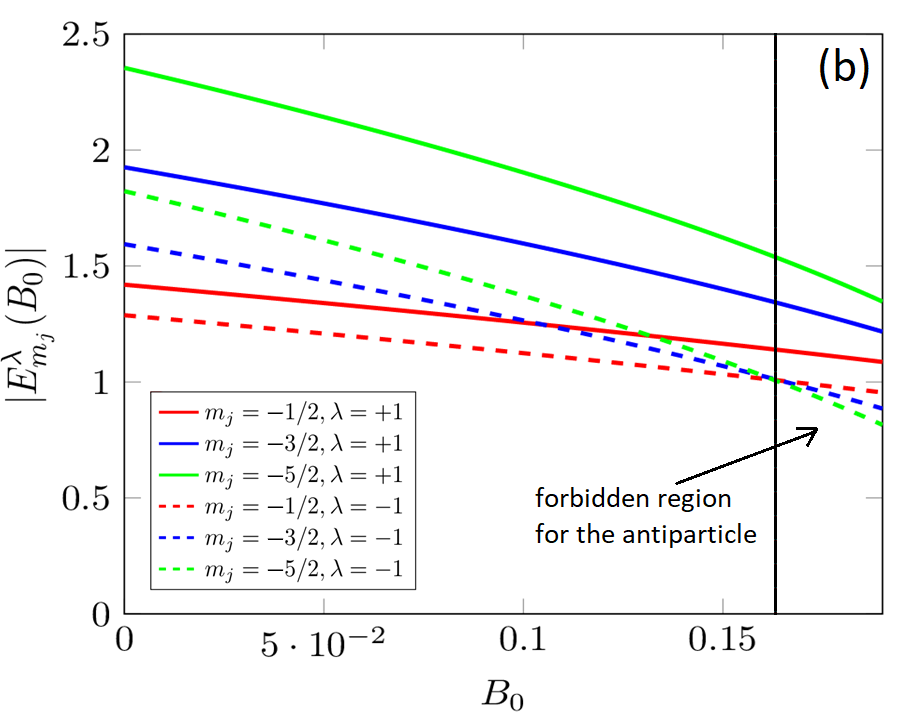}
    }
    \caption{Behavior of $\vert E_{m_j}^\lambda (B_0)\vert$ vs. $B_0$ for three different values of $m_j$ (with $n=0$), where (a) is for $m_j=+1/2,+3/2,+5/2$ and $s=-1$ (with $0\leq B_0\leq 0.6$), and (b) is for $m_j=-1/2,-3/2,-5/2$ and $s=+1$ (with $0\leq B_0\leq 0.19$).}
    \label{fig2}
\end{figure}

Already in Figs. \ref{fig2}-(a) and \ref{fig2}-(b), we have the behavior of $\vert E_{m_j}^\lambda (B_0)\vert$ vs. $B_0$ for three different values of $m_j$ (with $n=0$), where (a) is for $m_j=+1/2,+3/2,+5/2$ and $s=-1$, with $0\leq B_0\leq 0.6$, and (b) is for $m_j=-1/2=-3/2,-5/2$ and $s=+1$, with $0\leq B_0\leq 0.19$ (we use $\Lambda=\Omega=0.05$, and $\omega=0.1$). So, we see that the behavior of the energies presents more similarities than differences with those of Figs. \ref{fig1}-(a) and \ref{fig1}-(b). Indeed, the difference (with the exception of $m_j=\pm 1/2$) is basically in the energy values (or difference/spacing between them), consequently, we see that the effect generated by the variation of the quantum number $m_j$ is greater than the effect generated by the variation of the quantum number $n$. In other words, we can say that Figs. \ref{fig2}-(a) and \ref{fig2}-(b) are the ``enlargement'' (or ``enlarged version'') of Figs. \ref{fig1}-(a) and \ref{fig1}-(b), i.e. they only have higher energies. In addition, it is also important to comment that in both Figs. \eqref{fig2}-(a) and \eqref{fig2}-(b), the condition $\left(\omega-s\frac{\omega_c}{2}+s\frac{\sqrt{2\Lambda}\sigma\Omega E}{m_0}\right)>0$ is also satisfied (as it should be). Now, comparing (a) with (b), we see that the energies are greater for (a), that is, when the particle/antiparticle has positive angular momentum and negative spin.

\begin{figure}[ht]
    \centering
    \subfigure{%
        \includegraphics[width=0.49\textwidth]{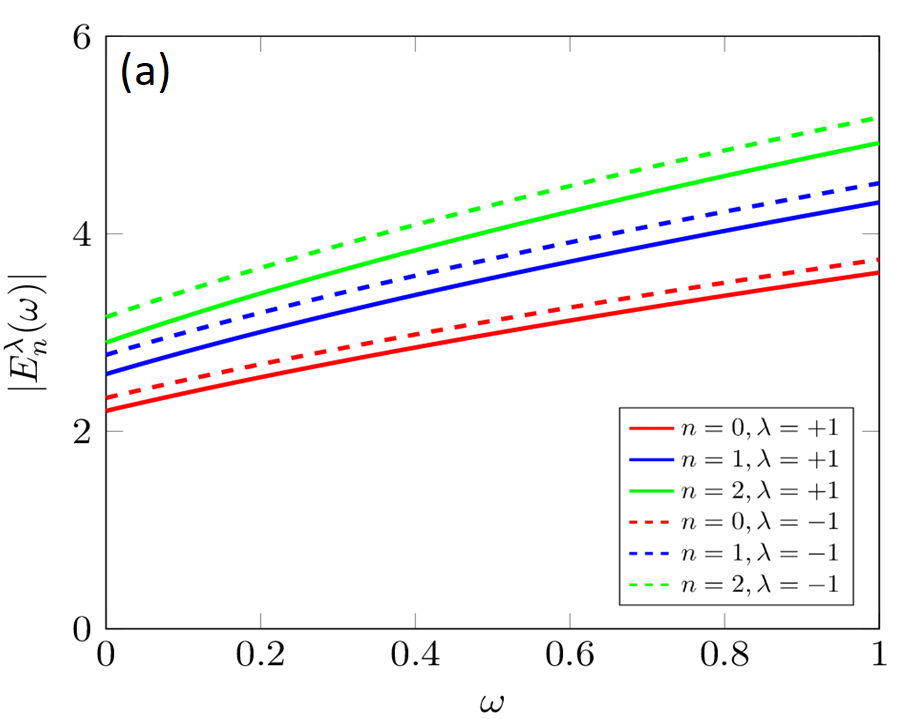}
    }\hfill
    \subfigure{%
        \includegraphics[width=0.49\textwidth]{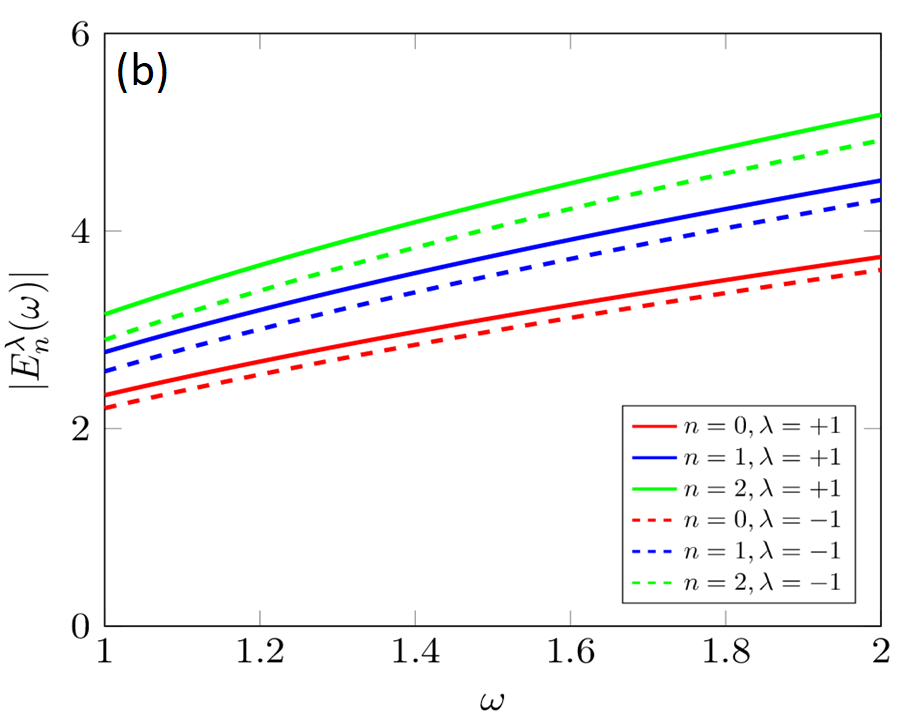}
    }
    \caption{Behavior of $\vert E_n^\lambda (\omega)\vert$ vs. $\omega$ for three different values of $n$, where (a) is for $m_j=+1/2$ and $s=-1$ (with $0\leq\omega\leq 1$), and (b) is for $m_j=-1/2$ and $s=+1$ (with $1\leq\omega\leq 2$).}
    \label{fig3}
\end{figure}

In Fig. \ref{fig3}, we have the behavior of $\vert E_n^\lambda (\omega)\vert$ vs. $\omega$ for three different values of $n$, where (a) is for $m_j=+1/2$ and $s=-1$, with $0\leq \omega\leq 1$, and (b) is for $m_j=-1/2$ and $s=+1$, with $1\leq\omega\leq 2$ (we use $\Lambda=\Omega=0.05$, and $B_0=1$). So, we see that (a) and (b) have more similarities than differences. For example, in both graphs, the energies increase with the increase of $n$ (as it should be), i.e. we have $\Delta E_n^\pm=\vert E_{n+1}^\pm \vert-\vert E_n^\pm \vert>0$, and also increase as a function of $\omega$ (both the energies of the particle and antiparticle increase with the increase of $\omega$), i.e. we have $\Delta E(\omega)=E(\omega+\delta \omega)-E(\omega)=E_{final}(\omega)-E_{initial}(\omega)>0$. Besides, in both graphs, the energy difference also increases as a function of $\omega$, i.e. we have $\Delta E_n^\pm(\omega+\delta \omega)>\Delta E_n^\pm(\omega)$. However, the (only) difference between (a) and (b) is that in (a), the energies of the antiparticle are greater than those of the particle ($\vert E^-_n\vert>\vert E^+_n \vert$), while in (b), the opposite happens, that is, the energies of the particle are greater than those of the antiparticle ($\vert E^+_n\vert>\vert E^-_n \vert$). With this, we see that the energies of the particle in (a) are equal (same values) to that of the antiparticle in (b), and the energies of the antiparticle in (a) are equal to those of the particle in (b), respectively (however, this only works when we have an equal range for $\omega$, as was the case). Indeed, one graph can be converted into another by exchanging the particle for the antiparticle and vice-versa (particle $\Leftrightarrow$ antiparticle), or even changing the value of $m_j$ and $s$ in such a form $(m_j>0,s=-1)\Leftrightarrow(m_j<0,s=+1)$. Therefore, while the energies of the particle are greater when it has negative angular momentum and positive spin, in the case of the antiparticle, the opposite happens, i.e. when it has positive angular momentum and negative spin, respectively. In addition, it is also important to comment that in both graphs, the condition $\left(\omega-s\frac{\omega_c}{2}+s\frac{\sqrt{2\Lambda}\sigma\Omega E}{m_0}\right)>0$ is also satisfied (as it should be).

\begin{figure}[ht]
    \centering
    \subfigure{%
        \includegraphics[width=0.49\textwidth]{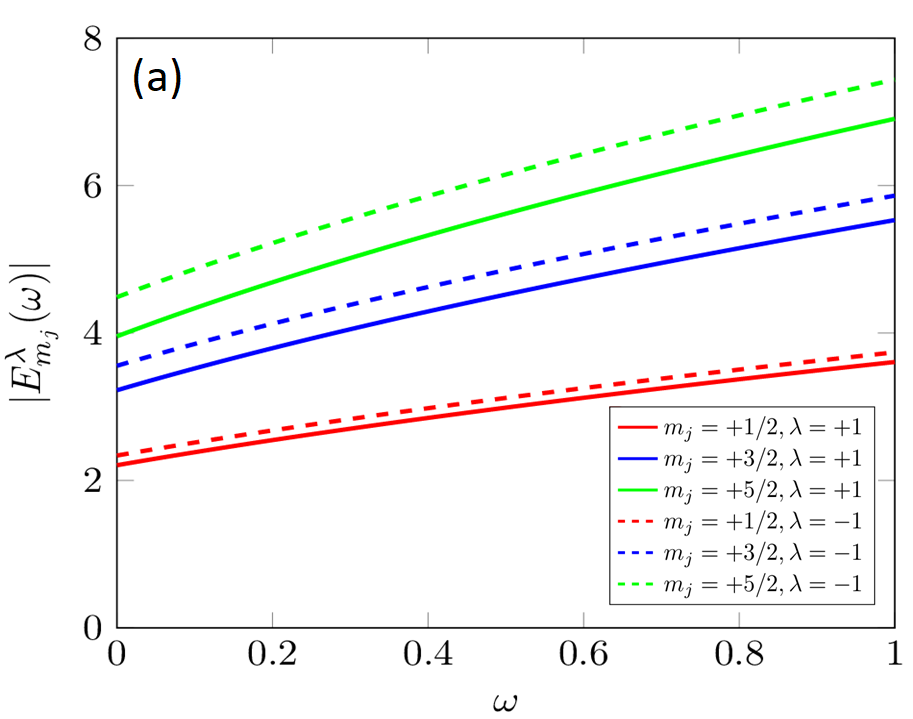}
    }\hfill
    \subfigure{%
        \includegraphics[width=0.49\textwidth]{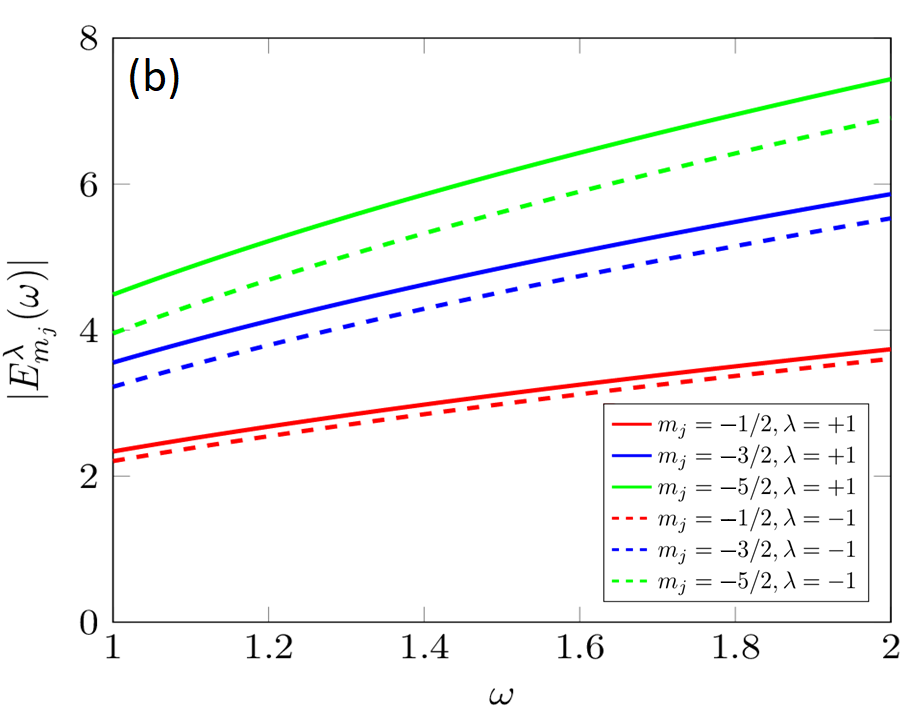}
    }
    \caption{Behavior of $\vert E_n^\lambda (\omega)\vert$ vs. $\omega$ for three different values of $m_j$ (with $n=0$), where (a) is for $m_j=+1/2,+3/2,+5/2$ and $s=-1$ (with $0\leq\omega\leq 1$), and (b) is for $m_j=-1/2,-3/2,-5/2$ and $s=+1$ (with $1\leq\omega\leq 2$).}
    \label{fig4}
\end{figure}

Already in Figs. \ref{fig4}-(a) and \ref{fig4}-(b), we have the behavior of $\vert E_{m_j}^\lambda (\omega)\vert$ vs. $\omega$ for three different values of $m_j$ (with $n=0$), where (a) is for $m_j=+1/2,+3/2,+5/2$ and $s=-1$, with $0\leq\omega\leq 1$, and (b) is for $m_j=-1/2,-3/2,-5/2$ and $s=+1$, with $1\leq\omega\leq 2$ (we use $\Lambda=\Omega=0.05$, and $B_0=1$). So, we see that the behavior of the energies presents more similarities than differences with those of Figs. \ref{fig3}-(a) and \ref{fig3}-(b). Indeed (such as occurs in Figs. \ref{fig1} and \ref{fig2}), the difference (with the exception of $m_j=\pm 1/2$) is basically in the energy values (or difference/spacing between them), consequently, we see that the effect generated by the variation of the quantum number $m_j$ is greater than the effect generated by the variation of the quantum number $n$. In other words, we can say that Figs. \ref{fig4}-(a) and \ref{fig4}-(b) are the ``enlargement'' (or ``enlarged version'') of Figs. \ref{fig3}-(a) and \ref{fig3}-(b), i.e. they only have higher energies. In addition, it is also important to comment that in both Figs. \eqref{fig4}-(a) and \eqref{fig4}-(b), the condition $\left(\omega-s\frac{\omega_c}{2}+s\frac{\sqrt{2\Lambda}\sigma\Omega E}{m_0}\right)>0$ is also satisfied (as it should be). Now, comparing (a) with (b), we see that the energies of the particle are greater for (b), that is, when it has negative angular momentum and positive spin, while in the case of the antiparticle, are greater for (a), that is, when it has positive angular momentum and negative spin, respectively.

\begin{figure}[ht]
    \centering
    \subfigure{%
        \includegraphics[width=0.49\textwidth]{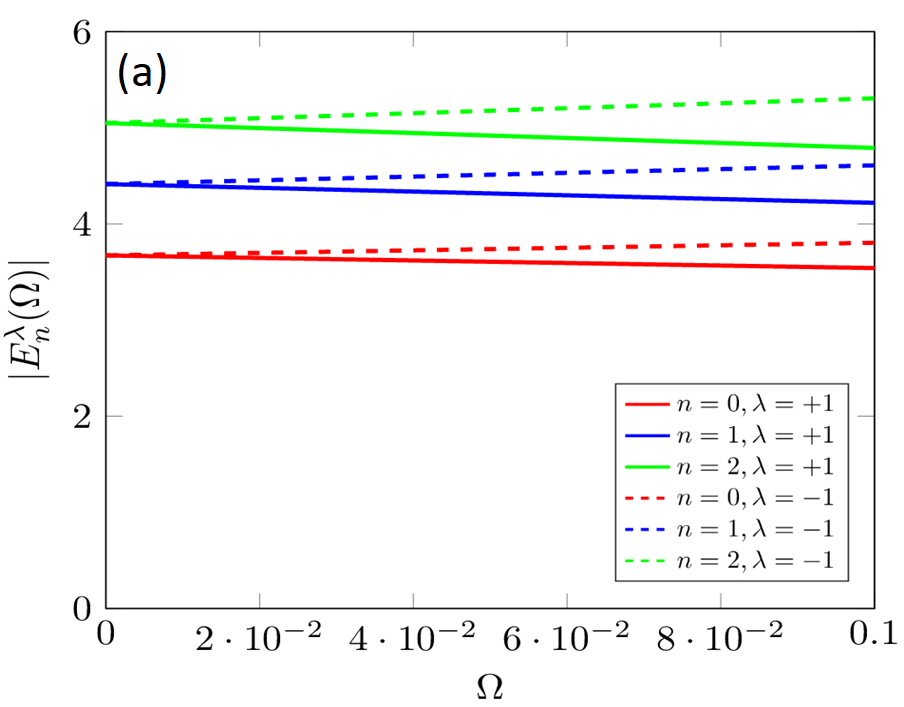}
    }\hfill
    \subfigure{%
        \includegraphics[width=0.49\textwidth]{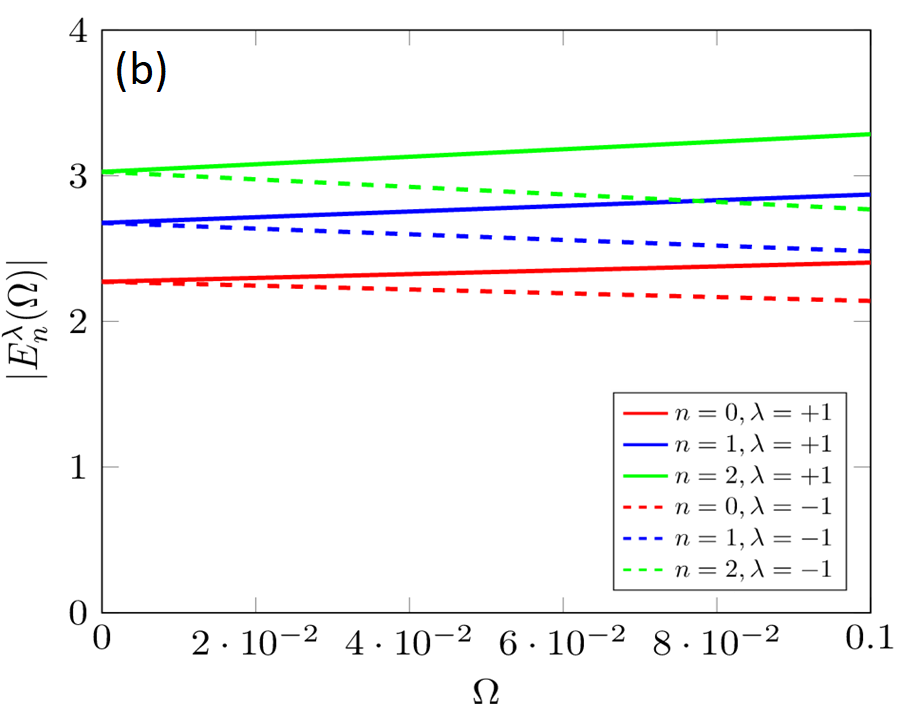}
    }
    \caption{Behavior of $\vert E_n^\lambda (\Omega)\vert$ vs. $\Omega$ for three different values of $n$ (with $0\leq\Omega\leq 0.1$), where (a) is for $m_j=+1/2$ and $s=-1$, and (b) is for $m_j=-1/2$ and $s=+1$.}
    \label{fig5}
\end{figure}

In Fig. \ref{fig5}, we have the behavior of $\vert E_n^\lambda (\Omega)\vert$ vs. $\Omega$ for three different values of $n$ with $0\leq\Omega\leq 0.1$, where (a) is for $m_j=+1/2$ and $s=-1$, and (b) is for $m_j=-1/2$ and $s=+1$ (we use $\Lambda=0.05$, and $\omega=B_0=1$). So, we see that (a) and (b) have some similarities and differences. For example, in both graphs, the energies increase with the increase of $n$ (as it should be), i.e. we have $\Delta E_n^\pm=\vert E_{n+1}^\pm\vert - \vert E_n^\pm \vert>0$. However, in (a) the energies of the antiparticle are greater than those of the particle ($\vert E^-_n \vert>\vert E^+_n \vert$), while in (b) the opposite happens, that is, the energies of the particle are greater than those of the antiparticle ($\vert E^+_n \vert>\vert E^-_n \vert$). Besides, in (a) the energies of the particle decrease as a function of $\Omega$ (i.e. $\Delta E(\Omega)=E(\Omega+\delta \Omega)-E(\Omega)=E_{final}(\Omega)-E_{initial}(\Omega)<0$) and the energies of the antiparticle increase (i.e. $\Delta E(\Omega)=E(\Omega+\delta \Omega)-E(\Omega)=E_{final}(\Omega)-E_{initial}(\Omega)>0$), while in (b) the opposite happens, that is, the energies of the particle increase as a function of $\Omega$ (i.e. $\Delta E(\Omega)=E(\Omega+\delta \Omega)-E(\Omega)=E_{final}(\Omega)-E_{initial}(\Omega)>0$) and the energies of the antiparticle decrease (i.e. $\Delta E(\Omega)=E(\Omega+\delta \Omega)-E(\Omega)=E_{final}(\Omega)-E_{initial}(\Omega)<0$), respectively. On the other hand, unlike the previous graphs, here (in both (a) and (b)) the energy difference as a function of $\Omega$ is always constant, i.e. $\Delta E_n^\pm (\Omega+\delta\Omega)=\Delta E_n^\pm (\Omega)=constant>0$. In other words, in any region (or value) of $\Omega$, the energy difference is always the same. In addition, it is also important to comment that in both graphs, the condition $\left(\omega-s\frac{\omega_c}{2}+s\frac{\sqrt{2\Lambda}\sigma\Omega E}{m_0}\right)>0$ is also satisfied (as it should be). Now, comparing (a) with (b), we see that the energies are greater for (a), that is, when the particle/antiparticle has positive angular
momentum and negative spin.

\begin{figure}[ht]
    \centering
    \subfigure{%
        \includegraphics[width=0.49\textwidth]{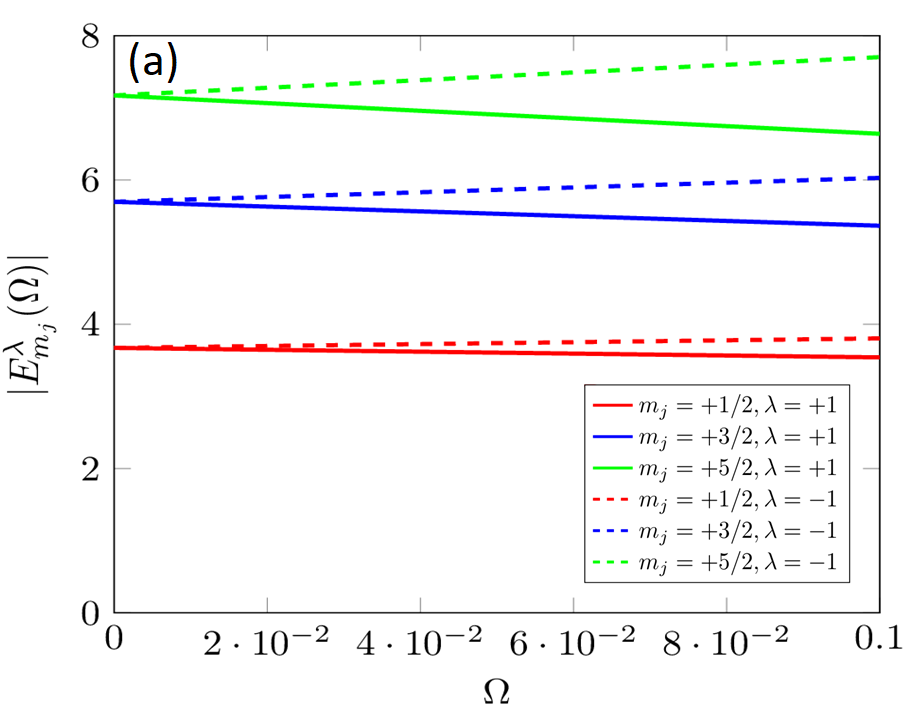}
    }\hfill
    \subfigure{%
        \includegraphics[width=0.49\textwidth]{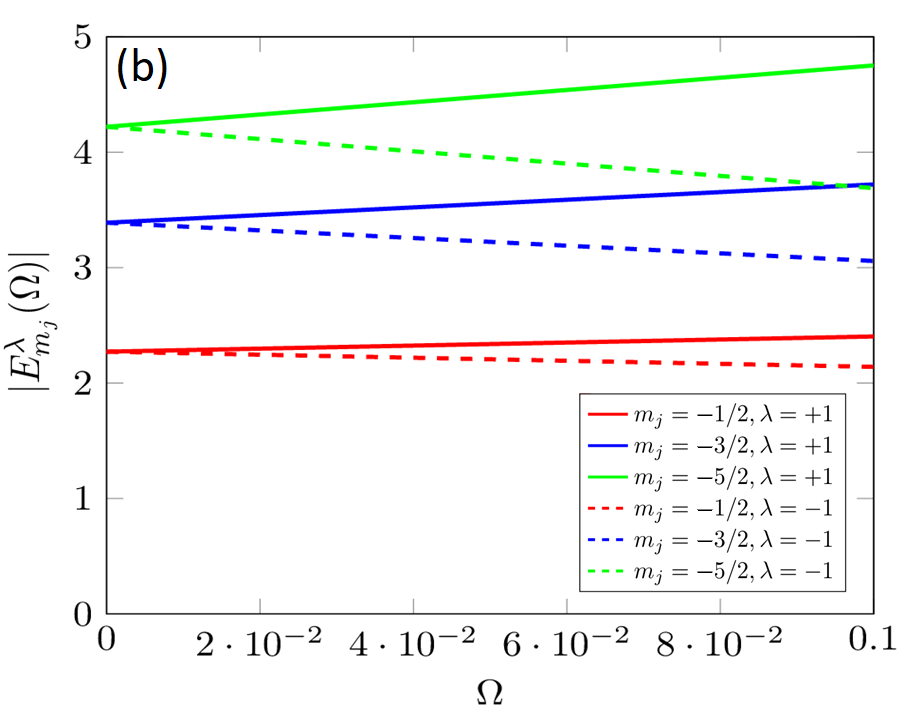}
    }
    \caption{Behavior of $\vert E_{m_j}^\lambda (\Omega)\vert$ vs. $\Omega$ for three different values of $m_j$ (with $n=0$ and $0\leq \Omega\leq 0.1$), where (a) is for $m_j=+1/2,+3/2,+5/2$ and $s=-1$, and (b) is for $m_j=-1/2,-3/2,-5/2$ and $s=+1$.}
    \label{fig6}
\end{figure}

Already in Figs. \ref{fig6}-(a) and \ref{fig6}-(b), we have the behavior of $\vert E_{m_j}^\lambda (\Omega)\vert$ vs. $\Omega$ for three different values of $m_j$ (with $n=0$ and $0\leq \Omega\leq 0.1$), where (a) is for $m_j=+1/2,+3/2,+5/2$ and $s=-1$, with $0\leq\omega\leq 1$, and (b) is for $m_j=-1/2,-3/2,-5/2$ and $s=+1$, with $1\leq\omega\leq 2$ (we use $\Lambda=0.05$, and $\omega=B_0=1$). So, we see that the behavior of the energies presents more similarities than differences with those of Figs. \ref{fig5}-(a) and \ref{fig5}-(b). Indeed (such as occurs in Figs. \ref{fig1}-\ref{fig4}), the difference (with the exception of $m_j=\pm 1/2$) is basically in the energy values (or difference/spacing between them), consequently, we see that the effect generated by the variation of the quantum number $m_j$ is greater than the effect generated by the variation of the quantum number $n$. In other words, we can say that Figs. \ref{fig6}-(a) and \ref{fig6}-(b) are the ``enlargement'' (or ``enlarged version'') of Figs. \ref{fig5}-(a) and \ref{fig5}-(b), i.e. they only have higher energies. In addition, it is also important to comment that in both Figs. \eqref{fig6}-(a) and \eqref{fig6}-(b), the condition $\left(\omega-s\frac{\omega_c}{2}+s\frac{\sqrt{2\Lambda}\sigma\Omega E}{m_0}\right)>0$ is also satisfied (as it should be). Now, comparing (a) with (b), we see that the energies of the particle are greater for (b), that is, when it has negative angular momentum and positive spin, while in the case of the antiparticle, are greater for (a), that is, when it has positive angular momentum and negative spin, respectively.

\begin{figure}[ht]
    \centering
    \subfigure{%
        \includegraphics[width=0.49\textwidth]{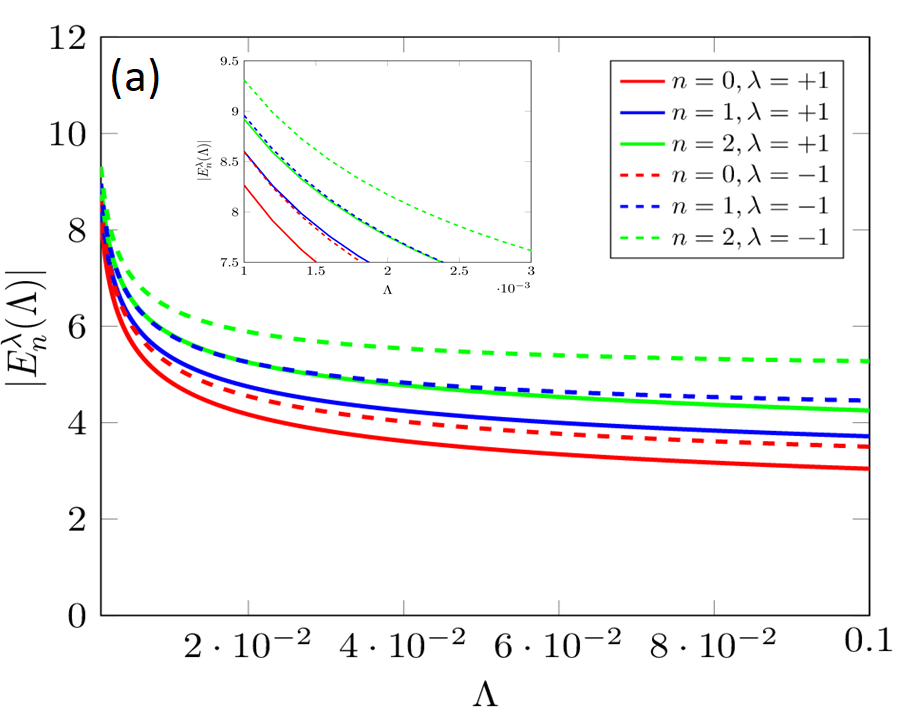}
    }\hfill
    \subfigure{%
        \includegraphics[width=0.49\textwidth]{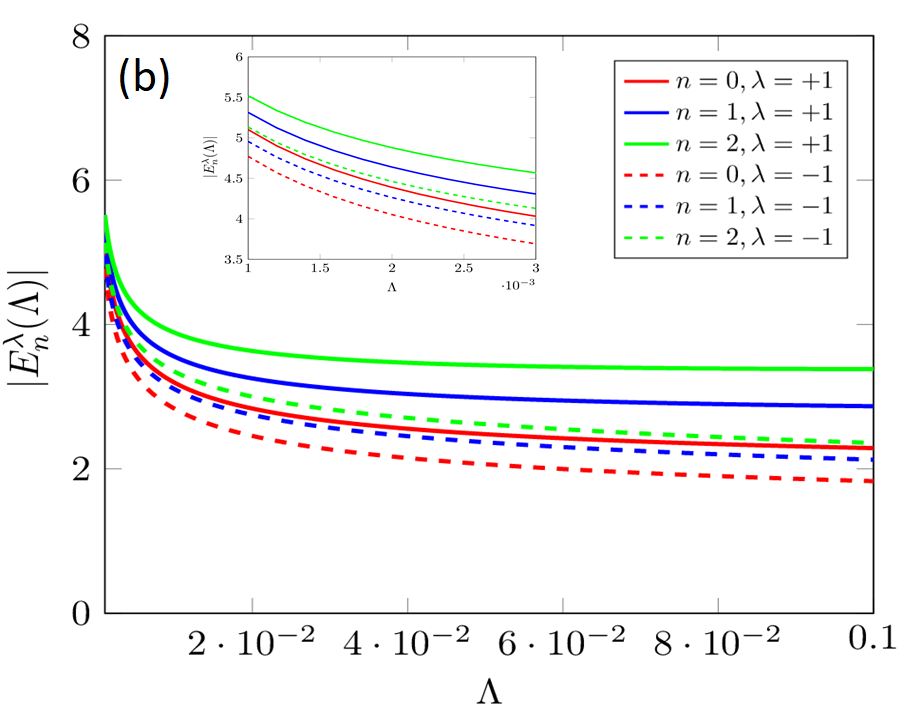}
    }
    \caption{Behavior of $\vert E_n^\lambda (\Lambda)\vert$ vs. $\Lambda$ for three different values of $n$ (with $0.001\leq\Lambda\leq 0.1$), where (a) is for $m_j=+1/2$ and $s=-1$, and (b) is for $m_j=-1/2$ and $s=+1$.}
    \label{fig7}
\end{figure}

In Fig. \ref{fig7}, we have the behavior of $\vert E_n^\lambda (\Lambda)\vert$ vs. $\Lambda$ for three different values of $n$ with $0.001\leq\Lambda\leq 0.1$, where (a) is for $m_j=+1/2$ and $s=-1$, and (b) is for $m_j=-1/2$ and $s=+1$ (we use $\Omega=0.05$, and $\omega=B_0=1$). So, we see that (a) and (b) have some similarities and differences. For example, in both graphs, the energies increase with the increase of $n$ (as it should be), i.e. we have $\Delta E_n^\pm=\vert E_{n+1}^\pm\vert -\vert E_n^\pm \vert>0$, and decrease as a function of $\Lambda$, i.e. $\Delta E(\Lambda)=E(\Lambda+\delta \Lambda)-E(\Lambda)=E_{final}(\Lambda)-E_{initial}(\Omega)<0$. However, as mentioned in the previous section, small values for $\Lambda$ transform the line element of Bonnor-Melvin-Lambda into a line element of the cosmic string-like where the (conical) curvature parameter is given by $\alpha_\Lambda=\sqrt{2\Lambda}\sigma$ (or better $\alpha_\Lambda=\sqrt{2\Lambda}$, since $\sigma=1$). In particular, this analogy implies that increasingly smaller values for $\Lambda$ result in increasingly larger values for the associated curvature. Therefore, we can say that the energies in (a) and (b) increase as a function of the curvature. With this, we see that for $\Lambda<0.002$, the energies become more significant (i.e. have higher values), while for $\Lambda>0.004$, are approximately constant (or vary very little). Besides, in (a) the energies of the antiparticle are greater than those of the particle ($\vert E^-_n \vert>\vert E^+_n \vert$), while in (b) the opposite happens, that is, the energies of the particle are greater than those of the antiparticle ($\vert E^+_n \vert>\vert E^-_n \vert$). However, in (a) we see that the energy of the particle for $n=2$ is practically equal to the energy of the antiparticle for $n=1$ for most of the range of $\Lambda$ (i.e. the solid green and dashed blue lines coincide/overlap). In addition, it is also important to comment that in both graphs, the condition $\left(\omega-s\frac{\omega_c}{2}+s\frac{\sqrt{2\Lambda}\sigma\Omega E}{m_0}\right)>0$ is also satisfied (as it should be). Now, comparing (a) with (b), we see that the energies are greater for (a), that is, when the particle/antiparticle has positive angular momentum and negative spin.

\begin{figure}[ht]
    \centering
    \subfigure{%
        \includegraphics[width=0.49\textwidth]{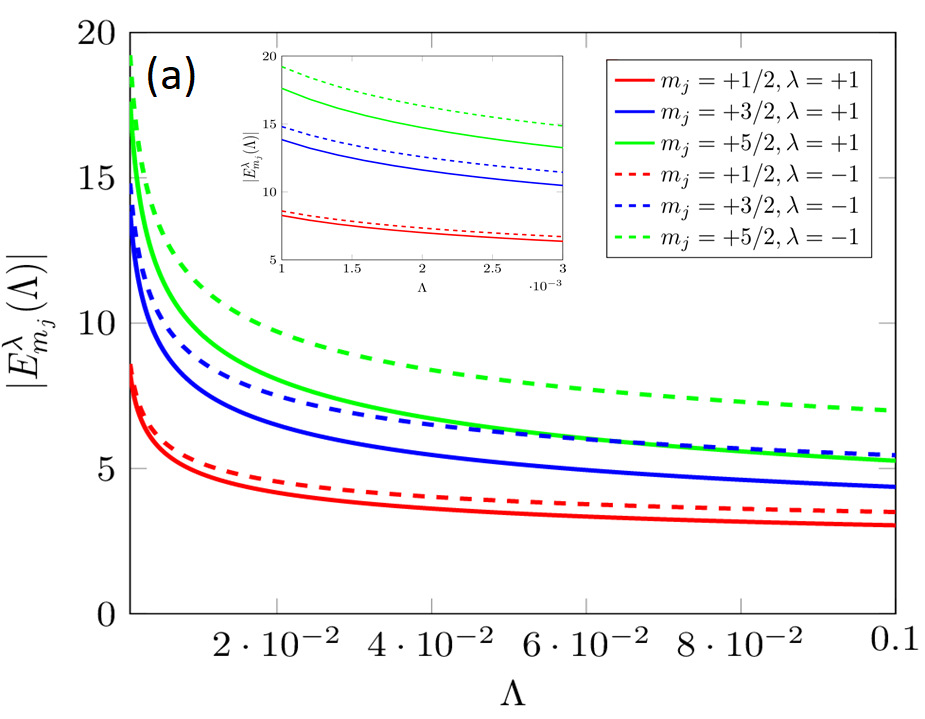}
    }\hfill
    \subfigure{%
        \includegraphics[width=0.49\textwidth]{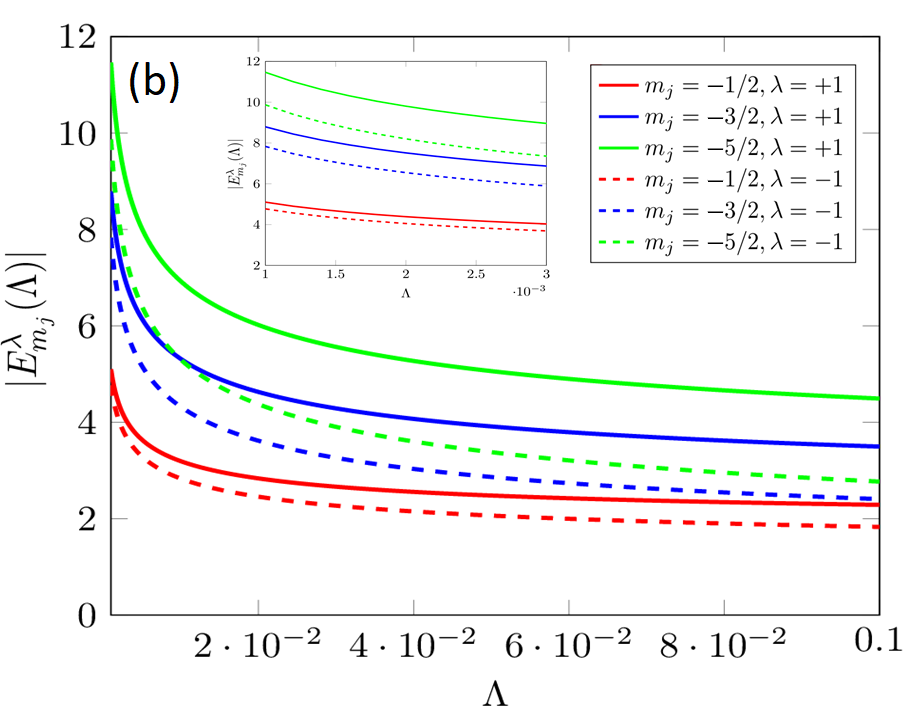}
    }
    \caption{Behavior of $\vert E_{m_j}^\lambda (\Lambda)\vert$ vs. $\Lambda$ for three different values of $m_j$ (with $n=0$ and $0.001\leq\Lambda\leq 0.1$), where (a) is for $m_j=+1/2,+3/2,+5/2$ and $s=-1$, and (b) is for $m_j=-1/2,-3/2,-5/2$ and $s=+1$.}
    \label{fig8}
\end{figure}

Already in Figs. \ref{fig8}-(a) and \ref{fig8}-(b), we have the behavior of $\vert E_{m_j}^\lambda (\Lambda)\vert$ vs. $\Lambda$ for three different values of $m_j$ (with $n=0$ and $0.001\leq\Lambda\leq 0.1$), where (a) is for $m_j=+1/2,+3/2,+5/2$ and $s=-1$, and (b) is for $m_j=-1/2,-3/2,-5/2$ and $s=+1$ (we use $\Omega=0.05$, and $\omega=B_0=1$). So, we see that the behavior of the energies presents more similarities than differences with those of Figs. \ref{fig7}-(a) and \ref{fig7}-(b). Indeed (such as occurs in all previous figures), the difference (with the exception of $m_j=\pm 1/2$) is basically in the energy values (or difference/spacing between them), consequently, we see that the effect generated by the variation of the quantum number $m_j$ is greater than the effect generated by the variation of the quantum number $n$. In other words, we can say that Figs. \ref{fig8}-(a) and \ref{fig8}-(b) are the ``enlargement'' (or ``enlarged version'') of Figs. \ref{fig7}-(a) and \ref{fig7}-(b), i.e. they only have higher energies. In addition, it is also important to comment that in both Figs. \eqref{fig8}-(a) and \eqref{fig8}-(b), the condition $\left(\omega-s\frac{\omega_c}{2}+s\frac{\sqrt{2\Lambda}\sigma\Omega E}{m_0}\right)>0$ is also satisfied (as it should be). Now, comparing (a) with (b), we see that the energies of the particle are greater for (b), that is, when it has negative angular momentum and positive spin, while in the case of the antiparticle, are greater for (a), that is, when it has positive angular momentum and negative spin, respectively.

Now, let us focus on the (functional) form of the Dirac spinor for the relativistic bound states of the DO in a rotating frame in the Bonnor-Melvin-Lambda spacetime. So, through the components \eqref{solution3}, we can have two possible configurations for the spinor \eqref{spinor}. However, this spinor (or configurations) does not ``serve'' our purposes; that is, to normalize the Dirac spinor later, we must first try to write it in terms of just one normalization constant (actually two, one for each value of $Ms$). Therefore, we should try to write the components \eqref{solution3} in terms of just one: $C^+_{Ms}$, or $C^-_{Ms}$ (otherwise, there will be one constant depending on the other and we will not find an explicit/well-defined form for any of them). For example, choosing $C^+_{Ms}$ (but it could also be $C^-_{Ms}$, however, the constant of the upper component is usually chosen \cite{Greiner,Strangebook}), we must then find the new lower components for the spinor \eqref{spinor}, that is, lower components that will also be written in terms of $C^+_{Ms}$. For this, we can use Eq. \eqref{EDO4}. However, it is convenient to write it in terms of the variable $w=m_0\bar{\Omega}r^2$. In this way, Eq. \eqref{EDO4} becomes
\begin{equation}\label{EDO5.1}
\psi^-(w)=\frac{\sqrt{m_0\bar{\Omega}}}{\left(\sqrt{\frac{\Lambda}{2}}\sigma\Omega+m_0+E\right)}\left(2\sqrt{w}\frac{d}{dw}+\sqrt{w}-\frac{Ms}{\sqrt{w}}\right)\psi^+(w), \ \  (with \ E=E_{u=+1}=E_+),
\end{equation}
or better
\begin{equation}\label{EDO5}
\psi_{Ms}^-(w)=\frac{\sqrt{m_0\bar{\Omega}}}{\left(\sqrt{\frac{\Lambda}{2}}\sigma\Omega+m_0+E\right)}\left(2\sqrt{w}\frac{d}{dw}+\sqrt{w}-\frac{Ms}{\sqrt{w}}\right)\psi_{Ms}^+(w).
\end{equation}

Therefore, using $\psi^+_{Ms}(w)$, given in \eqref{solution3}, we can obtain from \eqref{EDO5} two possible configurations for the lower component depending on the values of $Ms$. Explicitly, we have
\begin{equation}\label{soluion4}
\psi^-_{Ms}(w)=-\frac{2C^+_{Ms}\sqrt{m_0\bar{\Omega}}}{\left(\sqrt{\frac{\Lambda}{2}}\sigma\Omega+m_0+E\right)}e^{-\frac{w}{2}}\begin{cases}
w^{\frac{(Ms+1)}{2}}L_{n-1}^{Ms+\frac{1}{2}}(w), \ \ \ \ \ \ \ \ \ \ \ \ \ \ \ \ \ \ \ \ \ \ \ \ \ \ \ \ \ \ \ \ \ \ \ \ \ \ \ \ Ms>0,
\\
w^{\frac{(-Ms+2)}{2}}\left[\frac{\left(Ms-\frac{1}{2}\right)}{w}L_{n}^{-Ms+\frac{1}{2}}(w)+L_{n-1}^{-Ms+\frac{3}{2}}(w)\right], \ \ Ms<0,
\end{cases}
\end{equation}
where implies
\begin{equation}\label{solution5}
\psi^u_{Ms}(w)= \begin{cases}
\psi^+_{Ms}(w)=\begin{cases}
C^+_{Ms} e^{-\frac{w}{2}}w^{\frac{Ms}{2}}L^{Ms-\frac{1}{2}}_n(w), \ \ \ \ \ \ \ \ \ \ \ \ \ \ \ \ \ \ \ \ \ \ \ \ \ \ \ \ \ \ \ \ \ \ \ \ \ \ \ \ \ \ \ \ \ \ \ \ \ \ \ \ \ \ \ \ \ \ \ \ \ Ms>0, 
\\
C^+_{Ms} e^{-\frac{w}{2}}w^{\frac{(-Ms+1)}{2}}L^{-Ms+\frac{1}{2}}_n(w), \  \ \ \ \ \ \ \ \ \ \ \ \ \ \ \ \ \ \ \ \ \ \ \ \ \ \ \ \ \ \ \  \ \ \ \ \ \ \ \ \ \ \ \ \ \ \ \ \ \ \ \ \ Ms<0,
\end{cases}
\\
\psi^-_{Ms}(w)=\begin{cases}
-\frac{2C^+_{Ms}\sqrt{m_0\bar{\Omega}}}{\left(\sqrt{\frac{\Lambda}{2}}\sigma\Omega+m_0+E\right)}e^{-\frac{w}{2}}w^{\frac{(Ms+1)}{2}}L_{n-1}^{Ms+\frac{1}{2}}(w), \ \ \ \ \ \ \ \ \ \ \ \ \ \ \ \ \ \ \ \ \ \ \ \ \ \ \ \ \ \ \ \ \ \ \ \ \ \ \ \ Ms>0,
\\
-\frac{2C^+_{Ms}\sqrt{m_0\bar{\Omega}}}{\left(\sqrt{\frac{\Lambda}{2}}\sigma\Omega+m_0+E\right)}e^{-\frac{w}{2}}w^{\frac{(-Ms+2)}{2}}\left[\frac{\left(Ms-\frac{1}{2}\right)}{w}L_{n}^{-Ms+\frac{1}{2}}(w)+L_{n-1}^{-Ms+\frac{3}{2}}(w)\right], \ \ Ms<0.
\end{cases}
\end{cases}
\end{equation}

Consequently, the spinor \eqref{spinor} takes the form
\begin{equation}\label{spinor2}
\psi_{Ms}=\frac{e^{i(m_j\theta-Et)}}{\sqrt{r}}C^+_{Ms} e^{-\frac{w}{2}}\begin{cases}\left(
           \begin{array}{c}
            w^{\frac{Ms}{2}}L^{Ms-\frac{1}{2}}_n(w) \\
            -\frac{2i\sqrt{m_0\bar{\Omega}}}{\left(\sqrt{\frac{\Lambda}{2}}\sigma\Omega+m_0+E\right)}w^{\frac{(Ms+1)}{2}}L_{n-1}^{Ms+\frac{1}{2}}(w)\\
           \end{array}
         \right), \ \ \ \ \ \ \ \ \ \ \ \ \ \ \ \ \ \ \ \ \ \ \ \ \ \ \ \ \ \ \ \ \ \ \ \ \ \ \  Ms>0,
\\ \left(
           \begin{array}{c}
            w^{\frac{(-Ms+1)}{2}}L^{-Ms+\frac{1}{2}}_n(w) \\
            -\frac{2i\sqrt{m_0\bar{\Omega}}}{\left(\sqrt{\frac{\Lambda}{2}}\sigma\Omega+m_0+E\right)}w^{\frac{(-Ms+2)}{2}}\left[\frac{\left(Ms-\frac{1}{2}\right)}{w}L_{n}^{-Ms+\frac{1}{2}}(w)+L_{n-1}^{-Ms+\frac{3}{2}}(w)\right]\\
           \end{array}
         \right), \ \ Ms<0,
\end{cases}
\end{equation}
or yet
\begin{equation}\label{spinor3}
\psi_{Ms}=(m_0\bar{\Omega})^{\frac{1}{4}}C^+_{Ms} e^{i(m_j\theta-Et)}e^{-\frac{w}{2}}\begin{cases}\left(
           \begin{array}{c}
            w^{\frac{(2Ms-1)}{4}}L^{Ms-\frac{1}{2}}_n(w) \\
            -\frac{2i\sqrt{m_0\bar{\Omega}}}{\left(\sqrt{\frac{\Lambda}{2}}\sigma\Omega+m_0+E\right)}w^{\frac{(2Ms+1)}{4}}L_{n-1}^{Ms+\frac{1}{2}}(w)\\
           \end{array}
         \right), \ \ \ \ \ \ \ \ \ \ \ \ \ \ \ \ \ \ \ \ \ \ \ \ \ \ \ \ \ \ \ \ \ \ \ \ \ Ms>0,
\\ \left(
           \begin{array}{c}
            w^{\frac{(-2Ms+1)}{4}}L^{-Ms+\frac{1}{2}}_n(w) \\
            -\frac{2i\sqrt{m_0\bar{\Omega}}}{\left(\sqrt{\frac{\Lambda}{2}}\sigma\Omega+m_0+E\right)}w^{\frac{(-2Ms+3)}{4}}\left[\frac{\left(Ms-\frac{1}{2}\right)}{w}L_{n}^{-Ms+\frac{1}{2}}(w)+L_{n-1}^{-Ms+\frac{3}{2}}(w)\right]\\
           \end{array}
         \right), Ms<0,
\end{cases}
\end{equation}
where we use $r=\sqrt{w/m_0\bar{\Omega}}$. In particular, the spinor above will be of great use when we find the normalization constant.

Now, knowing that $\psi=e^{\frac{i\theta\Sigma_3}{2}}\Psi$ (or $\Psi=e^{-\frac{i\theta\Sigma_3}{2}}\psi$), we can then from this obtain the form of the Dirac spinor $\Psi$. However, since this exponential was originally written in the absence of $s$ (which is basically for $s=+1$) in (3+1)-dimensions \cite{Schluter}, we must then write it in terms of $s$ in $(2+1)$-dimensions (unfortunately, this was not done in \cite{Villalba1}). So, starting from the fact that $\Sigma_3$, $\gamma_1$ and $\gamma_2$ in $(2+1)$-dimensions are defined as $\Sigma_3=\sigma_3=i\gamma_1\gamma_2$, $\gamma_1=\sigma_3\sigma_1$ and $\gamma_2=s\sigma_3\sigma_2$ (with $\sigma_1\sigma_3=-\sigma_3\sigma_1$, $\sigma_3\sigma_3=1$ and $\sigma_1\sigma_2=i\sigma_3$), implies that: $e^{-\frac{i\theta\Sigma_3}{2}}=e^{\frac{\theta\gamma_1\gamma_2}{2}}=e^{-\frac{is\theta\sigma_3}{2}}$=diag$(e^{-\frac{is\theta}{2}},e^{\frac{is\theta}{2}})$. In this way, the two-component Dirac spinor $\Psi$ is written in the following form
\begin{equation}\label{spinor4}
\Psi_{Ms}=(m_0\bar{\Omega})^{\frac{1}{4}}C^+_{Ms} e^{-iEt}e^{-\frac{w}{2}}\begin{cases}\left(
           \begin{array}{c}
            e^{i(m_j-\frac{s}{2})\theta}w^{\frac{(2Ms-1)}{4}}L^{Ms-\frac{1}{2}}_n(w) \\
            -\frac{2i\sqrt{m_0\bar{\Omega}}}{\left(\sqrt{\frac{\Lambda}{2}}\sigma\Omega+m_0+E\right)}e^{i(m_j+\frac{s}{2})\theta}w^{\frac{(2Ms+1)}{4}}L_{n-1}^{Ms+\frac{1}{2}}(w)\\
           \end{array}
         \right), \ \ \ \ \ \ \ \ \ \ \ \ \ \ \ \ \ \ \ \ \ \ \ \ \ \ \ \ \ \ \ \ \ \ \ \ \ \ Ms>0,
\\ \left(
           \begin{array}{c}
            e^{i(m_j-\frac{s}{2})\theta}w^{\frac{(-2Ms+1)}{4}}L^{-Ms+\frac{1}{2}}_n(w) \\
            -\frac{2i\sqrt{m_0\bar{\Omega}}}{\left(\sqrt{\frac{\Lambda}{2}}\sigma\Omega+m_0+E\right)}e^{i(m_j+\frac{s}{2})\theta}w^{\frac{(-2Ms+3)}{4}}\left[\frac{\left(Ms-\frac{1}{2}\right)}{w}L_{n}^{-Ms+\frac{1}{2}}(w)+L_{n-1}^{-Ms+\frac{3}{2}}(w)\right]\\
           \end{array}
         \right), \ Ms<0,
\end{cases}
\end{equation}
or yet
\begin{equation}\label{spinor5}
\Psi_{Ms}=(m_0\bar{\Omega})^{\frac{1}{4}}C^+_{Ms} e^{-iEt}e^{-\frac{w}{2}}\begin{cases}\left(
           \begin{array}{c}
            e^{im_l\theta}w^{\frac{(2Ms-1)}{4}}L^{Ms-\frac{1}{2}}_n(w) \\
            -\frac{2i\sqrt{m_0\bar{\Omega}}}{\left(\sqrt{\frac{\Lambda}{2}}\sigma\Omega+m_0+E\right)}e^{i(m_l+s)\theta}w^{\frac{(2Ms+1)}{4}}L_{n-1}^{Ms+\frac{1}{2}}(w)\\
           \end{array}
         \right), \ \ \ \ \ \ \ \ \ \ \ \ \ \ \ \ \ \ \ \ \ \ \ \ \ \ \ \ \ \ \ \ \ \ \ \ \ \ Ms>0,
\\ \left(
           \begin{array}{c}
            e^{im_l\theta}w^{\frac{(-2Ms+1)}{4}}L^{-Ms+\frac{1}{2}}_n(w) \\
            -\frac{2i\sqrt{m_0\bar{\Omega}}}{\left(\sqrt{\frac{\Lambda}{2}}\sigma\Omega+m_0+E\right)}e^{i(m_l+s)\theta}w^{\frac{(-2Ms+3)}{4}}\left[\frac{\left(Ms-\frac{1}{2}\right)}{w}L_{n}^{-Ms+\frac{1}{2}}(w)+L_{n-1}^{-Ms+\frac{3}{2}}(w)\right]\\
           \end{array}
         \right), \ Ms<0,
\end{cases}
\end{equation}
where we use the relation $m_j=m_l+s/2$. So, comparing the angular part of the spinors above with the literature, we verified that \eqref{spinor4} for $s=+1$ is in accordance with Ref. \cite{Villalba2}, while the spinor \eqref{spinor5} for $s=+1$ and $s=\pm 1$ is in accordance with Refs. \cite{Chernodub,Andrade,Andrade2}. 

However, in terms of the original variable $r$ (since $w=m_0\bar{\Omega}r^2$), the spinor \eqref{spinor4} will be rewritten as
\begin{equation}\label{spinor6}
\Psi_{Ms}=C^+_{Ms}e^{-iEt}e^{-\frac{m_0\bar{\Omega}r^2}{2}}\begin{cases}\left(
           \begin{array}{c} \sqrt{(m_0\bar{\Omega})^{Ms}}e^{i\gamma\theta}r^{(Ms-\frac{1}{2})}L^{Ms-\frac{1}{2}}_n(m_0\bar{\Omega}r^2) \\
            -\frac{2i\sqrt{(m_0\bar{\Omega})^{(Ms+2)}}}{\left(\sqrt{\frac{\Lambda}{2}}\sigma\Omega+m_0+E\right)}e^{i\bar{\gamma}\theta}r^{(Ms+\frac{1}{2})}L_{n-1}^{Ms+\frac{1}{2}}(m_0\bar{\Omega}r^2)\\
           \end{array}
         \right), \ \ \ \ \ \ \ \ \ \ \ \ \ \ \ \ \ \ \ \ \ \ \ \ \ \ \ \ \ \ \ \ \ \  \ \ \ \ \ \ \ \ \ \ \ \ Ms>0,
\\ \left(
           \begin{array}{c}
            \sqrt{(m_0\bar{\Omega})^{(-Ms+1)}}e^{i\gamma\theta}r^{(-Ms+\frac{1}{2})}L^{-Ms+\frac{1}{2}}_n(m_0\bar{\Omega}r^2) \\
            -\frac{2i\sqrt{(m_0\bar{\Omega})^{(-Ms+3)}}}{\left(\sqrt{\frac{\Lambda}{2}}\sigma\Omega+m_0+E\right)}e^{i\bar{\gamma}\theta}r^{(-Ms+\frac{3}{2})}\left[\frac{\left(Ms-\frac{1}{2}\right)}{m_0\bar{\Omega}r^2}L_{n}^{-Ms+\frac{1}{2}}(m_0\bar{\Omega}r^2)+L_{n-1}^{-Ms+\frac{3}{2}}(m_0\bar{\Omega}r^2)\right]\\
           \end{array}
         \right), \ Ms<0.
\end{cases}
\end{equation}
where we define $\gamma\equiv m_j-s/2$ and $\bar{\gamma}\equiv m_j+s/2$.

So, according to Refs. \cite{Greiner,Strangebook,Dvornikov,Antoine,Pollock,Sousa,Falcone,Iyer,Pavlov}, the normalization (or orthonormalization) condition for the Dirac spinor in curved spacetimes is given by the following integral (in polar coordinates)
\begin{equation}\label{integral1}
\int J^t (x) dA_{curved} =\int_{0}^{2\pi}\int_{0}^{\infty}\sqrt{-g(x)}\bar{\Psi}_{Ms}\gamma^t (x)\Psi_{Ms}drd\theta = 1,
\end{equation}
where $g(x)$ is the determinant of the metric (i.e. $g(x)$=det$g_{\mu\nu}(x)<0$), $dA_{curved}=\sqrt{-g(x)}d^2 x=\sqrt{-g(x)}dr d\theta$ is the curved area/surface element, $J^t (x)=\rho (x)\geq 0$ is the curved probability density (positive definite quantity), which is the zero/temporal component of the curved conserved current or curved probability current density $J^\mu (x)=\bar{\Psi}\gamma^\mu (x)\Psi$ (or $J^\mu (x)=e^\mu_{\ a}(x)J^a$), and satisfies the curved continuity equation given by $\nabla_\mu (x) J^\mu (x)=0$, being $\bar{\Psi}=\Psi^\dagger \gamma^0$ the adjoint spinor (or Dirac adjoint), $\Psi^\dagger$ denotes the Hermitian conjugate of $\Psi=(\Psi_1,\Psi_2)^T$ (i.e. $\Psi^\dagger=(\Psi_1^\star,\Psi_2^\star)$), and $\gamma^t (x)=e^t_{\ a}(x)\gamma^a$ is the zero/temporal component of the curved gamma matrices $\gamma^\mu (x)$, respectively. In particular, we can call the integral above the ``positive definite inner product'' or the ``normalized Dirac inner product''.

On the other hand, we can still write the square root of \eqref{integral1} in terms of (inverse) tetrads, that is: $\sqrt{-g(x)}$=det$(e^a_{\ \mu}(x))$ \cite{Lawrie}. In this way, we have
\begin{equation}\label{integral2}
\int_{0}^{2\pi}\int_{0}^{\infty} det(e^a_{\ \mu}(x)) \bar{\Psi}_{Ms}\gamma^t (x)\Psi_{Ms}drd\theta = 1,
\end{equation}
or better
\begin{align}\label{integral3}
\int_{0}^{2\pi}\int_{0}^{\infty} det(e^a_{\ \mu}(x)) \Psi^{\dagger}_{Ms}\gamma^0 e^t_{\ a}(x)\gamma^a (x)\Psi_{Ms}drd\theta 
&=\int_{0}^{2\pi}\int_{0}^{\infty} det(e^a_{\ \mu}(x)) \Psi^{\dagger}_{Ms}\gamma^0[e^t_{\ 0}(x)\gamma^0+e^t_{\ 2}(x)\gamma^2] \Psi_{Ms}drd\theta=1, \nonumber\\
&= \int_{0}^{2\pi}\int_{0}^{\infty} det(e^a_{\ \mu}(x)) \Psi^{\dagger}_{Ms}[e^t_{\ 0}(x)\gamma^0\gamma^0+e^t_{\ 2}(x)\gamma^0\gamma^2] \Psi_{Ms}drd\theta=1, \nonumber\\
&= \int_{0}^{2\pi}\int_{0}^{\infty} det(e^a_{\ \mu}(x)) \Psi^{\dagger}_{Ms}[e^t_{\ 0}(x)-se^t_{\ 2}(x)\sigma_2] \Psi_{Ms}drd\theta=1,
\end{align}
where use $\gamma^0\gamma^0=\sigma_3\sigma_3=1$ and $\gamma^0\gamma^2=-\gamma_0\gamma_2=-\sigma_3s\sigma_3\sigma_2=-s\sigma_2$ \cite{Hagen1,Hagen2,Blum,Hagen3,Hagen4,Hagen5,Hagen6,CQG,Oli2,Sousa}.

Therefore, using the tetrads and their inverse given in \eqref{tetrads}, the integral above becomes
\begin{equation}\label{integral4}
\int_{0}^{2\pi}\int_{0}^{\infty} \left(\sigma\sin\left(\sqrt{2\Lambda}r\right)\right) \Psi^{\dagger}_{Ms}\left[\frac{1}{\sqrt{1-V^2}}-s\frac{V}{\sqrt{1-V^2}}\sigma_2\right] \Psi_{Ms}drd\theta=1, \ \ \left(V=\sigma\Omega\sin\left(\sqrt{2\Lambda}r\right)\right),
\end{equation}
or using the two approximations ($\Lambda\ll 1$ and $\Omega\ll 1$), such as
\begin{equation}\label{integral5}
\int_{0}^{2\pi}\int_{0}^{\infty} \sqrt{2\Lambda}\sigma r\Psi^{\dagger}_{Ms}\left[1-s\sqrt{2\Lambda}\sigma\Omega r\sigma_2\right] \Psi_{Ms}drd\theta=1.
\end{equation}

So, starting from the fact that $\Psi_{Ms}$ is given by $\Psi_{Ms}=e^{-\frac{i\theta\sigma_3}{2}}\psi_{Ms}$, where  $\left(e^{-\frac{i\theta\sigma_3}{2}}\right)^\dagger\left(e^{-\frac{i\theta\sigma_3}{2}}\right)=1$, $e^{\pm\frac{i\theta\sigma_3}{2}}=\cos{(\frac{\theta}{2})}\pm i\sigma_3\sin{(\frac{\theta}{2})}$, and $\sigma_2 e^{-\frac{i\theta\sigma_3}{2}}=e^{\frac{i\theta\sigma_3}{2}}\sigma_2$ (since $\sigma_2\sigma_3=-\sigma_3\sigma_2$), and using $w=m_0\bar{\Omega}r^2$ (where $rdr=dw/2m_0\bar{\Omega}$), implies that the integral above takes the following form (for simplicity/didactic reasons, we omit a little of the mathematical rigor)
\begin{align}\label{integral6}
\int_{0}^{2\pi}\int_{0}^{\infty}\sqrt{2\Lambda}\sigma r\Psi^{\dagger}_{Ms}\left[1-s\sqrt{2\Lambda}\sigma\Omega r\sigma_2\right] \Psi_{Ms}drd\theta 
&=\frac{\sqrt{2\Lambda}\sigma}{2m_0\bar{\Omega}}\int_{0}^{2\pi}\int_{0}^{\infty}\psi^{\dagger}_{Ms}\left[1-e^{i\theta\sigma_3}s\sqrt{2\Lambda}\sigma\Omega r\sigma_2\right] \psi_{Ms} dwd\theta, \nonumber\\
&= \frac{\sqrt{2\Lambda}\sigma}{2m_0\bar{\Omega}}\int_{0}^{\infty}\psi^{\dagger}_{Ms}\left[\int_{0}^{2\pi} d\theta-\int_{0}^{2\pi}[(\cos{\theta}+i\sigma_3\sin{\theta})d\theta]s\sqrt{2\Lambda}\sigma\Omega r\sigma_2\right]\psi_{Ms}dw, \nonumber\\
&= \frac{\sqrt{2\Lambda}\sigma}{2m_0\bar{\Omega}}\int_{0}^{\infty}\psi^{\dagger}_{Ms}\left[ [\theta]^{2\pi}_{0}-[\sin{\theta}-i\sigma_3\cos{\theta}]^{2\pi}_{0}s\sqrt{2\Lambda}\sigma\Omega r\sigma_2\right] \psi_{Ms} dw,  \nonumber\\
&= \frac{\sqrt{2\Lambda}\sigma}{2m_0\bar{\Omega}}\int_{0}^{\infty}\psi^{\dagger}_{Ms}[2\pi]\psi_{Ms} dw, \nonumber\\
&=\frac{\pi\sqrt{2\Lambda}\sigma}{m_0\bar{\Omega}}\int_{0}^{\infty}\psi^{\dagger}_{Ms}(w)\psi_{Ms}(w) dw=1.
\end{align}

In particular, we have to solve the integral above for both $Ms>0$ and $Ms<0$, where the expression for $\psi_{Ms}$ is given in \eqref{spinor3}. Therefore, starting first for $Ms>0$, we have
\begin{equation}\label{integral7}
\sigma\pi\sqrt{\frac{2\Lambda}{m_0\bar{\Omega}}} (C^+_{Ms})^2 \left[\int_{0}^{\infty} e^{-w} w^{Ms-\frac{1}{2}}\left[L^{Ms-\frac{1}{2}}_n (w)\right]^2 dw+\frac{4m_0\bar{\Omega}}{\left(\sqrt{\frac{\Lambda}{2}}\sigma\Omega+m_0+E\right)^2}\int_{0}^{\infty} e^{-w} w^{Ms+\frac{1}{2}}\left[L^{Ms+\frac{1}{2}}_{n-1}(w)\right]^2 dw\right]=1.
\end{equation}

According to Ref. \cite{Bhuiyan,Arfken}, the first integral in \eqref{integral7} can be easily solved through the following orthogonality condition that the associated Laguerre polynomials must obey
\begin{equation}\label{orthogonalitycondition1}
\int_{0}^{\infty} e^{-z} z^{\alpha} L^{\alpha}_n (z) L^{\alpha}_{n'} (z) dz=\frac{(n+\alpha)!}{n!}\delta_{nn'}, \ \ (\alpha>0),
\end{equation}
where $n!$ is the factorial of $n$, which can also be written in terms of the gamma function, i.e. $n!=\Gamma(n+1)>0$.

So, in our case, we have
\begin{equation}\label{integral8}
\int_{0}^{\infty} e^{-w} w^{Ms-\frac{1}{2}}\left[L^{Ms-\frac{1}{2}}_n (w)\right]^2 dw=\frac{\left(n+Ms-\frac{1}{2}\right)!}{n!}.
\end{equation}

Now, according to Ref. \cite{Li1999}, the second integral in \eqref{integral7} can be solved through the following general formula for the integration of the product of two associated Laguerre polynomials
\begin{equation}\label{orthogonalitycondition2}
\int_{0}^{\infty} e^{-z} z^{\lambda} L^{\mu}_n (z) L^{\mu'}_{n'} (z) dz=(-1)^{n+n'}\lambda ! \sum_{k} \binom{\lambda-\mu}{n-k}\binom{\lambda-\mu'}{n'-k}\binom{\lambda+k}{k}
\end{equation}
where 
\begin{equation}
\binom{n}{k}=C^k_n=\frac{n!}{k!(n-k)!},
\end{equation}
is the binomial coefficient (indexed by a pair of integers $n$ and $k$, with $n\geq k$ or $n<k$), and $\lambda!$ is the factorial of $\lambda$ (with $\lambda!=\Gamma(\lambda+1)$). So, using \eqref{orthogonalitycondition2}, we can written the second integral in \eqref{integral7} as follows
\begin{equation}\label{integral9}
\int_{0}^{\infty} e^{-w} w^{Ms+\frac{1}{2}}\left[L^{Ms+\frac{1}{2}}_{n-1}(w)\right]^2 dw=\left(Ms+\frac{1}{2}\right)! \sum_{k} \binom{0}{n-1-k}^2\binom{Ms+\frac{1}{2}+k}{k}, \ \ ((-1)^{2(n-1)}=+1).
\end{equation}

Analyzing the first coefficient binomial in \eqref{integral9} according to the values of $n$, that are: $n=k$, $n>k$ (we can assume that $n=k+1$), and $n<k$ (we can assume that $n=k-1$), we have
\begin{equation}
\binom{0}{n-1-k} = \begin{cases}
\binom{0}{-1}=0,  \ \  \ \ \ \ \ n=k, \\
\binom{0}{0}=1,  \ \  \ \ n=k+1, \\
\binom{0}{-2}=0,  \ \  n=k-1.
\end{cases}
\end{equation}

Therefore, we see that only the case $n>k$ (or $k=n-1$) has a non-zero value for the coefficient. With this, the integral \eqref{integral9} becomes
\begin{align}\label{integral10}
\int_{0}^{\infty} e^{-w} w^{Ms+\frac{1}{2}}\left[L^{Ms+\frac{1}{2}}_{n-1}(w)\right]^2 dw 
&=\left(Ms+\frac{1}{2}\right)! \sum_{k=0}^{n-1}\binom{Ms+\frac{1}{2}+k}{k}, \nonumber\\
&= \left(Ms+\frac{1}{2}\right)! \sum_{k=0}^{n-1}\frac{(Ms+\frac{1}{2}+k)!}{k!(Ms+\frac{1}{2})!}, \nonumber\\
&=\sum_{k=0}^{n-1}\frac{(Ms+\frac{1}{2}+k)!}{k!},
\nonumber\\
&=\frac{n(n+Ms+\frac{1}{2})!}{n!(Ms+\frac{3}{2})}.
\end{align}

In this way, using the results \eqref{integral8} and \eqref{integral10} in \eqref{integral7}, we obtain the following normalization constant for $Ms>0$ 
\begin{equation}\label{C1}
C^+_{Ms}=\sqrt{\frac{1}{\sigma\pi}\sqrt{\frac{m_0\bar{\Omega}}{2\Lambda}} \frac{n!}{\left[(n+Ms-\frac{1}{2})!+\frac{4nm_0\bar{\Omega}(n+Ms+\frac{1}{2})!}{\left(\sqrt{\frac{\Lambda}{2}}\sigma\Omega+m_0+E\right)^2(Ms+\frac{3}{2})}\right]}}.
\end{equation}

Now, let us determine the normalization constant for $Ms<0$. So, using $\psi_{Ms}$ given in \eqref{spinor3} in the normalization condition \eqref{integral6}, we have
\begin{equation}\label{integral11}
\sigma\pi\sqrt{\frac{2\Lambda}{m_0\bar{\Omega}}} (C^+_{Ms})^2 \left\{I_1+\frac{4m_0\bar{\Omega}}{\left(\sqrt{\frac{\Lambda}{2}}\sigma\Omega+m_0+E\right)^2}\left[\left(Ms-\frac{1}{2}\right)^2 I_2+ 2\left(Ms-\frac{1}{2}\right) I_3+I_4\right]\right\}=1,
\end{equation}
where $I_1$, $I_2$, $I_3$ and $I_4$ are integrals, defined as follows
\begin{equation}\label{I1}
I_1\equiv\int_{0}^{\infty} e^{-w} w^{-Ms+\frac{1}{2}}\left[L^{-Ms+\frac{1}{2}}_n (w)\right]^2 dw,
\end{equation}
\begin{equation}\label{I2}
I_2\equiv \int_{0}^{\infty} e^{-w} w^{-Ms-\frac{1}{2}}\left[L_{n}^{-Ms+\frac{1}{2}}(w)\right]^2 dw,
\end{equation}
\begin{equation}\label{I3}
I_3\equiv\int_{0}^{\infty} e^{-w} w^{-Ms+\frac{1}{2}} L_{n}^{-Ms+\frac{1}{2}}(w)L_{n-1}^{-Ms+\frac{3}{2}}(w) dw,
\end{equation}
\begin{equation}\label{I4}
I_4\equiv \int_{0}^{\infty} e^{-w} w^{-Ms+\frac{3}{2}}\left[L_{n-1}^{-Ms+\frac{3}{2}}(w)\right]^2 dw.
\end{equation}

However, based on what we saw/did in the previous case (see \eqref{integral8} and \eqref{integral10}), the solutions of $I_1$ and $I_4$ are given by
\begin{equation}\label{I5}
I_1\equiv\int_{0}^{\infty} e^{-w} w^{-Ms+\frac{1}{2}}\left[L^{-Ms+\frac{1}{2}}_n (w)\right]^2 dw=\frac{\left(n-Ms+\frac{1}{2}\right)!}{n!},
\end{equation}
\begin{equation}\label{I6}
I_4\equiv \int_{0}^{\infty} e^{-w} w^{-Ms+\frac{3}{2}}\left[L_{n-1}^{-Ms+\frac{3}{2}}(w)\right]^2 dw=\frac{n(n-Ms+\frac{3}{2})!}{n!(-Ms+\frac{5}{2})}.
\end{equation}

Besides, according to Ref. \cite{Li1999}, the integral $I_2$ can be easily solved through the following integral
\begin{equation}
\int_{0}^{\infty} e^{-z} z^{\lambda-1}\left[L_{n}^{\lambda}(z)\right]^2 dz=\frac{(n+\lambda)!}{n!\lambda}.
\end{equation}

With this, the solution of the integral $I_2$ is given by
\begin{equation}\label{I7}
I_2\equiv\int_{0}^{\infty} e^{-w} w^{-Ms+\frac{1}{2}-1}\left[L_{n}^{-Ms+\frac{1}{2}}(w)\right]^2 dw=\frac{(n-Ms+\frac{1}{2})!}{n!(-Ms+\frac{1}{2})}.
\end{equation}

Now, we just need to solve the integral $I_3$. So, using \eqref{orthogonalitycondition2}, we can written $I_3$ as follows
\begin{equation}\label{integral12}
I_3\equiv \int_{0}^{\infty} e^{-w} w^{-Ms+\frac{1}{2}} L_{n}^{-Ms+\frac{1}{2}}(w)L_{n-1}^{-Ms+\frac{3}{2}}(w) dw=-\left(-Ms+\frac{1}{2}\right)! \sum_{k} \binom{0}{n-k}\binom{-1}{n-1-k}\binom{-Ms+\frac{1}{2}+k}{k},
\end{equation}
where we use $(-1)^{2n-1}=-1$.

Analyzing the first coefficient binomial in \eqref{integral12} according to the values of $n$ ($n=k$, $n>k$, and $n<k$), we have
\begin{equation}
\binom{0}{n-k} = \begin{cases}
\binom{0}{0}=1, \ \ \ \ \ \ \ \ \ n=k, \\
\binom{0}{1}=1,  \ \  \ \ n=k+1, \\
\binom{0}{-1}=0,  \ \  n=k-1,
\end{cases}
\end{equation}
while for the second coefficient, we have
\begin{equation}
\binom{-1}{n-1-k} = \begin{cases}
\binom{-1}{-1}=1, \ \ \ \ \ \ \ \ \ n=k, \\
\binom{-1}{0}=1,  \ \ \ \ n=k+1, \\
\binom{-1}{-2}=-1, \ \ n=k-1.
\end{cases}
\end{equation}

Therefore, we see that only the case $n\geq k$ has a non-zero value for the product of the two coefficients. With this, the integral \eqref{integral12} becomes
\begin{align}\label{integral13}
I_3\equiv \int_{0}^{\infty} e^{-w} w^{-Ms+\frac{1}{2}} L_{n}^{-Ms+\frac{1}{2}}(w)L_{n-1}^{-Ms+\frac{3}{2}}(w) dw 
&=-\left(-Ms+\frac{1}{2}\right)! \sum_{k=0}^{n}\binom{-Ms+\frac{1}{2}+k}{k}, \nonumber\\
&= -\left(-Ms+\frac{1}{2}\right)! \sum_{k=0}^{n}\frac{(-Ms+\frac{1}{2}+k)!}{k!(-Ms+\frac{1}{2})!}, \nonumber\\
&=-\sum_{k=0}^{n}\frac{(-Ms+\frac{1}{2}+k)!}{k!},
\nonumber\\
&=\frac{(n-Ms+\frac{3}{2})!}{n!(Ms+\frac{3}{2})}.
\end{align}

In this way, using the results \eqref{I5}, \eqref{I6}, \eqref{I7}, and \eqref{integral13} in \eqref{integral11}, we obtain the following normalization constant for $Ms<0$
\begin{equation}\label{C2}
C^+_{Ms}=\sqrt{\frac{1}{\sigma\pi}\sqrt{\frac{m_0\bar{\Omega}}{2\Lambda}} \frac{n!}{\left[(n-Ms+\frac{1}{2})!+\frac{4m_0\bar{\Omega}f_{n,Ms}}{\left(\sqrt{\frac{\Lambda}{2}}\sigma\Omega+m_0+E\right)^2}\right]}},
\end{equation}
where we define
\begin{equation}
f_{Ms}\equiv\left[\left(-Ms+\frac{1}{2}\right)\left(n-Ms+\frac{1}{2}\right)!+\frac{2\left(Ms-\frac{1}{2}\right)(n-Ms+\frac{3}{2})!}{(Ms+\frac{3}{2})}+\frac{n(n-Ms+\frac{3}{2})!}{(-Ms+\frac{5}{2})}\right].
\end{equation}

Therefore, using the normalization constants for $Ms>0$ and $Ms<0$ in \eqref{spinor6}, we have the following normalized Dirac spinor for the relativistic bound states
\begin{equation}\label{spinor7}
\Psi_{Ms}(t,r,\theta)=\begin{cases}\left(
           \begin{array}{c}
            K_{Ms}e^{i(\gamma\theta-Et)}e^{-\frac{m_0\bar{\Omega}r^2}{2}}r^{(Ms-\frac{1}{2})}L^{Ms-\frac{1}{2}}_n(m_0\bar{\Omega}r^2) \\
            -\frac{2im_0\bar{\Omega}K_{Ms}}{\left(\sqrt{\frac{\Lambda}{2}}\sigma\Omega+m_0+E\right)}e^{i(\bar{\gamma}\theta-Et)}e^{-\frac{m_0\bar{\Omega}r^2}{2}}r^{(Ms+\frac{1}{2})}L_{n-1}^{Ms+\frac{1}{2}}(m_0\bar{\Omega}r^2)\\
           \end{array}
         \right), \ \ \ \ \ \ \ \ \ \ \ \ \ \ \ \ \ \ \ \ \ \ \ \ \ \ \ \ \ \ \ \ \ \ \ \  \ \ \  \ \ \ \ \  Ms>0,
\\ 
\left(
           \begin{array}{c}
            K_{Ms}e^{i(\gamma\theta-Et)}e^{-\frac{m_0\bar{\Omega}r^2}{2}}r^{(-Ms+\frac{1}{2})}L^{-Ms+\frac{1}{2}}_n(m_0\bar{\Omega}r^2) \\
            -\frac{2im_0\bar{\Omega}K_{Ms}}{\left(\sqrt{\frac{\Lambda}{2}}\sigma\Omega+m_0+E\right)}e^{i(\bar{\gamma}\theta-Et)}e^{-\frac{m_0\bar{\Omega}r^2}{2}}r^{(-Ms+\frac{3}{2})}\left[\frac{\left(Ms-\frac{1}{2}\right)}{m_0\bar{\Omega}r^2}L_{n}^{-Ms+\frac{1}{2}}(m_0\bar{\Omega}r^2)+L_{n-1}^{-Ms+\frac{3}{2}}(m_0\bar{\Omega}r^2)\right]\\
           \end{array}
         \right), Ms<0,
\end{cases}
\end{equation}
where we define the constant $K_{Ms}$ in the form
\begin{equation}
K_{Ms}\equiv\begin{cases}
C^+_{Ms}(m_0\Omega)^{\frac{Ms}{2}}=\sqrt{\frac{(m_0\bar{\Omega})^{Ms+\frac{1}{2}}}{\pi\sqrt{2\Lambda}\sigma}\frac{n!}{\left[(n+Ms-\frac{1}{2})!+\frac{4nm_0\bar{\Omega}(n+Ms+\frac{1}{2})!}{\left(\sqrt{\frac{\Lambda}{2}}\sigma\Omega+m_0+E\right)^2(Ms+\frac{3}{2})}\right]}}, \ \ \ Ms>0, \\
C^+_{Ms}(m_0\Omega)^{\frac{-Ms+1}{2}}=\sqrt{\frac{(m_0\bar{\Omega})^{-Ms+\frac{3}{2}}}{\pi\sqrt{2\Lambda}\sigma}\frac{n!}{\left[(n-Ms+\frac{1}{2})!+\frac{4m_0\bar{\Omega}f_{Ms}}{\left(\sqrt{\frac{\Lambda}{2}}\sigma\Omega+m_0+E\right)^2}\right]}}, \ \ \ \ \ Ms<0.
\end{cases}
\end{equation}

%-------------------------------------------------------------------------
\section{Radial probability density \label{sec4}}

Here, let us graphically analyze the behavior of the radial probability density $P(r)$ (or simply, probability density, which is the probability per unit area) as a function of the radial coordinate $r$ for six different values of the quantum number $m_j$, magnetic field $B_0$ ($0\leq B_0<\infty$), angular frequency $\omega$ ($0\leq\omega<\infty$), angular velocity $\Omega$ ($0\leq\Omega\leq 0.06$), and of the cosmological constant $\Lambda$ ($0<\Lambda\leq 0.05$), with $n$ fixed (obs: in all graphs, the condition $0\leq r<r_0=1/\sqrt{2\Lambda}\Omega$ is satisfied, as it should be). That is, unlike the previous graphs, and in order to simplify the calculations as much as possible, here, we will consider only the state ground ($n=0$) for a particle ($E=E^+$) with $m_j>0$ and $s=-1$ (or $Ms<0$), whose spectrum is the first configuration in \eqref{spectrum9}. In fact, as we start from the upper component to reach the lower one (see \eqref{EDO5.1}), it was expected to choose one of the spectra for $u=+1$. In other words, we can say that we are going to analyze the probability density for the lowest Landau level (LLL), such as was done/discussed in Refs. \cite{Bhuiyan,Schattschneider}. However, with respect to the graphs of probability density for three different values of $B_0$, $\omega$, $\Omega$, and $\Lambda$, we will fixe both $n$ and $m_j$, such as $n=0$ and $m_j=+1/2$. So, through the of spinor \eqref{spinor7}, we can easily find the expression for $P(r)$, since $P(r)=P_{n,m_j}(r)=\Psi^\dagger_{Ms}\Psi_{Ms}\geq 0$ (or $P(r)=\Psi^\dagger_{LLL}\Psi_{LLL}\geq 0$). Therefore, knowing that $M=m_j/\sqrt{2\Lambda}\sigma$, we have the following expression for $P(r)$
\begin{equation}\label{P1}
P(r)=P_{0,m_j}(r)=K_{Ms}^2e^{-m_0\bar{\Omega}r^2}\left[r^{\left(2\frac{m_j}{\sqrt{2\Lambda}\sigma}+1\right)}+\frac{4m_0^2\bar{\Omega}^2}{\left(\sqrt{\frac{\Lambda}{2}}\sigma\Omega+m_0+E\right)^2}r^{\left(2\frac{m_j}{\sqrt{2\Lambda}\sigma}+3\right)}\left[\frac{\left(-\frac{m_j}{\sqrt{2\Lambda}\sigma}-\frac{1}{2}\right)}{m_0\bar{\Omega}r^2}\right]^2\right],
\end{equation}
or better (with $m_0=e=\sigma=1$)
\begin{equation}\label{P2}
P(r)=K_{Ms}^2e^{-\bar{\Omega}r^2}r^{\left(2\frac{m_j}{\sqrt{2\Lambda}}-1\right)}\left[r^2+\frac{\left(2\frac{m_j}{\sqrt{2\Lambda}}+1\right)^2}{(1+E)^2}\right],
\end{equation}
with
\begin{equation}
K_{Ms}=\sqrt{\frac{\bar{\Omega}^{\left(\frac{m_j}{\sqrt{2\Lambda}}+\frac{3}{2}\right)}}{\pi\sqrt{2\Lambda}}\frac{1}{\left[\left(\frac{m_j}{\sqrt{2\Lambda}}+\frac{1}{2}\right)!+\frac{4\bar{\Omega}}{(1+E)^2}\left(\frac{m_j}{\sqrt{2\Lambda}}+\frac{1}{2}\right)\left[\left(\frac{m_j}{\sqrt{2\Lambda}}+\frac{1}{2}\right)!+2\frac{\left(\frac{m_j}{\sqrt{2\Lambda}}+\frac{3}{2}\right)!}{\left(\frac{m_j}{\sqrt{2\Lambda}}-\frac{3}{2}\right)}\right]\right]}},
\end{equation}
\begin{equation}
\bar{\Omega}=\left(\omega-\frac{B_0}{2}+\sqrt{2\Lambda}\Omega E\right),
\end{equation}
\begin{equation}
E=E_{0,m_j}=-\sqrt{8\Lambda}\Omega\left(\frac{1}{2}+\frac{m_j}{\sqrt{2\Lambda}}\right)+\sqrt{1+4\left(\omega+\frac{B_0}{2}\right)\left(\frac{1}{2}+\frac{m_j}{\sqrt{2\Lambda}}\right)},
\end{equation}
where we consider $\left(\sqrt{\frac{\Lambda}{2}}\sigma\Omega+m_0+E\right)^2\approx (m_0+E)^2$ (such as we did to get \eqref{spectrum9}), and
we use the fact that $L^{\left(\frac{m_j}{\sqrt{2\Lambda}\sigma}+\frac{1}{2}\right)}_0 (m_0\bar{\Omega}r^2)=1$ and $L_{-1}^{\left(\frac{m_j}{\sqrt{2\Lambda}\sigma}+\frac{3}{2}\right)}=0$ (since $L_{n-1}^{k+1}(x)=-\frac{d}{dx}L_n^k(x)$ \cite{Pavlov,Arfken}). As we will see below, an expression in the form \eqref{P2} will exhibit a Gaussian function-like behavior (or Gaussian-like curves). Indeed, it is not a purely Gaussian behavior (as stated in \cite{Bhuiyan}) because the expression \eqref{P2} also carries two polynomial terms, such as $r^{\left(2\frac{m_j}{\sqrt{2\Lambda}}+1\right)}$ and $r^{\left(2\frac{m_j}{\sqrt{2\Lambda}}-1\right)}$, respectively.

However, we can simplify the effective frequency $\bar{\Omega}$. So, using $\Lambda\ll 1$ and $\Omega\ll 1$, we have
\begin{align}
\bar{\Omega} 
&=\left(\omega-\frac{B_0}{2}+\sqrt{2\Lambda}\Omega\left(-\sqrt{8\Lambda}\Omega\left(n+\frac{1}{2}+\frac{m_j}{\sqrt{2\Lambda}}\right)+\sqrt{1+4\left(\omega+\frac{B_0}{2}\right)\left(n+\frac{1}{2}+\frac{m_j}{\sqrt{2\Lambda}}\right)}\right)\right), \nonumber\\
&=\left(\omega-\frac{B_0}{2}+\sqrt{2\Lambda}\Omega\sqrt{1+4\left(\omega+\frac{B_0}{2}\right)\left(n+\frac{1}{2}+\frac{m_j}{\sqrt{2\Lambda}}\right)}\right), \ \ (4\Lambda \Omega^2\approx 0).
\end{align}

\begin{figure}[!h]
\centering
\includegraphics[width=9.0cm]{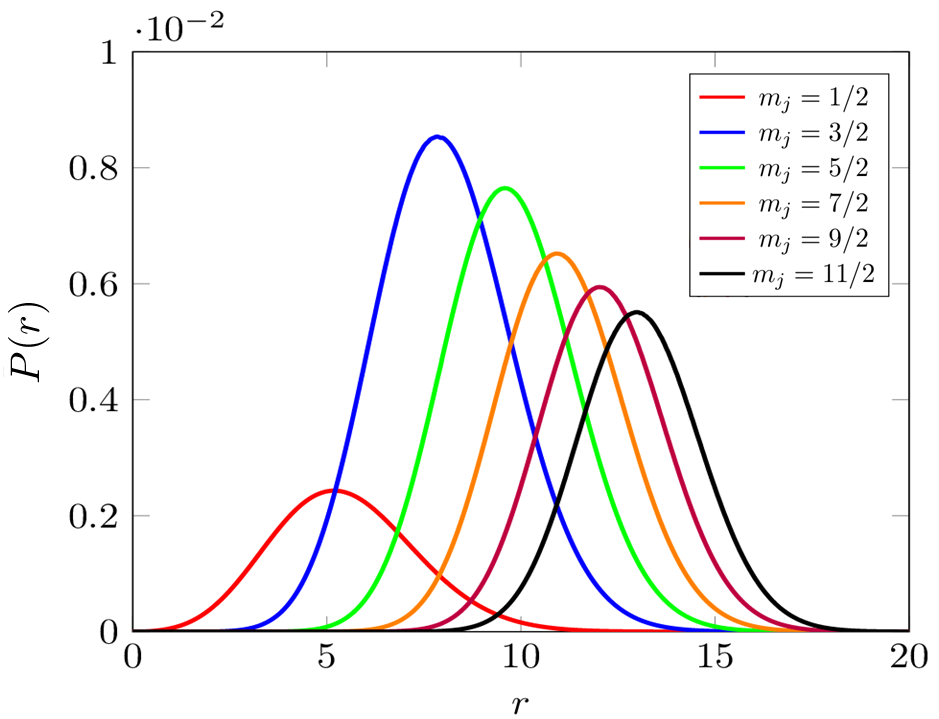}
\caption{Behavior of $P(r)$ vs. $r$ for six different values of $m_j$.}
\label{fig9}
\end{figure}

Therefore, in Fig. \ref{fig9} we have the behavior of $P(r)$ vs. $r$ for six different values of $m_j$, where we use $\Lambda=\Omega=0.05$, and $\omega=B_0=0.1$. So, according to the figure, we see that the probability (or their maximum values) can decrease or increase with the increase of $m_j$, that is, it increases (substantially) from $m_j=1/2$ to $m_j=3/2$, and then decreases with the increase of $m_j$ (i.e. for $m_j>3/2$). Therefore, we have $P_{m_j=3/2}(r)>P_{m_j=5/2}(r)>P_{m_j=7/2}(r)>P_{m_j=9/2}(r)>P_{m_j=11/2}(r)>P_{m_j=1/2}(r)$; consequently, this implies that the highest probability (or highest peak) of finding/locating the particle (in $r\approx 7.85$) is when its angular momentum has the value of $m_j=3/2$. In other words, the height (or peak value) of the Gaussian-like curves increases from $m_j=1/2$ to $m_j=3/2$, and decreases from $m_j=3/2$ to $m_j=11/2$, respectively. Furthermore, these curves move to the right (or further and further away from the origin) as $m_j$ increases. However, the width of these curves has practically the same value ($\Delta r\approx 11$).

\begin{figure}[!h]
\centering
\includegraphics[width=9.0cm]{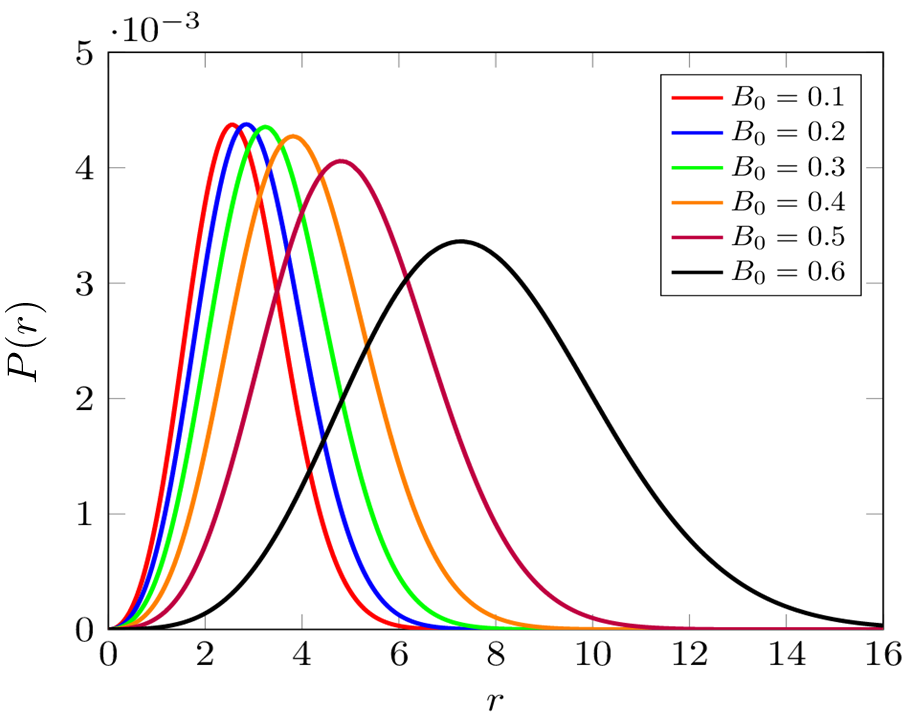}
\caption{Behavior of $P(r)$ vs. $r$ for six different values of $B_0$.}
\label{fig10}
\end{figure}

In Fig. \ref{fig10}, we have the behavior of $P(r)$ vs. $r$ for six different values of $B_0$, where we use $\Lambda=\Omega=0.05$, $\omega=0.3$, and $m_j=1/2$. So, according to the figure, we see that the probability decreases with the increase of $B_0$ (perhaps with the exception of the first two values). With this, we have $P(B_0=0.1)\approx P(B_0=0.2)>P(B_0=0.3)>P(B_0=0.4)>P(B_0=0.5)>P(B_0=0.6)$; consequently, this implies that the highest probability of finding the particle (in $r\approx 2.56$ or $r\approx 2.85$) is when the magnetic field has the value of $B_0=0.1$ or $B_0=0.2$. In other words, the height of the Gaussian-like curves decreases as $B_0$ increases. Furthermore, these curves move to the right (or further and further away from the origin) as $B_0$ increases. However, the width of these curves increases as $B_0$ increases, that is, $\Delta r (B_0=0.6)>\Delta r (B_0=0.5)>\Delta r (B_0=0.4)>\Delta r (B_0=0.3)>\Delta r (B_0=0.2)>\Delta r (B_0=0.1)$.

\begin{figure}[!h]
\centering
\includegraphics[width=9.0cm]{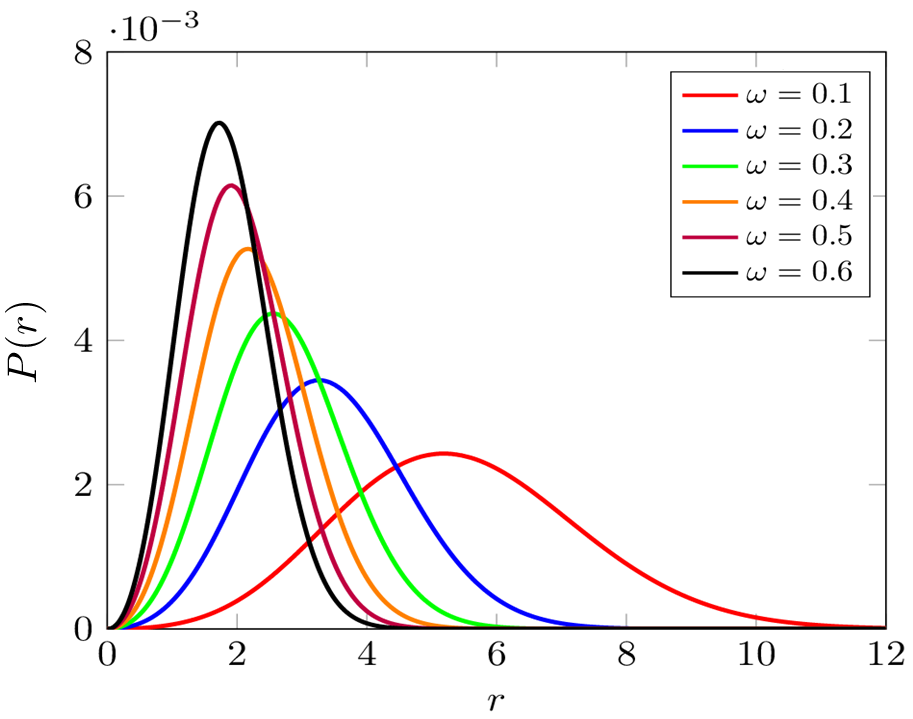}
\caption{Behavior of $P(r)$ vs. $r$ for six different values of $\omega$.}
\label{fig11}
\end{figure}

In Fig. \ref{fig11}, we have the behavior of $P(r)$ vs. $r$ for six different values of $\omega$, where we use $\Lambda=\Omega=0.05$, $B_0=0.1$, and $m_j=1/2$. So, according to the figure, we see that the probability increases with the increase of $\omega$. Therefore, we have $P(\omega=0.6)>P(\omega=0.5)>P(\omega=0.4)>P(\omega=0.3)>P(\omega=0.2)>P(\omega=0.1)$; consequently, this implies that the highest probability of finding the particle (in $r\approx 1.72$) is when the (or its) angular frequency has the value of $\omega=0.6$. In other words, the height of the Gaussian-like curves increases as $\omega$ increases. Furthermore, these curves move to the left (or ever closer to the origin) as $\omega$ increases. However, the width of these curves decreases as $\omega$ increases, that is, $\Delta r (\omega=0.1)>\Delta r (\omega=0.2)>\Delta r (\omega=0.3)>\Delta r (\omega=0.4)>\Delta r (\omega=0.5)>\Delta r (\omega=0.6)$.

\begin{figure}[!h]
\centering
\includegraphics[width=9.0cm]{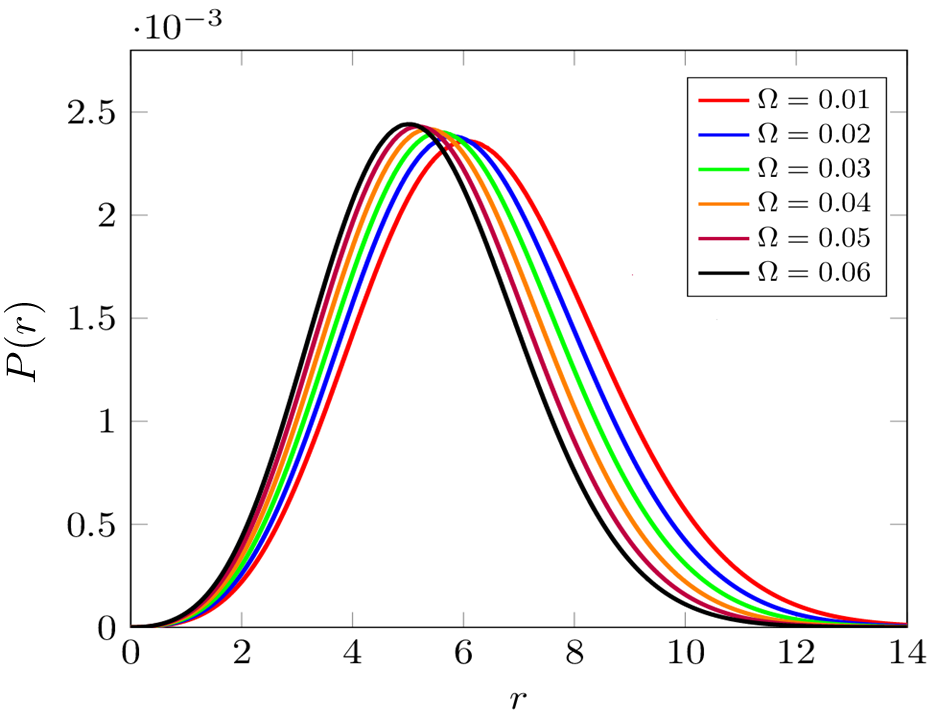}
\caption{Behavior of $P(r)$ vs. $r$ for six different values of $\Omega$.}
\label{fig12}
\end{figure}

In Fig. \ref{fig12}, we have the behavior of $P(r)$ vs. $r$ for six different values of $\Omega$, where we use $\Lambda=0.05$, $\omega=B_0=0.1$, and $m_j=1/2$. So, according to the figure, we see that the probability increases with the increase of $\Omega$. Therefore, we have $P(\Omega=0.06)>P(\Omega=0.05)>P(\Omega=0.04)>P(\Omega=0.03)>P(\Omega=0.02)>P(\Omega=0.01)$; consequently, this implies that the highest probability of finding the particle (in $r\approx 5.02$) is when the angular velocity has the value of $\Omega=0.06$. In other words, the height of the Gaussian-like curves increases as $\Omega$ increases. Furthermore, these curves move to the left (or ever closer to the origin) as $\Omega$ increases. However, the width of these curves decreases as $\Omega$ increases, that is, $\Delta r (\Omega=0.01)>\Delta r (\Omega=0.02)>\Delta r (\Omega=0.03)>\Delta r (\Omega=0.04)>\Delta r (\Omega=0.05)>\Delta r (\Omega=0.06)$.

\begin{figure}[!h]
\centering
\includegraphics[width=9.0cm]{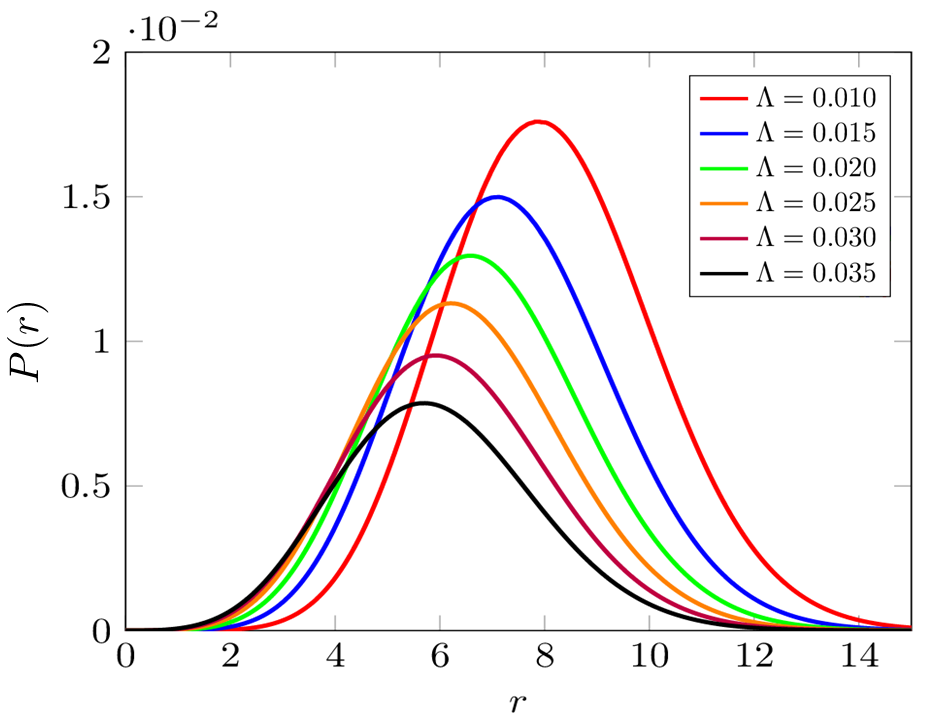}
\caption{Behavior of $P(r)$ vs. $r$ for six different values of $\Lambda$.}
\label{fig13}
\end{figure}

Already in Fig. \ref{fig13}, we have the behavior of $P(r)$ vs. $r$ for six different values of $\Lambda$, where we use $\Omega=0.05$, $\omega=B_0=0.1$, and $m_j=1/2$. So, according to the figure, we see that the probability decreases with the increase of $\Lambda$. Therefore, we have $P(\Lambda=0.010)>P(\Lambda=0.015)>P(\Lambda=0.020)>P(\Lambda=0.025)>P(\Lambda=0.030)>P(\Lambda=0.035)$; consequently, this implies that the highest probability of finding the particle (in $r\approx 8$) is when the cosmological constant has the value of $\Lambda=0.010$. In other words, the height of the Gaussian-like curves decreases as $\Lambda$ increases. Furthermore, these curves move to the left (or ever closer to the origin) as $\Lambda$ increases. However, the width of these curves decreases as $\Lambda$ increases, that is, $\Delta r (\Lambda=0.010)>\Delta r (\Lambda=0.015)>\Delta r (\Lambda=0.020)>\Delta r (\Lambda=0.025)>\Delta r (\Lambda=0.030)>\Delta r (\Lambda=0.035)$.

%-------------------------------------------------------------------------

\section{Conclusions}\label{sec5}

In this paper, we determine the relativistic bound-state solutions for the charged DO in a rotating frame in the Bonnor-Melvin-Lambda spacetime in $(2+1)$-dimensions, where such solutions are given by the two-component Dirac spinor and by the relativistic energy spectrum or relativistic LLs (i.e. the solutions of an eigenvalues equation). With the normalized spinor in hand, we were also able to analyze another important result (often neglected in the literature), which is the radial probability density. To achieve our goal, we work with the curved DO with minimal coupling in relativistic polar coordinates, where the formalism used to write the DO in a curved spacetime was the tetrad formalism of GR. In addition, we also consider the ``spin'' of the (2D) planar fermion, described by a parameter $s$, called the spin parameter. To analytically solve our problem, we consider two approximations, where the first is that the cosmological constant is very small (i.e. we adopt a conical approximation), and the second is that the linear velocity of the rotating frame is much less than the speed of light (i.e. we adopted a slow rotation regime). 

So, by defining a stationary ansatz for the spinor, we obtain two coupled first-order differential equations, i.e. differential equations that contain both two components of the spinor. Therefore, by substituting one equation into the other and vice-versa (i.e. decoupling the equations), we obtain a second-order differential equation for each component, where each equation depends on a quantity given by $Ms$, being $M$ an effective total magnetic quantum number (whose origin comes from the angular part of the spinor). In particular, $Ms>0$ can be ``interpreted'' as a particle with both positive angular momentum and spin ($M>0$ and $s=+1$), or with both negative angular momentum and spin ($M<0$ and $s=-1$), while $Ms<0$ can be ``interpreted'' as a particle with positive angular momentum and negative spin ($M>0$ and $s=-1$), or with negative angular momentum and positive spin ($M<0$ and $s=+1$), respectively. In other words, all our results depend on the value/sign of $Ms$. Thus, to analytically solve the second-order differential equations, we use a procedure based on a change of variable and asymptotic behavior. Consequently, we obtain from this a generalized Laguerre equation (whose solutions are the generalized Laguerre polynomials) as well as the relativistic energy spectrum for the DO (or bound states of the system).

With respect to the spectrum, we note that in addition to describing the positive-energy states/solutions (electron or DO) as well as the negative-energy states/solutions (positron or anti-DO), it is quantized in terms of the radial and total magnetic quantum numbers $n=0,1,2,\ldots$ and $m_j=\pm 1/2,\pm 3/2,\pm 5/2,\ldots$, and explicitly depends on the angular frequency $\omega$ (describes the DO), cyclotron frequency $\omega_c=eB_0/m_0$ (describes the external magnetic field), angular velocity $\Omega$ (describes the rotating frame), spin parameter $s=\pm 1$ (describes the ``spin''), spinorial parameter $u=\pm 1$ (describes the two components of the spinor), effective rest mass $m_{eff}=m_0+\sqrt{\frac{\Lambda}{2}}\sigma\Omega$ (describes the rest mass modified by the spin-rotation coupling term), and on the curvature parameter $\sigma$ and cosmological constant $\Lambda$ (describes the Bonnor-Melvin-Lambda spacetime), respectively. Additionally, the energies of the particle and antiparticle are not equal, i.e. the spectrum is asymmetrical (a direct consequence due to the presence of $\Omega$). Therefore, this asymmetry in energy levels does
not emphasize the equilibrium between particle and antiparticle in the system. Furthermore, the presence of $s$ allows defining two ranges for $\omega_c$ and $\omega$ (this avoids complex energies), that is, for $s=+1$ we have $0\leq\omega_c<2\omega$ and $\omega_c/2<\omega<\infty$, and for $s=-1$, we have $0\leq\omega_c<\infty$ and $0<\omega<\infty$.

So, analyzing the spectrum according to the values of $Ms$ (i.e. each component of the spinor has two associated spectra depending on the values of $Ms$), we note that for $Ms<0$, the two spectra generated are the same, that is, the spectrum is the same for both two components (or is independent of the chosen component). In this case, as the parameter $\sigma$ (or $\sqrt{2\Lambda}\sigma$) is ``tied/linked'' to the quantum number $m_j$, we say that such spectra have their degeneracies broken (such as happens in the case of cosmic strings or Gödel-type spacetime). In other words, the presence of $\sigma$ (or $\sqrt{2\Lambda}\sigma$) breaks/destroys the degeneracy of all the energy levels (i.e. there is no longer a well-defined degeneracy). Already for $Ms>0$, we note that the two spectra generated now are not the same, that is, the spectrum is different for each component (or dependent on the chosen component). In this case, the spectrum for $Ms>0$ and $u=+1$ is the only one whose ground state ($n=0$) does not depend on $m_j$, $\omega$, and $\omega_c$. However, this ground state should not be accepted here because it implies in a non-normalizable solution. Besides, comparing our spectrum with other works/papers, we verified that it generalizes several (particular) cases in the literature \cite{Strange,Villalba1,Villalba2,Haldane,Schakel,Miransky,Bermudez,Bruckmann}.

Subsequently, we graphically analyze the behavior of the spectrum as a function of $B_0$, $\omega$, $\Omega$, and $\Lambda$ for three different values of $n$ and $m_j$, where $B_0$ is the strength of the magnetic field. In particular, while $n$ varies ($n=0,1,2$), $m_j$ remains fixed ($m_j=\pm 1/2$), i.e. we have our first scenario, and while $m_j$ varies ($m_j=\pm 1/2,\pm 3/2,\pm 5/2$), $n$ remains fixed ($n = 0$), i.e. we have our second scenario, respectively. In other words, the behavior of the spectrum was analyzed for both scenarios, where the first scenario is for $n=0,1,2$ with $m_j=+1/2$ and $s=-1$, and $m_j=-1/2$ and $s=+1$, while the second scenario is for $n=0$ with $m_j=+1/2,+3/2,+5/2$ and $s=-1$, and $m_j=-1/2,-3/2,-5/2$ and $s=+1$. For example, starting with $\vert E^\pm\vert$ vs. $B_0$ (energy versus magnetic field), we note that the energies increase with the increase of both $n$ and $m_j$ (Figs. \ref{fig1} and \ref{fig2}). However, in the case of the antiparticle this does not always happen, that is, there is a forbidden region for it when $m_j<0$ and $s=+1$ (Figs. \ref{fig1}-(b) and \ref{fig2}-(b)). Indeed, this happens because their energies decrease with the increase of $n$ (quantumly incoherent). Furthermore, for $m_j>0$ and $s=-1$ (Figs. \ref{fig1}-(a) and \ref{fig2}-(a)), the energies of the antiparticle are greater than those of the particle, while for $m_j<0$ and $s=+1$ (Figs. \ref{fig1}-(b) and \ref{fig2}-(b)), the opposite occurs. Now, for $m_j>0$ and $s=-1$ (Figs. \ref{fig1}-(a) and \ref{fig2}-(a)), the energies (both particle and antiparticle) increase as a function of $B_0$ (i.e. $B_0$ aims to increase energies), while for $m_j<0$ and $s=+1$ (Figs. \ref{fig1}-(b) and \ref{fig2}-(b)), the opposite occurs (i.e. $B_0$ aims to decrease energies).

In the graph $\vert E^\pm\vert$ vs. $\omega$ (energy versus angular frequency), we note that the energies (in both scenarios) increase with the increase of both $n$ and $m_j$ (Figs. \ref{fig3} and \ref{fig4}), and also as a function of $\omega$ (i.e. $\omega$ aims to increase energies). However, for $m_j>0$ and $s=-1$ (Figs. \ref{fig3}-(a) and \ref{fig4}-(a)), the energies of the antiparticle are greater than those of the particle, while for $m_j<0$ and $s=+1$ (Figs. \ref{fig3}-(b) and \ref{fig4}-(b)), the opposite occurs. Already in the graph $\vert E^\pm\vert$ vs. $\Omega$ (energy versus angular velocity), we note that the energies (in both scenarios) increase with the increase of both $n$ and $m_j$ (Figs. \ref{fig5} and \ref{fig6}). Furthermore, for $m_j>0$ and $s=-1$ (Figs. \ref{fig5}-(a) and \ref{fig6}-(a)), the energies of the antiparticle are greater than those of the particle, while for $m_j<0$ and $s=+1$ (Figs. \ref{fig5}-(b) and \ref{fig6}-(b)), the opposite occurs. Now, for $m_j>0$ and $s=-1$ (Figs. \ref{fig5}-(a) and \ref{fig6}-(a)), the energies of the particle decrease as a function of $\Omega$ (i.e. $\Omega$ aims to decrease energies of the particle), while the energies antiparticle increase (i.e. $\Omega$ aims to increase energies of the antiparticle). On the other hand, for $m_j<0$ and $s=+1$ (Figs. \ref{fig5}-(b) and \ref{fig6}-(b)), the energies of the particle increase as a function of $\Omega$, while the energies antiparticle decrease. In the graph $\vert E^\pm\vert$ vs. $\Lambda$ (energy versus cosmological constant), we note that the energies (in both scenarios) increase with the increase of both $n$ and $m_j$ (Figs. \ref{fig7} and \ref{fig8}), however, decrease as a function of $\Lambda$ (i.e. $\Lambda$ aims to increase energies). However, since $\Lambda\ll 1$ allows us to define a cosmic string-like curvature parameter, it implies that the energies increase with the curvature (i.e. increase with increasingly smaller values of $\Lambda$). For $m_j>0$ and $s=-1$ (Figs. \ref{fig7}-(a) and \ref{fig8}-(a)), the energies of the antiparticle are greater than those of the particle, while for $m_j<0$ and $s=+1$ (Figs. \ref{fig7}-(b) and \ref{fig8}-(b)), the opposite occurs. 

Finally, we conclude our work through the graphical analysis of the behavior of the probability density $P(r)$ (as a function of $r$) for six different values of $m_j$, $B_0$, $\omega$, $\Omega$, and $\Lambda$, with $n=0$ (ground state). So, we note that the probability can decrease or increase with the increase of $m_j$ (Fig. \ref{fig9}), that is, it increases from $m_j=1/2$ to $m_j=3/2$, and then decreases with the increase of $m_j$. In particular, the highest probability of finding the particle is when its angular momentum has the value of $m_j=3/2$. Furthermore, the Gaussian-like curves move further and further away from the origin as $m_j$ increases. On the other hand, we note that the probability increases with the decreases of $B_0$ and $\Lambda$ (Figs. \ref{fig10} and \ref{fig13}), however, increases with the increase of $\omega$ and $\Omega$ (Figs. \ref{fig11} and \ref{fig12}). In particular, the highest probability of finding the particle is for $B_0=0.1,0.2$, $\Lambda=0.010$, $\omega=0.6$, and $\Omega=0.06$. Besides, the curves move further and further away from the origin as $B_0$ and $\Lambda$ increase, and ever closer to the origin as $\omega$ and $\Omega$ increase.

%-------------------------------------------------------------------------
\section*{Acknowledgments}

\hspace{0.5cm}

The author would like to thank the anonymous referees for the careful reading of the paper as well as for the remarkable suggestions and constructive comments that substantially helped in improving the quality of the paper. In addition, the author would like to thank the Conselho Nacional de Desenvolvimento Cient\'{\i}fico e Tecnol\'{o}gico (CNPq) for financial support through the postdoc grant No. 175392/2023-4, and also to the Department of Physics at the Universidade Federal da Paraíba (UFPB) for hospitality and support.

%-------------------------------------------------------------------------
\section*{Data availability statement}

\hspace{0.5cm} There is no data because this paper is a totally/purely theoretical work based on analytical calculations of the relativistic bound-state solutions (two-component Dirac spinor and energy spectrum) for the charged DO in a rotating frame in the Bonnor-Melvin-Lambda spacetime.] 

%-------------------------------------------------------------------------

\appendix
\section{Nonrelativistic limit}\label{sec6}

Here, we will take the nonrelativistic limit (or low-energy limit/regime) of the DO and, consequently, we will obtain the Schrödinger oscillator with a strong spin-orbit coupling. In fact, the Schrödinger-Pauli oscillator, since we will still have the total magnetic quantum number $m_j=\pm 1/2,\pm 3/2,\pm 5/2,\ldots$, which in turn depends on $m_l=0,\pm 1,\pm 2,\ldots$ and $m_s=\pm 1/2=u/2$, what are the orbital and spin magnetic quantum numbers. That is, to really have the Schrödinger oscillator, we should only have $m_l$, but $\sqrt{2\Lambda}\sigma$ prevents this from happening, as occurs in the case of the cosmic string \cite{Oli2,Oli3}. Furthermore, unlike the $(3+1)$-dimensional case, where the spin-orbit coupling is given by $\vec{S}\cdot\vec{L}$ \cite{Moshinsky} (see (7a) and (7b) in this paper), here, this does not happen because our two coupled first-order differential equations do not depend on the Pauli matrix $\vec{\sigma}$ (see Eqs. \eqref{EDO1} and \eqref{EDO2}), i.e. because we are working in $(2+1)$-dimensions, or better, now in 2D, that is for the nonrelativistic case (however, in the case of the Zeeman Hamiltonian, we will have $\vec{\sigma}$). So, according to Refs. \cite{Greiner,Moshinsky,Oli2,Oli3,Lawrie,Strange,Andrade,Andrade2}, the nonrelativistic limit of DO can be investigated by considering only the spectrum of the particle and written as $E=m_0+\epsilon$, where $m_0\gg \epsilon$, i.e. most of the particle's energy is concentrated in its rest energy, being $\epsilon$ the nonrelativistic energy spectrum (or nonrelativistic LLs). Besides, it is also necessary to consider that the rest energy is much greater than the energy originated by noninertial/rotation effects, that is: $m_0\gg s\sqrt{8\Lambda}\sigma\Omega N$ and $m_0\gg \sqrt{\frac{\Lambda}{2}}\sigma\Omega$ \cite{Oli2,Oli3}. Therefore, from these conditions, we have
\begin{eqnarray}
&& E^2=(m_0+\epsilon)^2=m^2_0+2m_0\epsilon+\epsilon^2\approx m^2_0+2m_0\epsilon,
\nonumber\\
&& \left(m_0+\sqrt{\frac{\Lambda}{2}}\sigma\Omega\right)^2=m^2_0+2m_0\sqrt{\frac{\Lambda}{2}}\sigma\Omega+\frac{\Lambda}{2}\sigma^2\Omega^2\approx m^2_0+2m_0\sqrt{\frac{\Lambda}{2}}\sigma\Omega,
\nonumber\\
&& \bar{\Omega}=\left(\omega-s\frac{\omega_c}{2}+s\frac{\sqrt{2\Lambda}\sigma\Omega E}{m_0}\right)=\left(\omega-s\frac{\omega_c}{2}+s\frac{\sqrt{2\Lambda}\sigma\Omega(m_0+\epsilon)}{m_0}\right)\approx \left(\omega-s\frac{\omega_c}{2}+s\sqrt{2\Lambda}\sigma\Omega\right).
\end{eqnarray}

Consequently, the nonrelativistic limit of Eq. \eqref{dirac9} is given by (with $\psi^u_{Ms}(r)\to \sqrt{r}\varphi^u_{Ms}(r)$)
\begin{equation}
\left[\frac{d^2}{dr^2}+\frac{1}{r}\frac{d}{dr}-\frac{(M-\frac{su}{2})^2}{r^2}+(m_0\omega_1 r)^2+2m_0\epsilon-2m_0\sqrt{\frac{\Lambda}{2}}\sigma\Omega+m_0\left(\omega-s\frac{\omega_c}{2}+s\sqrt{2\Lambda}\sigma\Omega\right)(2Ms+u)\right]\varphi^u_{Ms}(r)=0,
\end{equation}
or better
\begin{equation}\label{NR}
\left[\frac{d^2}{dr^2}+\frac{1}{r}\frac{d}{dr}-\frac{l_u^2}{r^2}+(m_0\omega_1 r)^2+2m_0\epsilon-2m_0\sqrt{\frac{\Lambda}{2}}\sigma\Omega+2sm_0\omega_2 (l_u+su)+2m_0\sqrt{2\Lambda}\sigma\Omega (l_u+su)\right]\varphi^u_{Ms}(r)=0,
\end{equation}
where we define
\begin{equation}
\omega_1 \equiv \left(\omega-s\frac{\omega_c}{2}+s\sqrt{2\Lambda}\sigma\Omega\right), \ \ \omega_2 \equiv\left(\omega-s\frac{\omega_c}{2}\right), \ \ l_u\equiv M-\frac{su}{2},
\end{equation}
being $l_u=l_{eff}$ a new effective quantum number (with $M=m_j/\sqrt{2\Lambda}\sigma$).

So, we can obtain from Eq. \eqref{NR} the following Schrödinger-Pauli oscillator with a strong spin-orbit coupling term in a rotating frame in the Bonnor-Melvin-Lambda spacetime (or better, 2D Bonnor-Melvin-Lambda space)
\begin{equation}
\left[H^{2D}_{QHO}+H^{2D}_{Zeeman}+H^{2D}_{spin-orbit}+H^{2D}_{Mashhoon}+H^{2D}_{Page-Werner}-\omega \sigma_3\right]\Psi_{Pauli}=i\frac{\partial\Psi_{Pauli}}{\partial t},
\end{equation}
where we define
\begin{eqnarray}
&& H^{2D}_{QHO}\equiv-\frac{1}{2m_0}\left(\frac{d^2}{dr^2}+\frac{1}{r}\frac{d}{dr}-\frac{L_z^2}{r^2}\right)-\frac{1}{2}m_0\omega^2_1 r^2, \ \ H^{2D}_{Zeeman}\equiv\frac{\omega_c}{2}L_z+s\frac{\omega_c}{2}\sigma_3=-(\vec{\mu}_l+s\vec{\mu}_s)\cdot\vec{B}, \ \ L_z=-i\partial_\theta,
\nonumber\\
&& H^{2D}_{spin-orbit}\equiv-s\omega L_z, \ \ H^{2D}_{Mashhoon}\equiv-\sqrt{\frac{\Lambda}{2}}\sigma\Omega, \ \ H^{2D}_{Page-Werner}\equiv-\sqrt{2\Lambda}\sigma\Omega L_z=-\sqrt{2\Lambda}\sigma\vec{\Omega}\cdot\vec{L},
\end{eqnarray}
being $H^{2D}_{QHO}$ the Hamiltonian of the 2D quantum
harmonic oscillator (2DQHO) (in a uniform magnetic field) \cite{Andrade,Andrade2}, $H^{2D}_{Zeeman}$ is the (anomalous) Zeeman Hamiltonian (where $\vec{\mu}_l=-\frac{e}{2m_0}\vec{L}$ and $\vec{\mu}_s=-\frac{e}{m_0}\vec{S}$ are the orbital and spin magnetic dipole moment vectors, with $\vec{L}$ and $\vec{S}=\frac{1}{2}\vec{\sigma}$ being the orbital angular momentum and spin operators/vectors \cite{Griffiths}), $H^{2D}_{spin-orbit}$ is the Hamiltonian of the spin-orbit coupling or simply the spin-orbit Hamiltonian \cite{Andrade,Andrade2} (in 3D, would be $H^{3D}_{spin-orbit}=-2\omega\vec{S}\cdot\vec{L}$ \cite{Moshinsky,Strangebook}), $H^{2D}_{Mashhoon}$ (in 3D, would be $H^{3D}_{Mashhoon}=-\vec{\Omega}\cdot\vec{S}$ \cite{Hehl}) and $H^{2D}_{Page-Werner}$ are the Mashhoon and Page-Werner Hamiltonians (describe the ``nonrelativistic spin-rotation coupling'' and the ``nonrelativistic Sagnac effetc'' \cite{Hehl}), and $\Psi_{Pauli}(t,r,\theta)=e^{-i\epsilon t}e^{il_u\theta}(\varphi^+_{Ms},\varphi^-_{Ms})^T=e^{-i\epsilon t}(e^{il_+\theta}\varphi^+_{Ms},e^{il_-\theta}\varphi^-_{Ms})^T$ is the Pauli spinor or Schrödinger-Pauli wave function (where satisfy $\sigma_3\varphi^u_{Ms}=u\varphi^u_{Ms}$, being $u=+1$ for a particle with spin up and $u=-1$ for a particle with spin down), respectively. In particular, we see that of all Hamiltonians, the only ones not affected/modified by parameters of the Bonnor-Melvin-Lambda space (i.e. $\Lambda$ and $\sigma$) are the spin-orbit and Zeeman Hamiltonians. Furthermore, taking $\Lambda\to 1/2$ and $\sigma\to 1$ (absence of the Bonnor-Melvin-Lambda space) as well as $\Omega\to 0$ (absence of the rotating frame) and $\omega_c\to 0$ (absence of the magnetic field or Zeeman effect), and defining $l_u=m_j-su/2=m_l+u/2-su/2=m_l+u(1-s)/2\equiv m=0,\pm 1,\pm 2,\ldots$ (in fact, it would be exactly equal to $m$ for $s=+1$), we obtain the Schrödinger-Pauli oscillator of Refs. \cite{Andrade,Andrade2}.

\end{document}